\documentclass{lmcs}
\graphicspath{{figures/}} 

 \usepackage[nocaptions]{vntex}

\usepackage{amsmath, amssymb} \usepackage{stmaryrd} \usepackage{wasysym}
\usepackage{pifont}
\usepackage{thm-restate}

\usepackage{hyperref}
\usepackage{cleveref}
\crefname{thm}{Theorem}{Theorems}
\Crefname{thm}{Theorem}{Theorems}
\crefname{lem}{Lemma}{Lemmas}
\Crefname{lem}{Lemma}{Lemmas}
\crefname{cor}{Corollary}{Corollaries}
\Crefname{cor}{Corollary}{Corollaries}
\crefname{prop}{Proposition}{Propositions}
\Crefname{prop}{Proposition}{Propositions}
\crefname{defi}{Definition}{Definitions}
\Crefname{defi}{Definition}{Definitions}
\crefname{exa}{Example}{Examples}
\Crefname{exa}{Example}{Examples}

\usepackage{multirow}
\usepackage[justification=centering]{subcaption}

\usepackage{quiver}
\usetikzlibrary{decorations.markings}
\tikzset{kleisli/.style={ postaction={decorate, decoration={markings, mark= at position 0.5 with {
        \draw circle[radius=1.5pt]; }}
}}}
\usetikzlibrary{calc}
\makeatletter
\newcounter{tikzcdequation}
\renewcommand*\thetikzcdequation{(\arabic{tikzcdequation})}
\newif\iftikzcd@eq@customtag
\tikzcdset{
  number/.code=\pgfqkeys{/tikz/commutative diagrams/number}{
    /tikz/path only,normal tag,normal at,every number,#1,@do},
  number/every number/.code=,
  number/label/.initial=,
  number/tag/.code=\tikzcd@eq@customtagtrue\def\tikzcd@eq@tag{#1},
  number/normal tag/.code=
    \tikzcd@eq@customtagfalse\def\tikzcd@eq@tag{\thetikzcdequation},
  number/normal  at/.code=\def\tikzcd@eq@at{no at},
  number/at/matrix  center/.code=\def\tikzcd@eq@at{matrix center},
  number/at/diagram center/.code=\def\tikzcd@eq@at{diagram center},
  number/at/.code=\pgfkeysifdefined{/tikz/commutative diagrams/number/at/#1/.@cmd}
      {\pgfkeysvalueof{/tikz/commutative diagrams/number/at/#1/.@cmd}\pgfeov}
      {\def\tikzcd@eq@at{at={#1}}},
  number/@at no at/.code=, number/@at matrix  center/.style={
    commutative diagrams/number/@at at=(\tikzcdmatrixname.center)},
  number/@at diagram center/.style={
    commutative diagrams/number/@at at=(current bounding box.center)},
  number/@at at/.style={pos=,to path=\tikztonodes,nodes={at={#1}}},
  number/@do/.style={
    /tikz/every to/.append style/.expanded={
      commutative diagrams/number/@at \tikzcd@eq@at,
      edge node={
        node[commutative diagrams/eq node/.try,
             commutative diagrams/number/@@do\iftikzcd@eq@customtag*={\unexpanded\expandafter{\tikzcd@eq@tag}}\else=\fi
            {\pgfkeysvalueof{/tikz/commutative diagrams/number/label}}]{}}}},
  number/@@do/.style={
    node contents={
      \refstepcounter{tikzcdequation}\thetikzcdequation
      \IfBlankF{#1}{\@ifpackageloaded{cleveref}{\cref@label{#1}}{\ltx@label{#1}}}}},
  number/@@do*/.style n args={2}{
    /utils/exec=\def\thetikzcdequation{(#1)},commutative diagrams/number/@@do={#2}}}
\makeatother
\newcommand{\diagramlabel}[4][]{\ifx\\#1\\\arrow[from=#2, to=#3, number={label=cd:#4}]\else \arrow[from=#2, to=#3, number={label=cd:#4, tag=#1}]\fi }
\crefname{tikzcdequation}{diagram}{diagrams}
\Crefname{tikzcdequation}{Diagram}{Diagrams}

\newcommand{\diagramref}[4][]{\arrow[#4{description, #1}, draw=none, from=#2, to=#3]}
 \newcommand{\CallCommand}[1]{\csname#1\endcsname}
\newcommand{\ApplyCommand}[3][]{\expandafter\providecommand\csname#2#3\endcsname{\CallCommand{#2}{\CallCommand{#1#3}}}}
\newcommand{\ApplyCommandOption}[3][]{\expandafter\providecommand\csname#2#3\endcsname{\CallCommand{#2}[\CallCommand{#1#3}]}}
\newcommand{\ComposeCommands}[3][]{\expandafter\providecommand\csname#1#2#3\endcsname{\CallCommand{#1#2}\CallCommand{#1#3}}}

\newcommand{\DeclareEnvOfStringDiagrams}[2][]{\newenvironment{#2ofstringdiagrams#1}{\vspace{-10pt}\begincsname#2#1\endcsname}{\begincsname end#2#1\endcsname}\vspace{-50pt}}
\DeclareEnvOfStringDiagrams{equation}
\DeclareEnvOfStringDiagrams[*]{equation}
\DeclareEnvOfStringDiagrams{align}
\DeclareEnvOfStringDiagrams[*]{align}

\makeatletter
\DeclareRobustCommand{\crefnosort}[1]{\begingroup\@cref@sortfalse\cref{#1}\endgroup
}
\DeclareRobustCommand{\Crefnosort}[1]{\begingroup\@cref@sortfalse\Cref{#1}\endgroup
}
\makeatother

\DeclareFontEncoding{LGR}{}{}
\DeclareSymbolFont{sfgreek}{LGR}{cmss}{bx}{n}
\SetSymbolFont{sfgreek}{bold}{LGR}{cmss}{bx}{n}
\DeclareMathSymbol{\betasf}{\mathord}{sfgreek}{`b}
\newcommand{\cmark}{\ding{51}}
\newcommand{\xmark}{\ding{55}}

\newcommand{\complices}{\mathbb{C}}
\newcommand{\naturals}{\mathbb{N}}

\DeclareMathSymbol{:}{\mathpunct}{operators}{"3A}
\newcommand{\mathand}{\mathrel{\mathrm{and}}}
\newcommand{\suchthat}[2]{\left\{ #1 ~\middle|~ #2 \right\}}

\newcommand{\id}{\mathrm{id}}

\newcommand{\cat}[1]{\mathsf{#1}}
\newcommand{\namedCat}[1]{{\boldsymbol{\cat{#1}}}}
\newcommand{\DeclareCat}[1]{\expandafter\newcommand\csname#1\endcsname{\namedCat{#1}}}
\newcommand{\DeclareCatLetter}[1]{\expandafter\newcommand\csname cat#1\endcsname{\cat{#1}}}

\newcommand{\Rel}[1][]{\namedCat{Rel}\ifx\\#1\\\else\mathopen{}\left(#1\right)\mathclose{}\fi}
\DeclareCat{Cat}
\DeclareCat{Surj}
\DeclareCat{FinFib}

\newcommand{\EM}[1]{\namedCat{EM}\mathopen{}\left(#1\right)\mathclose{}}

\newcommand{\Kl}[1]{\namedCat{Kl}\mathopen{}\left(#1\right)\mathclose{}}
\newcommand{\DeclareRegCat}[2][]{\expandafter\providecommand\csname#1#2\endcsname{\namedCat{#2}}\expandafter\providecommand\csname Rel#2\endcsname{\Rel[\namedCat{#2}]}\expandafter\providecommand\csname Grph#2\endcsname{\Grph[\namedCat{#2}]}}
\newcommand{\DeclareRegCatLetter}[1]{\expandafter\providecommand\csname cat#1\endcsname{\cat{#1}}\expandafter\providecommand\csname Rel#1\endcsname{\Rel[\cat{#1}]}\expandafter\providecommand\csname Grph#1\endcsname{\Grph[\cat{#1}]}}
\DeclareRegCatLetter{C}
\DeclareRegCatLetter{D}
\DeclareRegCatLetter{E}
\DeclareRegCat{Set}
\DeclareRegCat{KHaus}
\DeclareRegCat{JSL}
\DeclareRegCat{Conv}
\DeclareRegCat{Mon}
\DeclareRegCat{CMon}
\DeclareCat{Pos}
\DeclareCatLetter{M}
\DeclareCat{Mono}
\DeclareCat{Open}
\newcommand{\catGamma}{\cat{\Gamma}}
\newcommand{\catDelta}{\cat{\Delta}}

\newcommand{\func}[1]{\cat{#1}}
\newcommand{\namedFunc}[1]{\namedCat{#1}}
\newcommand{\DeclareFunc}[1]{\expandafter\newcommand\csname#1\endcsname{\func{#1}}}
\newcommand{\DeclareFuncLetter}[1]{\expandafter\newcommand\csname func#1\endcsname{\func{#1}}}
\DeclareFunc{Id}
\newcommand{\Grph}[1][]{\namedFunc{Graph}_{#1}}
\DeclareFuncLetter{F}
\DeclareFuncLetter{G}
\DeclareFuncLetter{H}
\DeclareFuncLetter{U}
\DeclareFuncLetter{K}
\newcommand{\ext}[2][]{\underline{#2}_{#1}}
\newcommand{\lift}[2][]{\overline{#2}^{#1}}
\newcommand{\semilift}[2][]{\dashover{#2}^{#1}}

\newcommand{\unit}[1][]{\eta^{#1}}
\newcommand{\mult}[1][]{\mu^{#1}}
\newcommand{\counit}[1][]{\varepsilon^{#1}}

\newcommand{\rightEM}[1][]{\namedFunc{U}^{#1}}

\newcommand{\proj}[1][]{\pi^{#1}}
\newcommand{\inc}[1][]{\iota^{#1}}
\newcommand{\DeclareMonad}[3][]{\expandafter\providecommand\csname#1#2\endcsname{#3}
  \ComposeCommands[#1]{#2}{#2}\ComposeCommands[#1]{#2#2}{#2}\ComposeCommands[#1]{#2#2#2}{#2}\ApplyCommandOption[#1]{unit}{#2}\ApplyCommandOption[#1]{mult}{#2}\ApplyCommandOption[#1]{counit}{#2}\ApplyCommand[#1]{monad}{#2}\ApplyCommand[#1]{semimonad}{#2}\ApplyCommand[#1]{Kl}{#2}\ApplyCommand[#1]{EM}{#2}\ApplyCommand[#1]{SemiEM}{#2}\ApplyCommand[#1]{Alg}{#2}\ApplyCommandOption[#1]{leftKl}{#2}\ApplyCommandOption[#1]{rightKl}{#2}\ApplyCommandOption[#1]{leftEM}{#2}\ApplyCommandOption[#1]{leftSemiEM}{#2}\ApplyCommandOption[#1]{rightEM}{#2}\ApplyCommandOption[#1]{rightSemiEM}{#2}\ApplyCommandOption[#1]{Semi}{#2}\ApplyCommandOption[#1]{Split}{#2}\ApplyCommandOption[#1]{compare}{#2}\ApplyCommandOption[#1]{close}{#2}\ApplyCommandOption[#1]{proj}{#2}\ApplyCommandOption[#1]{inc}{#2}}
\newcommand{\DeclareMonadLetter}[1]{\DeclareMonad[func]{#1}{\func{#1}}}
\DeclareMonadLetter{D}
\renewcommand{\funcD}{\namedFunc{D}}
\DeclareMonadLetter{L}
\renewcommand{\funcL}{\namedFunc{L}}
\DeclareMonadLetter{M}
\renewcommand{\funcM}{\namedFunc{M}}
\DeclareMonadLetter{P}
\renewcommand{\funcP}{\namedFunc{P}}
\DeclareMonadLetter{S}
\DeclareMonadLetter{R}
\DeclareMonadLetter{T}
\DeclareMonadLetter{V}
\renewcommand{\funcV}{\namedFunc{V}}
\DeclareMonad{Ult}{\scalebox{.95}{$\betasf$}}
\newcommand{\smallUlt}{\scalebox{.665}{$\betasf$}}
\renewcommand{\unitUlt}{\unit[\smallUlt]}
\renewcommand{\multUlt}{\mult[\smallUlt]}

\renewcommand{\rightEMUlt}{\rightEM[\smallUlt]}

\DeclareMonad{Ms}{\funcM_S}
\DeclareMonad{Rad}{\namedFunc{R}}
\newcommand{\alg}[2]{\left(#1,#2\right)}

\newcommand{\DeclareDistrLaw}[6][]{\expandafter\newcommand\csname distr#2#3name\endcsname{#4}\expandafter\newcommand\csname distr#2#3type\endcsname{#5#6 \Rightarrow #6#5}\expandafter\newcommand\csname distr#2#3\endcsname{\csname distr#2#3name\endcsname: \csname distr#2#3type\endcsname}\expandafter\newcommand\csname #1#3cmp#2\endcsname[1][]{#6 \bullet_{##1} #5}\DeclareMonad[#1]{#3#2}{#6#5}\DeclareMonad{ext#2}{\ext{#5}}\DeclareMonad{lift#3}{{\lift{#6}}}\DeclareMonad{semilift#3}{{\semilift{#6}}}\ComposeCommands[#1]{#2}{#3}\ComposeCommands[#1]{#2#2}{#3}\ComposeCommands[#1]{#2#3}{#3}\ComposeCommands[#1]{#2#3}{#2}\ComposeCommands[#1]{#3#3}{#2}\ComposeCommands[#1]{#3#2}{#2}\ComposeCommands[#1]{#3#2}{#3}}
\newcommand{\DeclareDistrLawLambda}[5][]{\DeclareDistrLaw[#1]{#2}{#3}{\lambda^{{#4}/{#5}}}{#4}{#5}}
\newcommand{\DeclareDistrLawLetter}[3]{\DeclareDistrLaw[func]{#1}{#2}{#3}{\func#1}{\func#2}}
\newcommand{\DeclareDistrLawLetterLambda}[2]{\DeclareDistrLawLambda[func]{#1}{#2}{\csname func#1\endcsname}{\csname func#2\endcsname}}
\newcommand{\DeclareCompDistrLaw}[7][]{\DeclareDistrLaw[#1]{#3#2}{#4}{\csname distr#3#4name\endcsname #5 \circ #6 \csname distr#2#4name\endcsname}{#6#5}{#7}\DeclareDistrLaw[#1]{#2}{#4#3}{#7 \csname distr#2#3name\endcsname \circ \csname distr#2#4name\endcsname #6}{#5}{#7#6}\ComposeCommands[#1]{#2#3}{#4}
  \ComposeCommands[#1]{#3#4}{#2}
  \ComposeCommands[#1]{#4#2}{#3}
  \DeclareDistrLaw{lift#3}{lift#4}{\lift{\csname distr#3#4name\endcsname}}{\csname lift#3\endcsname}{\csname lift#4\endcsname}
  \DeclareDistrLaw{semilift#3}{semilift#4}{\semilift{\csname distr#3#4name\endcsname}}{\csname semilift#3\endcsname}{\csname semilift#4\endcsname}
}
\newcommand{\DeclareCompDistrLawLetter}[3]{\DeclareCompDistrLaw[func]{#1}{#2}{#3}{\func#1}{\func#2}{\func#3}
}

\DeclareDistrLawLetterLambda{P}{P}
\DeclareDistrLawLetterLambda{D}{P}
\DeclareDistrLawLetterLambda{T}{P}
\DeclareDistrLawLetterLambda{S}{P}
\DeclareDistrLawLetterLambda{V}{V}
\DeclareDistrLawLetterLambda{L}{M}
\DeclareDistrLawLetterLambda{L}{P}
\DeclareDistrLawLetterLambda{L}{D}
\DeclareDistrLawLetterLambda{M}{M}
\DeclareDistrLawLetterLambda{M}{P}
\DeclareDistrLawLetterLambda{M}{D}
\DeclareDistrLawLetter{T}{S}{\rho}
\DeclareDistrLawLetter{T}{R}{\sigma}
\DeclareDistrLawLetter{S}{R}{\tau}
\DeclareDistrLawLambda{Ult}{P}{\Ult}{\funcP}
\DeclareDistrLawLambda{Rad}{V}{\Rad}{\funcV}
\DeclareDistrLawLambda{Rad}{Vstar}{\Rad}{\funcV_*}
\DeclareDistrLawLambda{V}{Vstar}{\funcV}{\funcV_*}
\DeclareDistrLawLambda{Ms}{P}{\Ms}{\funcP}
\DeclareDistrLawLambda{L}{Ms}{\funcL}{\Ms}
\DeclareDistrLawLambda{M}{Ms}{\funcM}{\Ms}
\DeclareDistrLawLambda{S}{liftP}{\funcS}{\liftP}
\renewcommand{\distrUltPname}{\lambda^{\smallUlt/\funcP}}
\renewcommand{\distrUltP}{\distrUltPname: \distrUltPtype}
\DeclareCompDistrLawLetter{T}{S}{R}
\DeclareCompDistrLawLetter{T}{S}{P}
\DeclareCompDistrLaw{Ult}{P}{P}{\Ult}{\funcP}{\funcP}

\newcommand{\klarrow}[1][->]{\mathrel{\tikz [line width=.11ex, double distance=.33ex]
    \draw[#1, kleisli]
    (0,0) -- (0.5,0);}}
\newcommand{\adjunction}[4]{#3: #1 \leftrightarrows #2 {{:}#4}
}
\newcommand{\onto}{\twoheadrightarrow}
\newcommand{\subto}{\hookrightarrow}

\newcommand{\abs}[1]{\left| #1 \right|}
\newcommand{\norm}[1]{\abs{\abs{#1}}}
\newcommand{\pullbackcorner}{\scalebox{1.5}{$\lrcorner$}}

\makeatletter
\newcommand*{\inlineequation}[3][]{\begingroup
\refstepcounter{equation}\@eqnnum~
    \label{#2}\ifx\\#1\\\else
      \tag{#1}\fi \relpenalty=10000 \binoppenalty=10000 \ensuremath{#3}\endgroup
}
\makeatother

\makeatletter
\newcommand{\dashover}[2][\mathop]{#1{\mathpalette\df@over{{\dashfill}{#2}}}}
\newcommand{\df@over}[2]{\df@@over#1#2}
\newcommand\df@@over[3]{\vbox{
    \offinterlineskip
    \ialign{##\cr
      #2{#1}\cr
      \noalign{\kern1pt}
      $\m@th#1#3$\cr
    }
  }}
\newcommand{\dashfill}[1]{\kern-.5pt
  \xleaders\hbox{\kern.5pt\vrule height.5pt width \dash@width{#1}\kern.5pt}\hfill
  \kern-.5pt
}
\newcommand{\dash@width}[1]{.9pt
}
\makeatother
\title[Monotone Weak Distributive Laws in Categories of Algebras]{Monotone Weak
Distributive Laws over the Lifted Powerset Monad in Categories of Algebras}

\author[Q.~Aristote]{Quentin Aristote\lmcsorcid{0009-0001-4061-7553} }
\address{Université Paris Cité, CNRS, Inria, IRIF, F-75013,
  Paris,
  France}
\email{quentin.aristote@irif.fr}


\thanks{For sometimes small but always fruitful discussions on topics
  related to this work, the author thanks Gabriella B\"{o}hm, Victor Iwaniack,
  Jean Goubault-Larrecq, Alexandre Goy, Daniela Petri\c{s}an and Sam van
  Gool.}  
\begin{document}

\maketitle
\begin{abstract}
  In both the category of sets and the category of compact Hausdorff spaces,
  there is a monotone weak distributive law that combines two layers of
  non-determinism. Noticing the similarity between these two laws, we study
  whether the latter can be obtained automatically as a weak lifting of the
  former. This holds partially, but does not generalize to other categories of
  algebras. We then characterize when exactly monotone weak distributive laws
  over powerset monads in categories of algebras exist, on the one hand
  exhibiting a law combining probabilities and non-determinism in compact
  Hausdorff spaces and showing on the other hand that such laws do not exist in
  a lot of other cases.
\end{abstract}

\section{Introduction}\label{sec:introduction}

In the study of the semantics of programming languages, and since the seminal
work of Moggi~\cite{moggiComputationalLambdacalculusMonads1989}, effectful
computations are usually modeled with monads: an effectful function of type $X
\klarrow Y$ is interpreted as a function of type $X \to \funcT Y$, where the
monadic structure on $\funcT$ allows for having identities and compositions of
such effectful functions. When considering several effects at the same time, a
natural question arises: given monads corresponding to two effects, is it
possible to construct a monad that corresponds to the combination of these two
effects? In particular, combining probabilities and non-determinism has been a
very popular subject of study in the topological and (dually) domain-theoretic
settings: see for instance the introduction
of~\cite{keimelMixedPowerdomainsProbability2017} for an extensive bibliography
on the topic.

The most straightforward way to combine two monads $\funcS$ and $\funcT$ would
be to compose their underlying functors, but unfortunately in general the
resulting endofunctor $\funcST$ may not carry the structure of a monad. For this
to hold, a sufficient condition is the existence of a \emph{distributive law} of
$\funcT$ over $\funcS$, i.e., a natural transformation $\distrTStype$ satisfying
four axioms involving the units and multiplications of the two
monads~\cite{beckDistributiveLaws1969}. Such a distributive law makes $\funcST$
into a monad, and its data is equivalent to the data of a \emph{lifting} of
$\funcS$ to the Eilenberg-Moore category $\EMT$ of $\funcT$-algebras or to the
data of an \emph{extension} of $\funcT$ to the Kleisli category $\KlS$ of free
$\funcS$-algebras.

Unfortunately, distributive laws turn out to be not so common: proving that some
specific pairs of monads do not admit any distributive law between them has been
the focus of several
works~\cite{klinIteratedCovariantPowerset2018,varaccaDistributingProbabilityNondeterminism2006,dahlqvistCompositionalSemanticsNew2018},
culminating in~\cite{zwartNoGoTheoremsDistributive2019} where general techniques
for proving the absence of distributive laws between monads on $\Set$, so-called
``no-go theorems'', are exhibited. Among the culprits are the powerset monad
$\funcP$ and the probability distributions monad $\funcD$: there is no
distributive law $\distrPPtype$ nor $\distrDPtype$.

A next step in combining monads is thus to weaken the requirements asked for by
distributive laws. In~\cite{bohmWeakTheoryMonads2010} a 2-categorical theory of
such weakened distributive laws, where the axioms relating to the units of the
monads are relaxed, is developed. This theory encompasses two orthogonal kinds
of weakened laws, both called weak distributive laws: those
of~\cite{streetWeakDistributiveLaws2009}, extensively studied
in~\cite{bohm2CategoriesWeakDistributive2011,bohmIdempotentSplittingsColimit2012,bohmIterationWeakWreath2012},
and those more recently put light on in~\cite{garnerVietorisMonadWeak2019} and
given further instances of
in~\cite{goyCombiningProbabilisticNondeterministic2020,goyPowersetLikeMonadsWeakly2021,bonchiConvexityWeakDistributive2022}.
This work focuses on the latter kind of laws: in the following a \emph{weak
distributive law} of $\funcT$ over $\funcS$ will thus be a natural
transformation of type $\distrTStype$ (again), where only three of the four
axioms of distributive laws are required. Such a weak distributive law need not
make $\funcST$ into a monad, but does so of a retract thereof when the category
is well-behaved. Weak laws $\distrTStype$ are equivalent to \emph{weak
extensions} of $\funcT$ to $\KlS$ --- extensions of the semi-monad underlying
$\funcT$, i.e., of its endofunctor and multiplication but not of its unit ---
or, when the category is well-behaved, \emph{weak liftings} of $\funcS$ to
$\EMT$ --- liftings of $\funcS$ up to a retraction.

Weak extensions are a precious tool for building weak distributive laws, because
\emph{monotone} extensions to $\Rel$ --- the category of sets and relations,
which also happens to be the category of free algebras of the powerset monad
$\funcP$ on $\Set$ --- admit a nice characterization: this makes it possible to
find weak distributive laws without having to guess their formul\ae{} anymore,
and the monotonicity of the extension is a strong indicator that the resulting
law will be semantically interesting. In the context of weak distributive laws,
this strategy was first used by Garner~\cite{garnerVietorisMonadWeak2019}, who
exhibited a law $\distrPPtype$, but also a law $\distrUltPtype$, combining the
ultrafilter monad $\Ult$ and the powerset monad, and whose corresponding weak
lifting turned out to be the Vietoris monad $\funcV$ of closed subsets on the
category $\KHaus$ of compact Hausdorff spaces (the algebras of
$\Ult$~\cite{manesTripleTheoreticConstruction1969}). The same strategy was then
used for instance in~\cite{goyCombiningProbabilisticNondeterministic2020}, where
a law $\distrDPtype$, combining probabilities and non-determinisim, is
described.

Most non-trivial weak distributive laws in the literature (so-called
\emph{trivial} ones are described in~\cite[\textsection
2.2]{goyCompositionalityMonadsWeak2021}) are laws of type $\distrTPtype$ built
using this strategy. In fact, it was recently shown
in~\cite{goncharovOnlyDistributiveLaw2026} that for any \emph{accessible} monad
$\funcT$ on $\Set$, there is at most one weak distributive law $\distrTPtype$,
and if it exists the corresponding weak extension is monotone. A weak
distributive law $\distrVVtype$, combining two layers of non-determinism in
$\KHaus$, was similarly built by constructing a monotone weak
extension~\cite{goyPowersetLikeMonadsWeakly2021}; this law's formula is in
particular very close to that of the law $\distrPPtype$ in $\Set$.
Non-determinism can also be combined with probabilities in $\KHaus$ and other
continuous settings: weak distributive laws for this purpose are constructed by
hand in a recent pre-print~\cite{goubault-larrecqWeakDistributiveLaws2024}.

The goal of the present work is to take over the program of finding
non-$\Set$-based weak distributive laws: we focus here on categories of
algebras, which fit in the general framework for monotone weak laws presented
in~\cite{goyPowersetLikeMonadsWeakly2021}. In particular, we notice in
\Cref{lem:weak-distr-law-VV-weak-lifting-PP} that the law $\distrVVtype$ is
not only very similar to the law $\distrPPtype$, but is also actually some sort
of weak lifting of it. We thus study whether there is a general framework for
not only weakly lifting monads (as weak distributive laws do), but also weakly
lifting weak distributive laws themselves: this framework should yield or
simplify the construction of the law $\distrVVtype$, and hopefully generalize it
to other categories of algebras, in particular $\EMP$ and $\EMD$ which have
weakly lifted powerset monads thanks to the laws $\distrPPtype$ and
$\distrDPtype$.

This question is largely related to the problem of composing weak distributive
laws, which was investigated in~\cite{goyWeakeningIteratingLaws2023} from the
point of view of the Yang-Baxter equation --- the usual tool for composing and
lifting plain distributive laws~\cite{chengIteratedDistributiveLaws2011}. We end
up getting a general ``no-go theorem'' for monotone weak laws over weakly lifted
powerset monads: in that sense this work is also close in spirit
to~\cite{zwartNoGoTheoremsDistributive2019}, where general no-go theorems for
(strict) distributive laws are given. While not restricted to monotone
distributive laws, these theorems are unlikely to generalize to our setting
because they are based on the correspondence between monads and algebraic
theories, which is mostly restricted to $\Set$ and does not have any obvious
generalization to semi-monads --- to which weak distributive laws are deeply
related.

This article is organized as follows. In \Cref{sec:preliminaries}, we recall
definitions and notations for and give examples of monads and weak distributive
laws. This culminates in \Cref{lem:weak-distr-law-VV-weak-lifting-PP}, where we
notice that the law $\distrVVtype$ is some sort of weak lifting of the law
$\distrPPtype$. The next sections focus on lifting weak distributive laws: in
\Cref{sec:lifting-laws}, we study the approach of the Yang-Baxter equation,
showing it indeed allows for weakly lifting weak laws but does not apply to the
examples we consider. To focus next on monotone weak distributive laws,
\Cref{sec:relational-extensions} recalls the theory of regular categories and
how to construct extensions to relations therein. \Cref{sec:lifted-kleisli} then
characterizes the Kleisli category of weakly lifted powerset monads as
subcategories of relations, and this is finally applied in
\Cref{sec:lifting-monotone-laws} to obtain a simple characterization for the
existence of monotone weak laws in categories of algebras. We conclude in
\Cref{sec:conclusion}.

Our main contributions are the following:
\begin{itemize}
\item we show that, while the law $\distrVVtype$ is a kind of weak lifting of
  the law $\distrPPtype$ (\Cref{lem:distrVV-weak-proj-lifting-distrPP}), this
  lifting does not come from a Yang-Baxter equation
  (\Cref{cor:no-go-weak-law-weak-lifting}), the usual approch to lifting laws:
  it is an instance of a general no-go theorem for Yang-Baxter equations
  involving the law $\distrPPtype$
  (\Cref{prop:no-go-weak-law-weak-inc-lifting});
\item we characterize the Kleisli category of the weak lifting $\liftS$ of a
  monad $\funcS$ to the algebras of a monad $\funcT$ as the category of
  $\funcT$-algebras and $\KlS$-morphisms between them
  (\Cref{prop:kleisli-category-lifted-monad});
\item we give a characterization of the Kleisli categories of weakly lifted
  powerset monads as subcategories of relations
  (\Cref{thm:kleisli-lifted-subpowerobject}), in which a certain class of
  \emph{open morphisms} play a central role: it follows that the
  corresponding categories of algebras have \emph{open subobject
    classifiers} (\Cref{prop:open-subobject-classifiers});
\item we characterize when monotone extensions of functors and natural
  transformations to such subcategories of relations
  exist~\Cref{thm:partial-relational-extensions},
  and deduce that monads must preserve these open morphisms to have monotone
  weak distributive laws over weakly lifted powerset monads --- this is in fact
  a sufficient condition for monads that are themselves weak liftings
  (\Cref{cor:monotone-weak-distr-laws-lifted-powerset});
\item concrete instances of this last result are then easily derived: we recover
  independently the law combining probabilities and non-determinism in compact
  Hausdorff spaces and recently exhibited
  in~\cite{goubault-larrecqWeakDistributiveLaws2024}
  (\Cref{thm:radon-monotone-weak-distr-law}), but we observe otherwise that
  monotone weak distributive laws over weakly lifted powerset monads in
  categories of algebras seem very rare
  (\Cref{tab:monotone-weak-distr-laws-weakly-lifted-powerset-algebras}).
\end{itemize}
The present article is an extended version of an article presented at STACS
2025~\cite{aristoteMonotoneWeakDistributive2025}, in which the discussion in
\Cref{sec:lifted-kleisli:subobject-classifier} and the proofs were not included.
The presentation has also been improved.

\section{Preliminary: Weak Distributive Laws}\label{sec:preliminaries}

In this first section we briefly recall the theory of weak distributive laws,
from~\cite{garnerVietorisMonadWeak2019,bohmWeakTheoryMonads2010}. The reader is
assumed to be familiar with the basics of
category theory, an introduction to which appearing for instance
in~\cite{borceuxHandbookCategoricalAlgebra1994}.

\subsection{Monads and (Weak) Distributive Laws}\label{sec:preliminaries:distributive-laws}

\begin{defi}[monad]\label{def:monad}
  A \emph{monad} on a category $\catC$ is the data $\monadT$, often abbreviated
  $\funcT$, of an endofunctor $\funcT : \catC \to \catC$ and natural
  transformations $\unitT : \Id \Rightarrow \funcT$ (the \emph{unit}) and
  $\multT: \funcTT \Rightarrow \funcT$ (the \emph{multiplication}) satisfying
  the axioms $\multT \circ \funcT \unitT = \id = \multT \circ \unitT \funcT$ and
  $\multT \circ \funcT \multT = \multT \circ \multT \funcT$.
\end{defi}
\begin{exa}[monads]\label{ex:monad} In this work we will be particularly concerned with the
  following monads on $\Set$. Thereafter $X$ and $Y$ are sets, $x$ is an
  element of $X$ and $f: X \to Y$ is a function.
  \begin{description}
  \item[The \emph{powerset} monad $\funcP$] $\funcP X = \{ \textrm{subsets of }
    X \}$; for $e \subseteq X$, $(\funcP f)(e) = f[e] = \\ \suchthat{ f(x') }{ x'
      \in e }$; $\unitP_X(x) = \{ x \}$; for $E \subseteq \funcP X$,
    $\multP_X(E) = \bigcup E$.
  \item[The \emph{probability distributions} monad $\funcD$] $\funcD X$ is the
    set of finitely-supported probability distributions on $X$, i.e., functions
    $\varphi: X \to [0,1]$ such that $\varphi^{-1}(0,1]$ is finite and $\sum_{x
        \in X} \varphi(x) = 1$; $\funcD f$ is the pushforward along $f$, given
 by $(\funcD f)(\varphi)(y) = \sum_{x \in
        f^{-1}(y)} \varphi(x)$ for $\varphi \in \funcD X$ and $y \in Y$;
      $\unitD_X(x)$ is the Dirac $\delta_x$ for $x \in X$, i.e., the probability
      distribution such that $\delta_x(x) = 1$; and $\multD$ computes the mean
      of a distribution of distributions, so that, for $\Phi \in \funcDD X$,
      $\multD(\Phi)(x) = \sum_{\varphi \in \funcD X} \Phi(\varphi) \cdot
      \varphi(x)$.
  \item[The \emph{ultrafilter monad} $\Ult$] $\Ult X$ is the set of maximal
    filters on $X$, where filters are sets $E \in \funcPP X$ that are non-empty,
    up-closed (for inclusion), stable under finite intersections and that do not
    contain the empty set; for an ultrafilter $E \in \Ult X$, $(\Ult f)(E) =
    \suchthat{ e' \supseteq f[e] }{ e \in E }$; $\unitUlt_X(x)$ is the principal
    filter $\suchthat{ e \in \funcP X }{ x \in e }$; and, for $\mathfrak{E} \in
    \UltUlt X$, $\multUlt_X(\mathfrak{E}) = \bigcup \suchthat{ \bigcap
      \mathfrak{e} }{ \mathfrak{e} \in \mathfrak{E} }$.
  \end{description}
  We will also be concerned with a topological analogue of the powerset monad,
  defined on the category $\KHaus$ of compact Hausdorff spaces and continuous
  functions. Thereafter $X$ and $Y$ are compact Hausdorff spaces, $x$ is an
  element of $X$, and $f: X \to Y$ is a continuous function.
  \begin{description}
  \item[The \emph{Vietoris monad} $\funcV$] $\funcV X$ is the space of closed
    subsets of $X$ equipped with the topology for which a subbase is given by
    the sets $\Box u = \suchthat{ c \in \funcV X }{ c \subseteq u }$ and
    $\lozenge u = \suchthat{ c \in \funcV X }{ c \cap u \neq \varnothing }$,
    where $u$ ranges among all open sets of $X$; $(\funcV f)(c) = f[c] =
    \suchthat{ f(x) }{ x \in c }$ for $c \in \funcV X$; $\unitV_X(x) = \{ x \}$;
    and $\multV_X(C) = \bigcup C$ for $C \in \funcVV X$.
  \end{description}
  We also write $\funcP_*$ and $\funcV_*$ for the non-empty powerset and
  Vietoris monads, respectively obtained by removing the empty set from each
  $\funcP X$ and $\funcV X$, and $\funcP_f$ for the monad of finite subsets on
  $\Set$. Other monads that will be considered as examples are the \emph{list}
  and \emph{multiset} monads on $\Set$, denoted $\funcL$ and $\funcM$, and whose
  precise definitions we do not recall as they are not crucial to the
  presentation.
\end{exa}

Weak distributive laws are a certain type of natural transformations involving
monads.

\begin{defi}[{(weak) distributive law~\cite[Definition
        9]{garnerVietorisMonadWeak2019}}]\label{def:weak-distr-laws} A monad
  $\funcT$ \emph{weakly distributes} over a monad $\funcS$ when there is a
  \emph{weak distributive law} of $\funcT$ over $\funcS$, i.e., a natural
  transformation $\distrTS$ such that \Crefnosort{cd:unit+,,cd:mult-,,cd:mult+}
  below commute. A \emph{(strict) distributive law} is a weak distributive law
  that moreover makes \cref{cd:unit-} below commute.
\[
    \begin{tikzcd}
      & \funcS \\
      {\funcT \funcS} && {\funcS \funcT}
      \arrow["{\unitT \funcS}"', from=1-2, to=2-1]
      \arrow["{\funcS \unitT}", from=1-2, to=2-3]
      \arrow[""'{name=2}, "\distrTSname"', from=2-1, to=2-3]
      \diagramlabel[${\unit}^{-}$]{1-2}{2}{unit-}
    \end{tikzcd}
  \qquad
    \begin{tikzcd}
      {\funcTTS} & {\funcTST} & {\funcSTT} \\
      {\funcTS} && {\funcST}
      \arrow["{\funcT \distrTSname}", from=1-1, to=1-2]
      \arrow["{\multT \funcS}"', from=1-1, to=2-1]
      \arrow["{\distrTSname \funcT}", from=1-2, to=1-3]
      \arrow["{\funcS \multT}", from=1-3, to=2-3]
      \arrow[""{name=2}, "\distrTSname"', from=2-1, to=2-3]
      \diagramlabel[${\mult}^{-}$]{1-2}{2}{mult-}
    \end{tikzcd}
  \]
  \[
    \begin{tikzcd}
      & \funcT \\
      {\funcT \funcS} && {\funcS \funcT}
      \arrow["{\funcT \unitS}"', from=1-2, to=2-1]
      \arrow["{\unitS \funcT}", from=1-2, to=2-3]
      \arrow[""'{name=2}, "\distrTSname"', from=2-1, to=2-3]
      \diagramlabel[${\unit}^{+}$]{1-2}{2}{unit+}
    \end{tikzcd}
  \qquad
    \begin{tikzcd}
      {\funcTSS} & {\funcSTS} & {\funcSST} \\
      {\funcTS} && {\funcST}
      \arrow["{\distrTSname \funcS}", from=1-1, to=1-2]
      \arrow["{\funcT \multS}"', from=1-1, to=2-1]
      \arrow["{\funcS \distrTSname}", from=1-2, to=1-3]
      \arrow["{\multS \funcT}", from=1-3, to=2-3]
      \arrow[""{name=2}, "\distrTSname"', from=2-1, to=2-3]
      \diagramlabel[${\mult}^{+}$]{1-2}{2}{mult+}
    \end{tikzcd}
  \]
\end{defi}

If $\distrTS$ is a distributive law, the endofunctor $\funcST$ and the natural
transformations $\unitST \coloneq \unitS \unitT$ and $\multST \coloneq \multS
\multT \circ \funcS \rho \funcT$ form a monad~\cite{beckDistributiveLaws1969}.
If $\distrTSname$ is only a weak distributive law, this is not true anymore:
$\multST \circ \unitST \funcST$ need not be an identity, and can only be proven
to be idempotent (its composite with itself is itself). But if this idempotent
is \emph{split} --- i.e., if there is a functor $\funcScmpT$ and natural
transformations $p: \funcST \Rightarrow \funcScmpT$ and $i: \funcScmpT
\Rightarrow \funcST$ such that $p \circ i = \id$ and $i \circ p = \multST \circ
\unitST \funcST$ --- then $\funcScmpT$ can be made into a
monad~\cite{garnerVietorisMonadWeak2019} with unit $\unit[\funcScmpT] = p \circ
\unitS \unitT$ and multiplication $\mult[\funcScmpT] = p \circ \multS \multT
\circ \funcS \distrTSname \funcT \circ ii$: the monad $\funcScmpT$ is called the
\emph{weak composite} of $\funcS$ and $\funcT$.

\begin{exa}[weak distributive laws]\label{ex:weak-distr-laws}
  In this work we will be particularly concerned with the following weak
  distributive laws:
  \begin{itemize}
  \item in $\Set$, the weak distributive law $\distrPP$ given by
    \Cref{eq:distrPP} below has for weak composite the monad $\funcPcmpP$ of
    sets of subsets closed under non-empty
    unions~\cite{garnerVietorisMonadWeak2019};
  \item in $\KHaus$, there is a weak distributive law $\distrVV$ given by
    \Cref{eq:distrVV} below~\cite{goyPowersetLikeMonadsWeakly2021};
  \item in $\Set$, the weak distributive law $\distrDP$ given by
    \Cref{eq:distrDP} below has for weak composite the monad $\funcPcmpD$
    of convex sets of finitely supported probability
    distributions~\cite{goyCombiningProbabilisticNondeterministic2020}.
  \end{itemize}
  \begin{align}
    \label{eq:distrPP}
    \distrPPname_X(E) &= \suchthat{ e' \in \funcP X }
                        { e' \subseteq \bigcup E
                        \mathand
                        \forall e \in E, e \cap e' \neq \varnothing } \\
    \label{eq:distrVV}
    \distrVVname_X(C) &= \suchthat{ c' \in \funcV X }
                        { c' \subseteq \bigcup C
                        \mathand
                        \forall c \in C, c \cap c' \neq \varnothing } \\
    \label{eq:distrDP}
    \distrDPname_X(\Phi) &=  \suchthat{ \left(\multD_X \circ \funcD f\right)(\Phi) }
                             { f: \funcP X \rightarrow \funcD X
                             \mathand
                             \forall e \in \funcP X,
                             \Phi(e) \neq 0 \Rightarrow f(e) \in \funcD e }
  \end{align}
  For distributive laws, a notable result is that monads corresponding to
  \emph{linear} algebraic theories distributive over \emph{commutative}
  monads~\cite[Theorem~4.3.4]{manesMonadCompositionsGeneral2007}, so that for
  example there are distributive laws $\distrMPtype$ and $\distrLPtype$ [ibid.,
  Examples~5.1.3 \& 5.1.8].
\end{exa}

At this point two questions may come to the curious reader: how do we actually
find these weak distributive laws --- the formul\ae{} in
\Cref{ex:weak-distr-laws} are not trivial --- and where does the weak composite
monad structure on $\funcScmpT$ come from. These two questions will respectively
be partially answered in the next two subsections, which give two other
equivalent presentations of weak distributive laws.

\subsection{Weak Extensions}\label{sec:preliminaries:extensions}

Given a monad $\funcS$, one may form its \emph{Kleisli category} $\KlS$, which
is intuitively the category of $\funcS$-effectful $\catC$-arrows: its objects
are those of $\catC$, its arrows $X \klarrow Y$ are those arrows $X \to \funcS
Y$ in $\catC$, the identity arrow $X \klarrow X$ is given by $\unitS_X : X \to
\funcS X$, and the composite of two arrows $f : X \klarrow Y$ and $g : Y
\klarrow Z$ is given by the composition $X \xrightarrow{f} \funcS Y
\xrightarrow{\funcS g} \funcSS Z \xrightarrow{\multS_Z} \funcS Z$ in $\catC$.
$\KlS$ comes with an adjunction $\adjunction{\catC}{\KlS}{\leftKlS}{\rightKlS}$,
such that $\rightKlS \leftKlS = \funcS$: the left adjoint $\leftKlS$ sends
objects on themselves and an arrow $f: X \to Y$ to the \emph{pure} effectful
arrow $X \xrightarrow{f} Y \xrightarrow{\unitS_Y} \funcS Y$, while the right
adjoint $\rightKlS$ sends an object $X$ on $\funcS X$ and an effectful arrow $f:
X \to \funcS Y$ to the arrow $\funcS X \xrightarrow{\funcS f} \funcSS Y
\xrightarrow{\multS} Y$. The unit of the adjunction is $\unitS$ and its counit
is the natural transformation $\counitS: \leftKlS \rightKlS \Rightarrow \Id$ (in
$\KlS$) with components the effectful arrows $\id_{\funcS X}: \funcS X \to
\funcS X$.

An endofunctor $\funcT: \catC \to \catC$ is said to \emph{extend} to $\KlS$ if
there is an endofunctor $\extT: \KlS \to \KlS$ such that $\extT \leftKlS =
\leftKlS \funcT$ (i.e., such that $\extT$ coincides with $\funcT$ on objects and
pure effectful arrows). If $\funcT$ and $\funcT'$ have extensions $\extT$ and
$\extT'$, a natural transformation $\alpha: \funcT \Rightarrow \funcT'$ is said
to extend to $\KlS$ if $\ext{\alpha}$, given by $\ext{\alpha} \leftKlS =
\leftKlS \alpha$, is a natural transformation $\extT \Rightarrow \extT'$.

\begin{defi}[extensions of monads]
  Let $\funcS$ be a monad on $\catC$. A monad $\monadT$ on $\catC$ \emph{weakly
    extends} to $\KlS$ when $\funcT$ and $\multT$ extend to $\KlS$. It
  \emph{extends} to $\KlS$ when $\unitT$ also extends to $\KlS$.
\end{defi}
Giving an extension of a monad $\funcT$ to $\KlS$ is equivalent to giving a
distributive law $\distrTS$, just like giving a weak extension thereof is
equivalent to giving a weak distributive law of the same type~\cite[Proposition
  11]{garnerVietorisMonadWeak2019}: in both cases, the law $\distrTSname$
induces the extension which sends $f: X \rightarrow \funcS Y$ to $\funcT X
\xrightarrow{\funcT f} \funcTS Y \xrightarrow{\rho} \funcST Y$, and is computed
from the extension as $\rho = \rightKlS \extT \counitS \leftKlS \circ \unitS
\funcTS$. In particular \Cref{ex:weak-distr-laws} also yields examples of weak
extensions:

\begin{exa}
  \label{ex:weak-extensions}
  Note that, to describe a specific weak extension $\semimonadextT$ of a monad
  $\monadT$, it is enough to only describe the action of the extended
  endofunctor $\extT$, since its action on objects and the multiplication
  $\multextT$ coincide with those of $\funcT$. From \Cref{ex:weak-distr-laws} we
  therefore get the following weak extensions to $\KlP$, which we equivalently
  see as the category $\Rel$ of sets and relations between them.
  \begin{itemize}
  \item~\cite[\textsection 4.3]{garnerVietorisMonadWeak2019} The weak
    distributive law $\distrPP$ on $\Set$ gives rise to a weak extension $\extP$
    of $\funcP$ to $\Rel$, given on a relation $R \subseteq X \times X'$ by
    \[ \extP r \coloneq \suchthat{(e,e') \in \funcP X \times \funcP X'}{ \forall
        x \in e, \exists x' \in e', (x,x') \in r \mathand \forall x' \in e',
        \exists x \in e, (x,x') \in R} \]($\extP r$ is also known as the \emph{Egli-Milner} relation induced by $r$).
    \smallskip

  \item~\cite[\textsection 3]{goyCombiningProbabilisticNondeterministic2020} The
    weak distributive law $\distrDP$ on $\Set$ gives rise to a weak extension
    $\extD$ of $\funcD$ to $\Rel$, given on a relation $R \subseteq X \times X'$
    by
    \[ \extD r \coloneq \suchthat{(\phi,\phi') \in \funcD X \times \funcD X'}{
        \exists \psi \in \funcD R, \phi = \sum_{x' \in X'} \psi(-,x') \mathand
        \phi' = \sum_{x \in X} \psi(x,-)} \]\end{itemize}
\end{exa}

In fact, all the weak distributive laws of \Cref{ex:weak-distr-laws} were
originally obtained by describing the corresponding weak extensions. The
explanation of how these weak extensions themselves were constructed is deferred
to \Cref{sec:relational-extensions}.

\subsection{Weak Liftings}
\label{sec:preliminaries:liftings}

Let $\funcT$ be a monad over a category $\catC$. Its category of algebras is the
category $\EMT$ whose objects are pairs $\alg{A}{a}$ of a $\catC$-object $A$ and
a $\catC$-arrow $a: \funcT A \to A$ satisfying $a \circ \unitT_A = \id_A$ and $a
\circ \funcT a = a \circ \multT a$, and whose arrows $\alg{A}{a} \to \alg{B}{b}$
are those $\catC$-arrows $A \to B$ such that $f \circ a = b \circ \funcT f$ --
the identity morphism is the one with the identity as its underlying
$\catC$-arrow, and composition of morphisms is done by composing the underlying
$\catC$-arrows. Just like for $\KlT$, there is an adjunction
$\adjunction{\catC}{\EMT}{\leftEMT}{\rightEMT}$ such that $\rightEMT \leftEMT =
\funcT$: the right adjoint $\rightEMT$ sends an algebra $\alg{A}{a}$ to its
carrier object $A$ and a morphism $\alg{A}{a} \to \alg{B}{b}$ to the underlying
arrow $A \to B$, while $\leftEMT$ sends an object $X$ to the free
$\funcT$-algebra on $X$, given by the pair $\alg{\funcT X}{\multT_X}$, and an
arrow $f: X \to Y$ to the morphism $\alg{\funcT X}{\multT_X} \to \alg{\funcT
  Y}{\multT_Y}$ with underlying $\catC$-arrow $\funcT f: \funcT X \to \funcT Y$.
The unit of the adjunction is $\unitT$ and its counit is the natural
transformation $\counitT: \leftEMT \rightEMT \Rightarrow \Id$ (in $\EMT$) given
by $\rightEMT \counitT_{\alg{A}{a}} = a: \funcT A \to A$.

\begin{defi}[{weak liftings~\cite[Definitions 4.1 and
    4.2]{bohmWeakTheoryMonads2010}}]
  \label{def:weak-liftings} A \emph{weak lifting} of an endofunctor $\funcS:
  \catC \to \catC$ to $\EMT$ is the data of an endofunctor $\liftS: \EMT \to
  \EMT$ along with natural transformations $\projT_\funcS: \funcS \rightEMT
  \Rightarrow \rightEMT \liftS$ and $\incT_\funcS: \rightEMT \liftS \Rightarrow
  \funcS \rightEMT$ -- also written $\proj_\funcS$ and $\inc_\funcS$ when not
  ambiguous -- such that $\projT_\funcS \circ \incT_\funcS = \id$. The
  composite $\incT_\funcS \circ \projT_\funcS$ is then written $\closeT_\funcS$.

  Let $\alpha: \funcS \Rightarrow \funcR$ be a natural transformation between
  two $\catC$-endofunctors with weak liftings
  $\left(\liftS,\proj_\funcS,\inc_\funcS\right)$ and
  $\left(\liftR,\proj_\funcR,\inc_\funcR\right)$. If a natural transformation
  $\lift{\alpha}: \liftS \Rightarrow \liftR$ makes \Cref{cd:weak-lifting:proj},
  \Cref{cd:weak-lifting:inc} or both
  \Cref{cd:weak-lifting:proj,cd:weak-lifting:inc} below commute, it is
  respectively a \emph{weak $\proj$-}, \emph{weak $\inc$-} or a \emph{weak
    lifting} of $\alpha$.
\[
  \begin{tikzcd}[sep=large]
    {\funcS \rightEMT} & {\rightEMT \liftS}  \\
    {\funcR \rightEMT} & {\rightEMT \liftR}
    \arrow["{\projT_\funcS}"{name=1}, from=1-1, to=1-2]
    \arrow["{\rightEMT \lift{\alpha}}", from=1-2, to=2-2]
    \arrow["{\alpha \rightEMT}"', from=1-1, to=2-1]
    \arrow["{\projT_\funcR}"'{name=2}, from=2-1, to=2-2]
    \diagramlabel[$\proj$]{1}{2}{weak-lifting:proj}
  \end{tikzcd}
  \qquad
  \begin{tikzcd}[sep=large]
    {\rightEMT \liftS} & {\funcS \rightEMT}  \\
    {\rightEMT \liftR} & {\funcR \rightEMT}
    \arrow["{\incT_\funcS}"{name=1}, from=1-1, to=1-2]
    \arrow["{\rightEMT \lift{\alpha}}"', from=1-1, to=2-1]
    \arrow["{\alpha \rightEMT}", from=1-2, to=2-2]
    \arrow["{\incT_\funcR}"'{name=2}, from=2-1, to=2-2]
    \diagramlabel[$\inc$]{1}{2}{weak-lifting:inc}
  \end{tikzcd}
\]
\end{defi}
As discussed in \cite{bohmWeakTheoryMonads2010}, weak $\proj$- and weak
$\inc$-liftings are necessarily unique and given by $\rightEMT \lift{\alpha} =
\projT_\funcR \circ \alpha \circ \incT_\funcS$ when they exist. The existence of
weak liftings is moreover fully characterized:
\begin{thm}[{\cite[Proposition 4.3 and Theorem 4.4]{bohmWeakTheoryMonads2010}}]
  \label{thm:weak-liftings}
  Suppose idempotents split in $\catC$. Then idempotents also split in $\EMT$,
  and having a weak lifting of $\funcS: \catC \to \catC$ to $\EMT$ is equivalent
  to having a law $\distrTS$ that makes \Cref{cd:mult-} commute.

  Fix now weak liftings $\liftS$ and $\liftR$ of $\funcS$ and $\funcR$ given by
  laws $\distrTS$ and $\distrTR$. Then, $\alpha: \funcS \Rightarrow \funcR$
  respectively has a weak $\inc$-, weak $\proj$- or weak lifting if and only if
  it makes \Cref{cd:weak-lifting:inc:condition},
  \Cref{cd:weak-lifting:proj:condition} or both
  \Cref{cd:weak-lifting:inc:condition,,cd:weak-lifting:proj:condition} below on
  the left commute. The latter case also holds if and only if
  \Cref{cd:weak-lifting:cond} below on the right commutes.

  Weakly lifting is functorial: the composition of the weak (resp. weak
  $\proj$-, weak $\inc$-) liftings of two functors (resp. natural
  transformations) is a weak (resp. weak $\proj$-, weak $\inc$-) lifting of
  their composite.

  \[
    \begin{tikzcd}[ampersand replacement=\&]
      \funcTS \&\&\&\& \funcST \\
      \funcTTS \& \funcTST \& \funcTRT \& \funcRTT \& \funcRT \\
      \funcTS \&\&\&\& \funcTR
      \arrow[""'{name=1}, "\distrTSname", from=1-1, to=1-5]
      \arrow["{\funcT \unitT \funcS}"', from=1-1, to=2-1]
      \arrow["{{\alpha \funcT}}", from=1-5, to=2-5]
      \arrow["{{\funcT \distrTSname}}", from=2-1, to=2-2]
      \arrow["{{\funcT \alpha \funcT}}", from=2-2, to=2-3]
      \arrow["{{\distrTRname \funcT}}", from=2-3, to=2-4]
      \arrow["{{\funcR \multT}}", from=2-4, to=2-5]
      \arrow["{\unitT \funcTS}", from=3-1, to=2-1]
      \arrow[""{name=3}, "{{\funcT\alpha}}"', from=3-1, to=3-5]
      \arrow["\distrTRname"', from=3-5, to=2-5]
      \diagramlabel{1}{2-3}{weak-lifting:inc:condition}
      \diagramlabel{2-3}{3}{weak-lifting:proj:condition}
    \end{tikzcd}
    \qquad
    \begin{tikzcd}
        \funcTS & \funcST \\
        \phantom{\funcT} \\
        \funcTR & \funcRT \arrow["\distrTSname", from=1-1, to=1-2]\arrow["{\funcT \alpha}"', from=1-1, to=3-1]\arrow["{\alpha \funcT}", from=1-2, to=3-2]\arrow["\distrTRname"', from=3-1, to=3-2]\diagramlabel{1-1}{3-2}{weak-lifting:cond}
      \end{tikzcd}
  \]\end{thm}

In~\cite{garnerVietorisMonadWeak2019}, Garner instantiates
\Cref{thm:weak-liftings} to give another presentation of weak distributive laws:
if idempotents split in $\catC$, a weak distributive law $\distrTStype$ is
equivalently given by a weak lifting of $\monadS$ to $\EMT$, i.e., by weak
liftings of $\funcS$, $\unitS$ and $\multS$, respectively coming from
\Crefnosort{cd:mult-,,cd:unit+,,cd:mult+}. That these two natural
transformations weakly lift respectively means that
\Cref{cd:proj-unit,,cd:inc-unit} and \Cref{cd:proj-mult,,cd:inc-mult} below
commute. When \Cref{cd:unit-} also commutes, $\rho$ is a (strict) distributive
law and this weak lifting is a (strict) lifting: $\proj_\funcS$ and
$\inc_\funcS$ are both the identity.
\[
  \begin{tikzcd}
    & \rightEMT \\
    { \funcS \rightEMT} && {\rightEMT \liftS}
    \arrow["{\unitS \rightEMT}"', from=1-2, to=2-1]
    \arrow["{\rightEMT \unitliftS}", from=1-2, to=2-3]
    \arrow[""'{name=2}, "\projT_\funcS"', from=2-1, to=2-3]
    \diagramlabel[$\proj \circ \unit$]{1-2}{2}{proj-unit}
  \end{tikzcd}
  \qquad
  \begin{tikzcd}
    {\funcSS \rightEMT} & {\funcS \rightEMT \liftS} & {\rightEMT \liftSliftS} \\
    {\funcS \rightEMT} && {\rightEMT \liftS}
    \arrow["{\funcS \projT_\funcS}", from=1-1, to=1-2]
    \arrow["{\multS \rightEMT}"', from=1-1, to=2-1]
    \arrow["{\projT_\funcS \liftS}", from=1-2, to=1-3]
    \arrow["{\rightEMT \multliftS}", from=1-3, to=2-3]
    \arrow[""{name=2}, "\projT_\funcS"', from=2-1, to=2-3]
    \diagramlabel[$\proj \circ \mult$]{1-2}{2}{proj-mult}
  \end{tikzcd}
\]\[
  \begin{tikzcd}
    & \rightEMT \\
    {\rightEMT \liftS} && { \funcS \rightEMT}
    \arrow["{\rightEMT \unitliftS}"', from=1-2, to=2-1]
    \arrow["{\unitS \rightEMT}", from=1-2, to=2-3]
    \arrow[""'{name=2}, "\incT_\funcS"', from=2-1, to=2-3]
    \diagramlabel[$\inc \circ \unit$]{1-2}{2}{inc-unit}
  \end{tikzcd}
  \qquad
  \begin{tikzcd}
    {\rightEMT \liftSliftS} & {\funcS \rightEMT \liftS} & {\funcSS \rightEMT} \\
    {\rightEMT \liftS} && {\funcS \rightEMT}
    \arrow["{\incT_\funcS \liftS}", from=1-1, to=1-2]
    \arrow["{\rightEMT \multliftS}"', from=1-1, to=2-1]
    \arrow["{\funcS \incT_\funcS}", from=1-2, to=1-3]
    \arrow["{\multS \rightEMT}", from=1-3, to=2-3]
    \arrow[""{name=2}, "\incT_\funcS"', from=2-1, to=2-3]
    \diagramlabel[$\inc \circ \mult$]{1-2}{2}{inc-mult}
  \end{tikzcd}
\]

The weak composite monad $\funcScmpT$ corresponding to a weak distributive law
$\distrTStype$ can be retrieved from the weak lifting of $\funcS$ to $\EMT$ by
setting $\funcScmpT = \rightEMT \liftS \leftEMT$, $\unit[\funcScmpT] = \rightEMT
\unitliftS \leftEMT \circ \unitT$ and $\mult[\funcScmpT] = \rightEMT \multliftS
\leftEMT \circ \rightEMT \liftS \counitT \liftS \leftEMT$. $\closeT_\funcS
\leftEMT$ is moreover the idempotent $\multST \circ \unitST \funcST: \funcST
\Rightarrow \funcST$, and its splitting is thus given by $\projT_\funcS
\leftEMT: \funcST \Rightarrow \funcScmpT$ and $\incT_\funcS \leftEMT: \funcScmpT
\Rightarrow \funcST$.

While the above discussion characterizes the existence of weak liftings
entirely, it does not explain how to actually construct them. This explanation
is deferred to \Cref{sec:lifted-kleisli}. Still, the weak distributive laws of
\Cref{ex:weak-distr-laws} give rise to weakly lifted monads that we now
describe.

\begin{exa}[weak liftings]
  \label{ex:weak-liftings}
  All idempotents split in $\Set$ (they factor through their image).
  \begin{itemize}
  \item The algebras of $\funcP$ are the \emph{complete join-semilattices} (we
    write $\EMP \cong \JSL$): the weak lifting corresponding to the law
    $\distrPPtype$ is the monad of \emph{subsets closed under non-empty
      joins}~\cite{goyPowersetLikeMonadsWeakly2021}. In this case $\proj_\funcP$
    computes the closure of a subset under non-empty joins, while $\inc_\funcP$
    embeds the set of closed subsets into the set of all subsets.

  \item The algebras of $\funcD$ are the \emph{barycentric algebras}, also
    called \emph{convex spaces} (we write $\EMD \cong \Conv$): the weak lifting
    corresponding to the law $\distrDPtype$ is the monad of \emph{convex-closed
      subsets}~\cite{goyCombiningProbabilisticNondeterministic2020}. In this
    case $\proj_\funcP$ computes the closure of a subset under convex
    combinations, while $\inc_\funcP$ embeds the set of closed subsets into the
    set of all subsets.

  \item Finally, there is also a weak distributive law
    $\distrUltPtype$~\cite[\textsection 5.1]{garnerVietorisMonadWeak2019}.
    Assuming the axiom of choice, the algebras of $\Ult$ are the compact
    Hausdorff spaces~\cite{manesTripleTheoreticConstruction1969} (we write
    $\EMUlt \cong \KHaus$): the corresponding weak lifting is the Vietoris monad
    $\funcV$~[loc. cit.]. In this last case $\proj_\funcP$ computes the
    topological closure of a subset, while $\inc_\funcP$ embeds the set of
    closed subsets into the set of all subsets.
  \end{itemize}
\end{exa}

\section{Weakly Lifting Weak Distributive Laws}
\label{sec:lifting-laws}

Recall from \Cref{ex:weak-liftings} that $\funcV$ is a weak lifting of $\funcP$
to $\KHaus \cong \EMUlt$, and from \Cref{ex:weak-distr-laws} that the two weak
distributive laws $\distrPPtype$~\labelcref{eq:distrPP} and
$\distrVVtype$~\labelcref{eq:distrVV} look very similar. In fact, this second
law seems to be some kind of weak lifting of the first one:

\begin{lem}
  \label{lem:weak-distr-law-VV-weak-lifting-PP}
  Diagram \labelcref{cd:weak-distr-law-VV-weak-lifting-PP} commutes.
  \[
    \begin{tikzcd}[ampersand replacement=\&, sep=large]
      {\rightEMUlt \funcVV} \& {\funcP \rightEMUlt \funcV} \& {\funcPP \rightEMUlt}  \\
      {\rightEMUlt \funcVV} \& {\funcP \rightEMUlt \funcV} \& {\funcPP \rightEMUlt}
      \arrow["{\rightEM \distrVVname}"', from=1-1, to=2-1]
      \arrow["\distrPPname", from=1-3, to=2-3]
      \arrow["{\inc_\funcP \funcV}", from=1-1, to=1-2]
      \arrow["{\funcP \inc_\funcP}", from=1-2, to=1-3]
      \arrow["{\proj_\funcP \funcV}", from=2-2, to=2-1]
      \arrow["{\funcP \proj_\funcP}", from=2-3, to=2-2]
      \diagramlabel{1-2}{2-2}{weak-distr-law-VV-weak-lifting-PP}
    \end{tikzcd}
  \]
\end{lem}
\begin{proof}
  Recall from~\cite{garnerVietorisMonadWeak2019} that, if $X$ is a compact
  Hausdorff space that we see as a $\Ult$-algebra, the $X$-component of
  $\proj_\funcP$ is the function $\funcP X \to \funcV X$ that takes a subset of
  $X$ and outputs its closure, while the $X$-component of $\inc_\funcP$ is the
  function $\funcV X \to \funcP X$ that embeds closed subsets in the set of all
  subsets. Hence for $C \in \funcVV X$ a closed set of closed sets,
  \begin{align*}
    \left( \funcP \proj_\funcP \circ
    \distrPPname \rightEMUlt \circ
    \funcP \inc_\funcP \circ \inc_\funcP \funcV
    \right)_X(C)
    &= \suchthat{ \overline{e} }{ e \subseteq \bigcup C
      \mathand \forall c \in C, c \cap e \neq \varnothing } \\
    \left( \inc_\funcP \funcV \circ \rightEMUlt \distrVVname \right)_X
    &= \suchthat{ c \in \funcV X }{ c \subseteq \bigcup C \mathand \forall c' \in C, c \cap c' \neq \varnothing }
  \end{align*}
  (where $\overline{e}$ denotes the closure of $e$). But if $e \subseteq \bigcup
  C$, then $\overline{e} \subseteq \bigcup C$ because $\bigcup C = \multV(C)$ is
  closed, and if $e \cap c \neq \varnothing$ then $\overline{e} \cap c \neq
  \varnothing$. Hence the two sets above are equal, and
  \[ \inc_\funcP \funcV \circ \rightEMUlt \distrVVname = \funcP \proj_\funcP
    \circ \distrPPname \rightEMUlt \circ \funcP \inc_\funcP \circ \inc_\funcP
    \funcV \]
  Because $\proj_\funcP \circ \inc_\funcP = \id$, we finally get
  \Cref{cd:weak-distr-law-VV-weak-lifting-PP} by post-composing the above
  equation with $\proj_\funcP \funcV$.
\end{proof}

Is this just a coincidence, or is this an instance of \Cref{thm:weak-liftings}?
And in the latter case, can the weak distributivity of $\distrVVname$ be
automatically derived from that of $\distrPPname$? Does this law
$\distrPPtype$ also weakly lift to laws on other categories of algebras where
the powerset monad weakly lifts, say $\EMP$ and $\EMD$?

In this section we thus consider three monads $\monadT$, $\monadS$ and $\monadR$
on a category $\catC$, and three weak distributive laws $\distrTS$, $\distrTR$
and $\distrSR$ (a mnemonic for which is which: $\distrTSname$ does not involve
$\funcR$, $\distrTRname$ does not involve $\funcS$ and $\distrSRname$ does not
involve $\funcT$), and study when $\distrSR$ weakly lifts to $\EMT$. We assume
that idempotents split in $\catC$, so that the corresponding weak composite and
weak liftings all exist.

\subsection{The Yang-Baxter Equation for Weak Distributive Laws}
\label{sec:lifting-laws:yang-baxter}

The standard way to lift (strict) distributive laws is to use the so-called
Yang-Baxter equation. This equation holds for the three laws $\distrTSname$, $
\distrTRname$ and $\distrSRname$ holds when \Cref{cd:YB} below commutes.
\[
  \begin{tikzcd}[row sep=tiny]
    & \funcTRS & \funcRTS \\
    \funcTSR &&& \funcRST \\
    & \funcSTR & \funcSRT
    \arrow[""{name=1}, "{\distrTRname \funcS}", from=1-2, to=1-3]
    \arrow["{\funcR \distrTSname}", from=1-3, to=2-4]
    \arrow["{\funcT \distrSRname}", from=2-1, to=1-2]
    \arrow["{\distrTSname \funcR}"', from=2-1, to=3-2]
    \arrow[""{name=3}, "{\funcS \distrTRname}"', from=3-2, to=3-3]
    \arrow["{\distrSRname \funcT}"', from=3-3, to=2-4]
    \diagramlabel[YB]{1}{3}{YB}
  \end{tikzcd}
\]
If these laws are strict distributive laws, then it is well-known since
\cite{chengIteratedDistributiveLaws2011} that the Yang-Baxter equation is enough
to show that $\distrTRS$ is a distributive law of $\funcT$ over $\funcRS$, and
that the distributive law $\distrSR$ lifts to a distributive law
$\distrliftSliftR$ in $\EMT$. The composition of weak distributive laws using
the Yang-Baxter equation was in turn investigated
in~\cite{goyWeakeningIteratingLaws2023}. A notable result is the following:

\begin{prop}[{\cite[Theorem 4.3]{goyWeakeningIteratingLaws2023}}]
  \label{prop:YB-compose-weak-laws}
  If the weak distributive laws $\distrTS$, $\distrTR$ and $\distrSR$ have
  \Cref{cd:YB} commute, then $\projS_\funcR \leftEMS \funcT \circ \distrTRSname
  \circ \funcT \incS_\funcR \leftEMS$ is a weak distributive law $\funcT
  (\funcRcmpS) \Rightarrow (\funcRcmpS) \funcT$.
\end{prop}

The Yang-Baxter equation thus allows for weakly lifting $\funcRcmpS$ to $\EMT$.
More importantly for our purpose, we show it also allows for weakly lifting the
weak distributive law $\distrSRtype$:

\begin{thm}
  \label{thm:YB-weakly-lift-weak-law}
  Weak distributive laws $\distrTS$, $\distrTR$ and $\distrSR$ satisfy the
  Yang-Baxter equation if and only if $\distrSR$ weakly lifts to $\EMT$, i.e., if
  there is a natural transformation $\distrliftSliftR$ such that
  \Cref{cd:weak-law-weak-lifting:proj,cd:weak-law-weak-lifting:inc} commute.
  \[
    \begin{tikzcd}[ampersand replacement=\&, sep=large]
      {\funcSR \rightEMT} \& {\funcS \rightEMT \liftR} \& {\rightEMT \liftSliftR} \\
      {\funcRS \rightEMT} \& {\funcR \rightEMT \liftS} \& {\rightEMT
        \liftRliftS}\arrow["{\distrSRname \rightEMT}"', from=1-1, to=2-1]\arrow["{\rightEMT \distrliftSliftRname}", from=1-3, to=2-3]\arrow["{\funcS \proj_\funcR}", from=1-1, to=1-2]\arrow["{\proj_\funcS \liftR}", from=1-2, to=1-3]\arrow["{\funcR \proj_\funcS}"', from=2-1, to=2-2]\arrow["{\proj_\funcR \liftS}"', from=2-2, to=2-3]\diagramlabel{1-2}{2-2}{weak-law-weak-lifting:proj}\end{tikzcd}\qquad
    \begin{tikzcd}[ampersand replacement=\&, sep=large]
      {\rightEMT \liftSliftR} \& {\funcS \rightEMT \liftR} \& {\funcSR \rightEMT} \\
      {\rightEMT \liftRliftS} \& {\funcR \rightEMT \liftS} \& {\funcRS
        \rightEMT}\arrow["{\rightEMT \distrliftSliftRname}"', from=1-1, to=2-1]\arrow["{\distrSRname \rightEMT}", from=1-3, to=2-3]\arrow["{\inc_\funcS \liftR}", from=1-1, to=1-2]\arrow["{\funcS \inc_\funcR}", from=1-2, to=1-3]\arrow["{\inc_\funcR \liftS}"', from=2-1, to=2-2]\arrow["{\funcR \inc_\funcS}"', from=2-2, to=2-3]\diagramlabel{1-2}{2-2}{weak-law-weak-lifting:inc}\end{tikzcd}\]
  If this holds, $\distrliftSliftRname$ is a weak distributive law, and the weak
  composite $\liftRcmpliftS$ and the weak lifting $\lift{\funcRcmpS}$ can be
  chosen to be equal (as monads).
\end{thm}
\begin{proof}
  We apply \Cref{thm:weak-liftings}: recall that $\funcSR$ and $\funcRS$ have
  weak liftings $\liftSliftR$ and $\liftRliftS$ thanks to the laws $\funcS
  \distrTRname \circ \distrTSname \funcR$ and $\funcR\distrTSname \circ
  \distrTRname \funcS$. Then $\distrSR$ weakly lifts to $\EMT$ as
  $\distrliftSliftR$ if and only if $\funcR \distrTSname \circ \distrTRname
  \funcS \circ \funcT \distrSRname = \distrSRname \funcT \circ \funcS
  \distrTRname \circ \distrTSname \funcR$: this is exactly the Yang-Baxter
  equation. Because weakly lifting is functorial (\Cref{thm:weak-liftings}),
  it preserves any equation satisfied by $\distrSRname$, $\unitS$, $\unitR$,
  $\multS$ and $\multR$ (they all weakly lift to $\EMT$): since $\distrSRname$
  is a weak distributive law, so is $\distrliftSliftRname$.

  We now prove that $\projS_\funcR \leftEMS: \funcRS \Rightarrow \funcRcmpS$ and
  $\incS_\funcR \leftEMS: \funcRcmpS \Rightarrow \funcRS$ also weakly lift to
  $\EMT$ as natural transformations $\liftRliftS \Rightarrow \lift{\funcRcmpS}$
  and $\lift{\funcRcmpS} \Rightarrow \liftSliftR$. It will then follow, again by
  functoriality, that these two liftings form a splitting of the idempotent
  $\liftRliftS \Rightarrow \liftRliftS$ and that $\lift{\funcRcmpS}$ can thus be
  chosen as the weak composite monad $\liftRcmpliftS$.

  We apply \Cref{thm:weak-liftings} again: $\funcRS$ and $\funcRcmpS$ have weak
  liftings $\liftRliftS$ and $\lift{\funcRcmpS}$ thanks to the laws
  $\funcR\distrTSname \circ \distrTRname \funcS$ and $\projS_\funcR \leftEMS
  \funcT \circ \distrTRSname \circ \funcT \incS_\funcR \leftEMS$ (see
  \Cref{prop:YB-compose-weak-laws} for the latter). Thus $\projS_\funcR$ weakly
  lifts to $\EMT$ if and only if
  \[ \projS_\funcR \leftEMS \funcT \circ \distrTRSname \circ \funcT
    \closeS_\funcR \leftEMS = \projS_\funcR \leftEMS \funcT \circ
    \funcR\distrTSname \circ \distrTRname \funcS \] Similarly, $\incS_\funcR$
  weakly lifts to $\EMT$ if and only if
  \[ \funcR\distrTSname \circ \distrTRname \funcS \circ \funcT \incS_\funcR
    \leftEMS = \incS_\funcR \leftEMS \funcT \circ \projS_\funcR \leftEMS \funcT
    \circ \distrTRSname \circ \funcT \incS_\funcR \leftEMS \] These two results
  are an immediate consequence of \cite[Lemma
  4.2]{goyWeakeningIteratingLaws2023} which states that $\closeS_\funcR \leftEMS
  \funcT \circ \distrTRSname = \distrTRSname \circ \funcT \closeS_\funcR
  \leftEMS$.

  Note that it is also possible but much more tedious to prove this result by
  hand, by checking that $\distrliftSliftRname$ as given by
  \Cref{cd:weak-law-weak-lifting:proj,,cd:weak-law-weak-lifting:inc} is indeed a weak
  distributive law.
\end{proof}

When \Cref{thm:YB-weakly-lift-weak-law} holds it immediately follows that $\rightEM
\distrliftSliftRname = \proj_\funcR \liftS \circ \funcR \proj_\funcS \circ
\distrSRname \rightEMT \circ \funcS \inc_\funcR \circ \inc_\funcS \liftR$, which
is exactly the result we got in \Cref{lem:weak-distr-law-VV-weak-lifting-PP}
for $\distrliftSliftRname = \distrVV$ and $\distrSRname = \distrPP$. It would
thus be a reasonable conjecture that $\distrUltP$, $\distrUltP$ and $\distrPP$
satisfy the Yang-Baxter equation, from which we would immediately retrieve the
weak distributivity of $\distrVV$ but also learn that the weak composite
$\funcVcmpV$ is a weak lifting of $\funcPcmpP$.
Unfortunately, the Yang-Baxter equation does not hold in that case. A rather
simple way to see this is to notice that $\distrVVname$ does not make
\Cref{cd:weak-law-weak-lifting:inc} commute, i.e., it is not a $\inc$-lifting of
$\distrPPname$. More generally, we show the following no-go theorem for weak
$\inc$-liftings:

\begin{prop}
  \label{prop:no-go-weak-law-weak-inc-lifting}
  Let $\distrTP$ be a weak distributive law with corresponding weak lifting
  $\liftP$, and, given a $\funcT$-algebra $\alg{A}{a}$, write $\liftPliftP A
  \subseteq \funcPP A$ for the carrier set of the algebra $\liftPliftP
  \alg{A}{a}$, seen as a set of subsets. Formally:
  \[ \liftPliftP A \coloneq \left(\funcP \iota_\funcP \circ \iota_\funcP
      \liftP\right)\left[ \rightEMT \liftPliftP (A,a) \right] \]If there is an $\alg{A}{a}$ such that $\{ A \} \in \liftPliftP A$ and
  $\funcP_* A \notin \liftPliftP A$, then $\distrPP$ does not have a weak
  $\inc$-lifting to $\EMT$.
\end{prop}
\begin{proof}
  Suppose there is such an algebra, and consider any natural transformation
  $\distrliftPliftP$. Then, by \Cref{eq:distrPP},
  \[ \left(\distrPPname \circ \funcP \iota_\funcP \circ \iota_\funcP
      \liftP\right)(\{A\}) = \distrPPname (\{A\}) = \funcP_*A \]But $\funcP_*A \notin \liftPliftP A$, hence necessarily $\left(\funcP
    \iota_\funcP \circ \iota_\funcP \liftP \circ \rightEMT
    \distrliftPliftPname\right)(\{A\}) \in \liftPliftP A$ must be different from
  $\funcP_* A$: \Cref{cd:weak-law-weak-lifting:inc} does not commute.
\end{proof}

\begin{cor}
  \label{cor:no-go-weak-law-weak-lifting}
  There is no weak $\inc$-lifting (let alone weak liftings) of $\distrPPname$ to
  $\KHaus$, $\JSL$ or $\Conv$.
\end{cor}
\begin{proof} \hfill
  \begin{itemize}
  \item Given a compact Hausdorff space seen as a $\Ult$-algebra $\alg{A}{a}$,
    $\liftPliftP A$ is the set of all closed sets (in the Vietoris topology) of
    closed sets of $\alg{A}{a}$. $\{ A \}$ is such a closed set of closed sets
    (all singletons and the whole set are always closed in a compact Hausdorff
    space) but $\funcP_* A$ is not one in general because it contains all
    non-empty sets, in particular non-closed sets if there are any (which is the
    case for the unit interval, for instance).
  \item Consider a $\funcP$-algebra $\alg{A}{a}$, i.e., a complete
    join-semilattice where $a: \funcP A \to A$ computes the join of a subset.
    Then $\liftPliftP A$ is the set of all non-empty-joins-closed sets of
    non-empty-joins-closed subsets of $A$. $\{ A \}$ is such a
    non-empty-joins-closed set of non-empty-joins-closed subsets, but
    $\funcP_*A$ is not one in general because it contains all non-empty sets, in
    particular non-join-closed subsets if there are any (which is the case for
    $(\funcP 2, {\cup})$ for instance).
  \item Consider a $\funcD$-algebra $\alg{A}{a}$, i.e., a barycentric algebra
    where $a: \funcD A \to A$ computes convex combinations of elements. Then
    $\liftPliftP A$ is the set of all convex sets of convex subsets of $A$. $\{
    A \}$ is such a convex set of convex subsets, but $\funcP_*A$ is not one in
    general because it contains all non-empty sets, in particular non-convex
    subsets if there are any (which is the case for $\funcD 2$ for instance).
    \qedhere
  \end{itemize}
\end{proof}

\begin{rem}
  It is not hard to show that the Yang-Baxter equation holding for $\distrTS$,
  $\distrTR$ and $\distrSR$ is also equivalent to $\distrTS$ having an extension
  $\ext{\rho}: \extT \extS \Rightarrow \extS \extT$ to $\KlR$. In the case of
  $\distrUltP$, $\ext{\Ult}$ is in fact a lax monad on $\Rel$ whose algebras are
  the topological spaces~\cite{barrRelationalAlgebras1970}. Unfortunately, we
  have just shown that the Yang-Baxter equation does not hold for $\distrUltP$,
  $\distrPP$ and $\distrPP$, and so do not get a way to weakly lift
  $\ext{\funcP}$ to topological spaces for free.
\end{rem}

\Cref{thm:YB-weakly-lift-weak-law} does have some concrete instances:
in~\cite{goyWeakeningIteratingLaws2023}, to provide examples to
\Cref{prop:YB-compose-weak-laws}, Goy gives a substantial number of examples of
triples of weak distributive laws for which the Yang-Baxter equation holds. But
these examples all involve at least one strictly distributive law out of the
three.

\subsection{The \texorpdfstring{$\proj$}{π}-Yang-Baxter Equation}
\label{sec:lifting-laws:proj-yang-baxter}

We still do not have an explanation for why $\distrVVname$ looks so much like
$\distrPPname$. But $\distrVVname$ being a weak lifting of $\distrPPname$ is not
a necessary condition for retrieving
\Cref{lem:weak-distr-law-VV-weak-lifting-PP}: in fact, $\distrVVname$ being
only a weak $\inc$- or $\proj$-lifting of $\distrPPname$ would be enough.
We saw in \Cref{cor:no-go-weak-law-weak-lifting} that the weak $\inc$-lifting
hypothesis was a dead-end: how about $\distrVVname$ being a weak $\proj$-lifting
of $\distrPPname$? This turns out to be true, although the proof is of course
more involved than that of the weaker
\Cref{lem:weak-distr-law-VV-weak-lifting-PP}.

\begin{lem}
  \label{lem:distrVV-weak-proj-lifting-distrPP}
  $\distrVV$ is a weak $\proj$-lifting of $\distrPP$.
\end{lem}

\begin{proof}
  We want to show that \Cref{cd:weak-law-weak-lifting:proj} commutes, i.e., that
  for any compact Hausdorff space $X$ and $E \in \funcPP \rightEMUlt X$, the two
  sets below coincide.
  \begin{equation}
\label{eq:proj-proj-distrPP}
    \overline{\suchthat{\overline{e}}{e \subseteq \bigcup E
        \mathand \forall e' \in E, e \cap e' \neq \varnothing}}
  \end{equation}
  \begin{equation}
\label{eq:distrVV-proj-proj}
    \suchthat{c \in \funcV X}{c \subseteq \bigcup
      \overline{\suchthat{\overline{e}}{e \in E}} \mathand \forall c' \in
      \overline{\suchthat{\overline{e}}{e \in E}}, c \cap c' \neq
      \varnothing}
  \end{equation}
  Notice first that, since $\multV$ is a weak $\proj$-lifting of $\multP$, we
  have that $\bigcup \overline{\suchthat{\overline{e}}{e \in E}} =
  \overline{\bigcup E}$.

  \begin{description}
  \item[$\eqref{eq:proj-proj-distrPP} \subseteq \eqref{eq:distrVV-proj-proj}$]
    Consider some $e \subseteq \bigcup E$ such that, for all $e' \in
    E$, $e \cap e' \neq \varnothing$.
    \begin{itemize}
    \item Then,
      \[ \overline{e} \subseteq \overline{\bigcup E} = \bigcup
        \overline{\suchthat{\overline{e}}{e \in E}} \]\item Moreover, for every $e' \in E$, $e \cap e' \neq \varnothing$ by
      assumption. Hence $\overline{e} \cap \overline{e'} \neq \varnothing$ and thus
      $\suchthat{ \overline{e'}}{e' \in E} \subseteq \lozenge
      \overline{e}$. But this latter subset of $\funcV X$ is closed therein,
      therefore $\overline{\suchthat{ \overline{e'}}{e' \in E}} \subseteq
        \lozenge\overline{e}$ as well and
      \[ \forall c' \in \overline{\suchthat{ \overline{e'}}{e' \in E}},
        \overline{e} \cap c' \neq \varnothing \]\end{itemize}
    All-in-all:
    \[ \suchthat{\overline{e}}{e \subseteq E \mathand \forall e' \in E, e \cap
        e' \neq \varnothing } \subseteq \eqref{eq:distrVV-proj-proj} \]Because this latter set is closed in $\funcVV X$,
    $\eqref{eq:proj-proj-distrPP} \subseteq \eqref{eq:distrVV-proj-proj}$.

  \item[$\eqref{eq:distrVV-proj-proj} \subseteq \eqref{eq:proj-proj-distrPP}$]
    Consider some $c \in \eqref{eq:distrVV-proj-proj}$, i.e., such that
    \[ c \subseteq \bigcup \overline{\suchthat{\overline{e}}{e \in E}} =
      \overline{\bigcup E} \]and, for every $c' \in \overline{\suchthat{\overline{e}}{e \in E}}$, $c \cap
    c' \neq \varnothing$. Let moreover $U = \Box u_0 \cap \bigcap_{i = 1}^n
    \lozenge u_i$ be a basic open set of $\funcV X$ that contains $c$: $c
    \subseteq u_0$ and, for every $1 \le i \le n$, $c \cap u_i \neq
    \varnothing$. We now construct some
    \[ \overline{f} \in U \cap \suchthat{\overline{e}}{e \subseteq E \mathand
        \forall e' \in E, e \cap e' \neq \varnothing } \]By regularity (see, e.g., \cite[Lemma~21]{goyPowersetLikeMonadsWeakly2021})
    and compactness of $X$, there is an open subset $u_0'$ of $X$ such that $c
    \subseteq u_0'$ and $\overline{u_0'} \subseteq u_0$. We let $f \coloneq u_0'
    \cap \bigcup E$.
    \begin{itemize}
    \item By definition, $\overline{f} \subseteq \overline{u_0'} \subseteq u_0$.
      Fix now some $1 \le i \le n$. Then, $u_i \cap c \neq \varnothing$ and $c
      \subseteq u_0'$, hence $(u_i \cap u_0') \cap c \neq \varnothing$. Since $c
      \subseteq \overline{\bigcup E}$, $(u_i \cap u_0') \cap
      \overline{\bigcup E} \neq \varnothing$ and thus, because $u_i \cap
      u_0'$ is open, $u_i \cap f = (u_i \cap u_0') \cap \bigcup E \neq
      \varnothing$. It follows that $u_i \cap \overline{f} \neq \varnothing$.

      Therefore $\overline{f} \in U$.

    \item By definition $f \subseteq \bigcup E$. Fix now some $e' \in E$. Then,
      $c \cap \overline{e'} \neq \varnothing$ hence $u_0' \cap \overline{e'}
      \neq \varnothing$ and thus $u_0' \cap e' \neq \varnothing$ as well since
      $u_0'$ is open. But $e' \subseteq \bigcup E$, therefore $f \cap
      e' \neq \varnothing$.

      It follows that $\overline{f} \in \suchthat{\overline{e}}{e \subseteq E
        \mathand \forall e' \in E, e \cap e' \neq \varnothing }$
    \end{itemize}
    Every open that contains $c$ intersects $\suchthat{\overline{e}}{e \subseteq
      E \mathand \forall e' \in E, e \cap e' \neq \varnothing }$, therefore $c$
    is in the closure of this latter set, namely \eqref{eq:proj-proj-distrPP}.
    \qedhere
  \end{description}
\end{proof}

A corollary of \Cref{lem:distrVV-weak-proj-lifting-distrPP} is that $\distrVV$
is a weak distributive law (this was already known). More generally,
\Cref{thm:YB-weakly-lift-weak-law} can be adapted to characterize the existence
of weak $\proj$-liftings of weak distributive laws.

\begin{prop}
  \label{prop:YB-weakly-proj-lift-weak-law}
  Weak distributive laws $\distrTS$, $\distrTR$ and $\distrSR$ satisfy the
  $\proj$-Yang-Baxter equation, given by \Cref{cd:pi-YB}, if and only if
  $\distrSR$ weakly $\proj$-lifts to $\EMT$, i.e., if there is a natural
  transformation $\distrliftSliftR$ such that
  \Cref{cd:weak-law-weak-lifting:proj} commutes.
  \[
    \begin{tikzcd}[ampersand replacement=\&]
      \funcTSR \& \funcTRS \& \funcRTS \& \funcRST \\
      \funcSTR \&\&\& {\funcRST\funcT} \\
      \funcSRT \& \funcRST \& {\funcT\funcRST} \& {\funcRT\funcST}
      \arrow["{\funcT \distrSRname}", from=1-1, to=1-2]
      \arrow["{\distrTSname \funcR}"', from=1-1, to=2-1]
      \arrow["{\distrTRname \funcS}", from=1-2, to=1-3]
      \arrow["{\funcR \distrTSname}", from=1-3, to=1-4]
      \arrow["{\funcS \distrTRname}"', from=2-1, to=3-1]
      \arrow["{\funcRS \multT}"', from=2-4, to=1-4]
      \arrow["{\distrSRname \funcT}"', from=3-1, to=3-2]
      \arrow["{\unitT \funcRST}"', from=3-2, to=3-3]
      \arrow["{\distrTRname \funcST}"', from=3-3, to=3-4]
      \arrow["{\funcR \distrTSname \funcT}"', from=3-4, to=2-4]
      \diagramlabel[$\proj$-YB]{1-2}{3-3}{pi-YB}
    \end{tikzcd}
  \]
  If this holds, $\distrliftSliftRname$ is a weak distributive law.
\end{prop}
\begin{proof}
  The proof is almost a copy-paste of that of
  \Cref{thm:YB-weakly-lift-weak-law}.

  We apply \Cref{thm:weak-liftings}: recall that $\funcSR$ and $\funcRS$ have
  weak liftings $\liftSliftR$ and $\liftRliftS$ thanks to the laws $\funcS
  \distrTRname \circ \distrTSname \funcR$ and $\funcR\distrTSname \circ
  \distrTRname \funcS$. Then $\distrSR$ weakly $\proj$-lifts to $\EMT$ as
  $\distrliftSliftR$ if and only if the corresponding instance of
  \Cref{cd:weak-lifting:proj:condition} commutes: this is exactly
  \Cref{cd:pi-YB} (up to commuting of natural transformations). Because the
  correspondence is functorial, it preserves any equation satisfied by
  $\distrSRname$, $\unitS$, $\unitR$, $\multS$ and $\multR$ (they all weakly
  $\proj$-lift to $\EMT$): since $\distrSRname$ is a weak distributive law, so
  is $\distrliftSliftRname$.
\end{proof}

The point of
\Cref{prop:YB-weakly-proj-lift-weak-law,prop:YB-weakly-proj-lift-weak-law} is that
they make it easier to exhibit weak distributive laws in a category of algebras:
to construct one there, it suffices to have one in the underlying category
$\catC$ and to check a single equation, the $\proj$-Yang-Baxter equation.

Still, working with the latter equation may be quite tedious, as it involves up
to four composed layers of functors. In fact, in
\Cref{lem:distrVV-weak-proj-lifting-distrPP} we did not use this
$\proj$-Yang-Baxter equation at all, and instead directly proved that
$\distrVVname$ was a weak $\proj$-lifting because we were already able to take
for granted that its components were morphisms of $\Ult$-algebras, i.e.,
continuous functions.

Another problem with \Cref{prop:YB-weakly-proj-lift-weak-law} is that even if we
manage to disprove its prerequisites, we only get as a result that there is no
weak $\proj$-lifting of the weak distributive law $\distrSR$ to $\EMT$, but we
do not learn anything about other possible meaningful weak distributive laws
$\distrliftSliftRtype$.

For these two reasons we do not try to apply this theorem to show the existence
or non-existence of weak $e$-liftings of $\distrPP$ to $\EMP$ and $\EMD$.

Instead, in the rest of this work, we will focus on weak distributive laws of
the shape $\distrSPtype$ and $\distrSliftPtype$, i.e., over powerset-like
monads. For every set $X$, $\funcP X$ is a poset (ordered with inclusion), and
similarly the carrier set of any $\liftP X$ in some category $\EMT$ is ordered
as a subset of $\funcP \rightEMT X$, via the inclusion $\incT_\funcP: \rightEMT
\liftP \Rightarrow \funcP \rightEMT$ (this function is a split monomorphism
hence an injection). $\KlP$- and $\KlliftP$-arrows with the same domain and
codomain can thus be ordered pointwise, and the weak distributive law $\distrPP$
enjoys the following monotonicity property:
\[ \forall f,g: X \rightrightarrows \funcP Y, f \le g \Rightarrow \distrPPname_Y
  \circ \funcP f \le \distrPPname_Y \circ \funcP g \]We will show in \Cref{sec:lifting-monotone-laws} that there are no weak
distributive laws that satisfy this property in $\JSL$ and $\Conv$. By the lemma
below, this will be enough to conclude that $\distrPP$ does not weakly
$\proj$-lift to $\EMP$ or $\EMD$.

\begin{lem}
  \label{lemma:weaklifting-preserves-monotonicity}
  In $\Set$, consider weak distributive laws $\distrTS$, $\distrTP$
  and $\distrSP$ such that both the mappings
  \begin{equation}
    \label{eq:wdl:extension}
    X \xrightarrow{f} \funcP Y \qquad \mapsto \qquad \funcS X
    \xrightarrow{\funcS f} \funcSP Y \xrightarrow{\distrSPname_Y} \funcPS Y
  \end{equation}
  and the components of $\incT_{\funcP}: \funcP \rightEMT \Rightarrow \funcP
  \rightEMT$ are monotone functions.

  Then, if $\distrliftSliftP$ is a weak $e$-lifting of $\distrSPname$, the
  mappings
  \[ \alg{X}{x} \xrightarrow{f} \liftP \alg{Y}{y} \qquad \mapsto \qquad \liftS
    \alg{X}{x} \xrightarrow{\liftS f} \liftSliftP \alg{Y}{y}
    \xrightarrow{\distrliftSliftPname_{\alg{Y}{y}}} \liftPliftS \alg{Y}{y} \]are also monotone functions.
\end{lem}

\begin{proof}
  Fix such a $\distrliftSliftPname$, and consider some $f,g: \alg{X}{x} \to \liftP
  \alg{Y}{y}$ in $\EMT$ such that $f \le g$, i.e., such that, in $\Set$,
  \[\begin{tikzcd}[row sep=tiny]
      & {\rightEMT \liftP Y} & \\
      {\rightEMT X} && {\funcP \rightEMT Y} \\
      & {\rightEMT \liftP Y}\arrow["{(\incT_{\funcP})_Y}", curve={height=-9pt}, from=1-2, to=2-3]\arrow["{\rightEMT g}", curve={height=-9pt}, from=2-1, to=1-2]\arrow["\rightEMT f"', curve={height=9pt}, from=2-1, to=3-2]\arrow["\le"{marking, allow upside down}, draw=none, from=3-2, to=1-2]\arrow["{(\incT_{\funcP})_Y}"', curve={height=9pt}, from=3-2, to=2-3]\end{tikzcd}\]

  By monotonicity of the mapping \eqref{eq:wdl:extension}, we deduce
  \[\begin{tikzcd}[row sep=tiny, column sep=large]
      & {\funcS \rightEMT \liftP Y} & {\funcSP \rightEMT Y} & \\
      {\funcS \rightEMT X} &&& {\funcPS \rightEMT Y} \\
      & {\funcS \rightEMT \liftP Y} & {\funcSP \rightEMT Y}\arrow[""{name=0, anchor=center, inner sep=0}, "{\funcS
        (\incT_{\funcP})_Y}", from=1-2, to=1-3]\arrow["{\distrSPname_{\rightEMT Y}}", curve={height=-9pt}, from=1-3,
      to=2-4]\arrow["{\funcS \rightEMT g}", curve={height=-9pt}, from=2-1, to=1-2]\arrow["{\funcS \rightEMT f}"', curve={height=9pt}, from=2-1, to=3-2]\arrow[""{name=1, anchor=center, inner sep=0}, "{\funcS
        (\incT_{\funcP})_Y}"', from=3-2, to=3-3]\arrow["{\distrSPname_{\rightEMT Y}}"', curve={height=9pt}, from=3-3,
      to=2-4]\arrow["\le"{marking, allow upside down}, draw=none, from=1, to=0]\end{tikzcd}\]Because the ordering is defined pointwise, this last inequality is preserved
  when pre-composing with $\incT_{\funcS}$. It is also preserved when
  post-composing with $\closeT_{\funcP}$ and $\funcP \projT_{\funcS}$, because
  these are monotone functions: the former one is so by assumption, and the
  latter one as the image of a function by $\funcP$. Hence:
  \[\begin{tikzcd}[sep=large]
      {\funcS \rightEMT X}&[-5pt] {\funcS \rightEMT \liftP Y} & {\funcSP \rightEMT Y} & {\funcPS \rightEMT Y}&[-5pt] {\funcP \rightEMT \liftS Y} \\
      {\rightEMT \liftS Y}&&&& {\funcP \rightEMT \liftS X} \\
      {\funcS \rightEMT X}& {\funcS \rightEMT \liftP Y} & {\funcSP \rightEMT Y} & {\funcPS \rightEMT Y} & {\funcP \rightEMT \liftS Y}\arrow["{\funcS \rightEMT g}", from=1-1, to=1-2]\arrow["{\funcS (\incT_{\funcP})_Y}", from=1-2, to=1-3]\arrow["{\distrSPname_{\rightEMT Y}}", from=1-3, to=1-4]\arrow["{\funcP (\projT_{\funcS})_Y}", from=1-4, to=1-5]\arrow["{(\closeT_{\funcP})_{\liftS Y}}", from=1-5, to=2-5]\arrow["{(\incT_{\funcS})_X}", from=2-1, to=1-1]\arrow["{(\incT_{\funcS})_X}"', from=2-1, to=3-1]\arrow["{\funcS \rightEMT f}"', from=3-1, to=3-2]\arrow["{\funcS (\incT_{\funcP})_Y}"', from=3-2, to=3-3]\arrow["\le"{marking, allow upside down}, draw=none, from=3-3, to=1-3]\arrow["{\distrSPname_{\rightEMT Y}}"', from=3-3, to=3-4]\arrow["{\funcP (\projT_{\funcS})_Y}"', from=3-4, to=3-5]\arrow["{(\closeT_{\funcP})_{\liftS Y}}"', from=3-5, to=2-5]\end{tikzcd}\]Moreover, the diagram below, for $f$, commutes, and so does the one does with
  $g$ in place of $f$.
  \[\begin{tikzcd}[sep=large]
      {\rightEMT \liftS X} &[-10pt] {\rightEMT \liftSliftP Y} & {\rightEMT \liftSliftP Y} & {\rightEMT \liftPliftS} & {\funcP \rightEMT \liftS Y} \\
      && {\funcS \rightEMT \liftP Y} \\
      {\funcS \rightEMT X} & {\funcS \rightEMT \liftP Y} & {\funcSP
        \rightEMT Y} & {\funcPS \rightEMT Y} & {\funcP \rightEMT \liftS
        Y}\arrow["{\rightEMT \liftS f}", from=1-1, to=1-2]\arrow["{(\incT_{\funcS})_X}"', from=1-1, to=3-1]\arrow[equals, from=1-2, to=1-3]\arrow["{(\incT_{\funcS})_{\liftP Y}}"{description}, from=1-2,
      to=3-2]\arrow["{\rightEMT \lift \distrSPname_Y}", from=1-3, to=1-4]\arrow["{(\incT_{\funcP})_{\liftS Y}}", from=1-4, to=1-5]\arrow["{(\projT_{\funcS})_{\liftP Y}}"{description}, from=2-3,
      to=1-3]\arrow["{\funcS \rightEMT f}"', from=3-1, to=3-2]\arrow["{\funcS (\incT_{\funcP})_X}"', from=3-2, to=3-3]\arrow["{\funcS (\projT_{\funcP})_Y}"{description}, from=3-3, to=2-3]\arrow["{\distrSPname_{\rightEMT Y}}"', from=3-3, to=3-4]\arrow["{\funcP (\projT_{\funcS})_Y}"', from=3-4, to=3-5]\arrow["{(\projT_{\funcP})_{\liftS Y}}"{description, pos=0.6,
        name=0}, curve={height=-12pt}, from=3-5, to=1-4]\arrow["{(\closeT_{\funcP})_{\liftS Y}}"', from=3-5, to=1-5]\diagramref{1-1}{3-2}{"\text{(nat.)}"}\diagramref{1-2}{3-3}{"\text{(splitting)}"}\diagramref{2-3}{0}{"\text{\ref{cd:weak-lifting:proj}}"}\diagramref{0}{1-5}{"\text{(splitting)}"}\end{tikzcd}\]
  Hence the last inequality above finally rewrites as
  \[\begin{tikzcd}[row sep=tiny, column sep=large]
      & {\rightEMT \liftSliftP Y} & {\rightEMT \liftPliftS Y} & \\
      {\rightEMT \liftS X} &&& {\funcP \rightEMT \liftS Y} \\
      & {\rightEMT \liftSliftP Y} & {\rightEMT \liftPliftS Y} & \qedhere \\
      \arrow[""{name=0, anchor=center, inner sep=0}, "{\rightEMT \lift
        \distrSPname_Y}", from=1-2, to=1-3]\arrow["{(\incT_{\funcP})_{\liftS Y}}", curve={height=-9pt},
      from=1-3, to=2-4]\arrow["{\rightEMT \liftS g}", curve={height=-9pt}, from=2-1,
      to=1-2]\arrow["{\rightEMT \liftS f}"', curve={height=9pt}, from=2-1,
      to=3-2]\arrow[""{name=1, anchor=center, inner sep=0}, "{\rightEMT \lift
        \distrSPname_Y}"', from=3-2, to=3-3]\arrow["{(\incT_{\funcP})_{\liftS Y}}"', curve={height=9pt},
      from=3-3, to=2-4]\arrow["\le"{marking, allow upside down}, draw=none, from=1, to=0]\end{tikzcd}\]\end{proof}

\section{Interliminary: Relational Extensions}\label{sec:relational-extensions}

When considering laws $\distrTStype$, it may happen that $\KlS$ carries more
structure than just that of a category, in which case it is natural to ask that
the corresponding extension of $\funcT$ to $\KlS$ preserve this additional
structure: intuitively, the more structure preserved, the more semantically
canonical the resulting law should be.

In particular, in \Cref{lemma:weaklifting-preserves-monotonicity} above we have
used the fact that sets of arrows of $\KlP \cong \Rel$ are ordered by inclusion,
and the monotonicity property of weak distributive laws studied there is
equivalent to the monotonicity of the corresponding extended functors (recall
from \Cref{sec:preliminaries:extensions} how the latter are constructed):
\begin{defi}[monotone (weak) distributive laws and extensions]
  A (weak) distributive law $\distrTStype$ and its (weak) extension $\extS$ to
  $\KlS$ are called \emph{monotone} when the sets of morphisms of $\KlS$ admit a
  canonical order that is preserved by $\extS$.
\end{defi}
It turns out that the existence of monotone extensions of functors and natural
transformations from $\Set$ to $\Rel$ can be fully characterized, and hence so
can be monotone weak distributive laws over the powerset monad.

In this section, we recall this characterization and extend it to monotone weak
distributive laws over certain powerset-like monads.
\Cref{sec:rel-exts:reg-cats} therefore starts by recalling the theory of
\emph{regular categories}: these are categories with a well-behaved notion of
relations, so that in particular monotone extensions to relations therein are
fully characterized. We then consider certain subcategories of relations in
\Cref{sec:rel-exts:transposable-subcats}, to which we generalize the latter
characterization in \Cref{sec:rel-exts:thms}. In particular, when these
subcategories of relations are also Kleisli categories, this characterizes
monotone weak distributive laws over the corresponding (powerset-like) monads.

\subsection{Regular Categories}\label{sec:rel-exts:reg-cats}

We start by briefly recalling the theory of regular categories: see for example
\cite[\textsection 2]{borceuxHandbookCategoricalAlgebra1994a} for a more
in-depth treatment.

Let $\catC$ be a finitely complete category. The \emph{kernel pair} of an arrow
$f: X \to Y$ is the pullback $p_1,p_2: X \times_Y X \rightrightarrows X$ of the
cospan $X \xrightarrow{f} Y \xleftarrow{f} X$. Assume $\catC$ has all the
coequalizers of these kernel pairs: these coequalizers are called \emph{regular
  epimorphisms}. $\catC$ is then called \emph{regular} if these conditions are
satisfied and if the regular epimorphisms are stable under pullbacks\footnote{a
  class of arrows is stable under pullbacks when for every pullback square $a
  \circ b = c \circ d$, if $a$ is in the class then so is $d$}. Regular
categories enjoy the fact that every arrow $f: X \to Y$ may be factored as a
regular epimorphism (denoted with $\onto$) followed by a monomorphism (denoted
with $\subto$), and this factorization is unique up to unique isomorphism -- one
should think of it as factoring an arrow through its image. $\Set$ is a
canonical example of a regular category (with surjections as regular
epimorphisms and injections as monomorphisms).

Regular categories are useful because they are categories where we can speak of
relations: if $\catC$ is a regular category, a $\catC$-relation between two
objects $X$ and $Y$ is a subobject $r: R \subto X \times Y$. Relations are
preordered: $r: R \subto X \times Y$ is smaller than $s: S \subto X \times Y$,
written $r \le s$, when there is a monomorphism $m: R \subto S$ such that $s
\circ m = r$. One may form the category $\RelC$ with objects those of $\catC$
and arrows $X \leftrightsquigarrow Y$ the equivalence classes of relations
between $X$ and $Y$. The identity on $X$ is the diagonal $\langle \id_X, \id_X
\rangle: X \subto X \times X$, and composition of relations $r = \langle r_X,
r_Y \rangle: R \subto X \times Y$ and $s = \langle s_Y, s_Z \rangle: S \subto Y
\times Z$, written $s \cdot r: X \leftrightsquigarrow Z$, is constructed as in
\Cref{fig:rel-composition}: by considering the pullback $R \times_Y S \to R
\times S$ of $R \xrightarrow{r_Y} Y \xleftarrow{s_Y} S$, and taking the
monomorphism in the factorization of $R \times_Y S \rightarrow R \times S
\xrightarrow{r_X \times s_Z} X \times Z$.

\begin{figure}[h]
  \centering
  \caption{Composition in $\RelC$, as a diagram.}
\begin{tikzcd}[column sep=scriptsize]
      && {R \times_Y S} \\
      & R & {} & S \\
      X && Y && Z
      \arrow[from=1-3, to=2-2, dotted]
      \arrow[from=1-3, to=2-4, dotted]
      \arrow["\pullbackcorner"{name=1, anchor=center, pos=.05, rotate=-45},
      draw=none, from=1-3, to=3-3, dotted]
      \arrow[two heads, from=1, to=2-3, dotted]
      \arrow["{r_X}"', from=2-2, to=3-1]
      \arrow["{r_Y}"{description}, from=2-2, to=3-3]
      \arrow[curve={height=6pt}, from=2-3, to=3-1]
      \arrow[curve={height=-6pt}, from=2-3, to=3-5]
      \arrow["{s_Y}"{description}, from=2-4, to=3-3]
      \arrow["{s_Z}", from=2-4, to=3-5]
    \end{tikzcd}
\label{fig:rel-composition}
\end{figure}

The graph functor $\GrphC: \catC \to \RelC$ sends objects to themselves and an
arrow $f: X \rightarrow Y$ to the relation $\langle \id_X, f \rangle: X \subto X
\times Y$. There is also a contravariant transpose functor $-^\dagger: \RelC \to
\RelC$ that sends objects on themselves and a relation $\langle r_X, r_Y
\rangle: R \subto X \times Y$ to $\langle r_Y, r_X \rangle: R \subto Y \times
X$. All in all, a relation $\langle f,g \rangle: R \subto X \times Y$ can also
be written as the composite $\GrphC g \cdot (\GrphC f)^\dagger$: we will often
omit $\Grph$ and write directly $g \cdot f^\dagger$.

The following recalls some useful properties of relations in regular categories
we may make use of implictly later~\cite[\textsection
1.4-7]{carboni2categoricalApproachChange1991}.

\newcounter{calculus-relations}
\begin{prop}\label{prop:calculus-relations}
  Let $\catC$ be a regular category. Consider first a commuting diagram
  \[\begin{tikzcd}[column sep=large]
        & Z & \\
        X & {Z'} & Y\arrow["f"', from=1-2, to=2-1]\arrow["h"{description}, from=1-2, to=2-2]\arrow["g", from=1-2, to=2-3]\arrow["{f'}", from=2-2, to=2-1]\arrow["{g'}"', from=2-2, to=2-3]\end{tikzcd}\]\begin{enumerate}
  \item \label{prop:calculus-relations:inequality-witness} If there is an arrow
    $h: Z \to Z'$ making the above diagram commute, then, $g \cdot f^\dagger \le
    g' \cdot {f'}^\dagger$.
  \item If $h$ is a regular epimorphism, the above inequality is an equality.
  \item \label{prop:calculus-relations:inequality-witness-iff} When $\langle
    f',g' \rangle: Z' \to X \times Y$ is a monomorphism, the above inequality
    holds if and only if such an $h: Z \to Z'$ exists, and it is an equality if
    and only if $h$ is a regular epimorphism.
    \setcounter{calculus-relations}{\value{enumi}}
  \end{enumerate}
  Next, consider a commuting diagram
  \[\begin{tikzcd}[column sep=large]
      & W & \\
      X && Y \\
      & Z
      \arrow["{g'}"', from=1-2, to=2-1]
      \arrow["{f'}", from=1-2, to=2-3]
      \arrow["f"', from=2-1, to=3-2]
      \arrow["g", from=2-3, to=3-2]
    \end{tikzcd}\] and let $h: W \to X \times_Z Y$ be the
  unique arrow given by the universal property of pullbacks.
  \begin{enumerate} \setcounter{enumi}{\value{calculus-relations}}
  \item Then, ${g'}^\dagger \cdot f' \le g^\dagger \cdot f$.\item\label{prop:calculus-relations:near-pullback} The above inequality is an
    equality if and only if $h$ is a regular epimorphism, in which case we say
    the diagram is a \emph{near pullback}.
  \item The diagram is a pullback if and only if $\langle g',f' \rangle$ is
    a monomorphism and the above inequality is an equality.
    \setcounter{calculus-relations}{\value{enumi}}
  \end{enumerate}
  Finally, let $f: X \to Y$ be any arrow.
  \begin{enumerate} \setcounter{enumi}{\value{calculus-relations}}
  \item \label{prop:calculus-relations:counit} $f \cdot f^\dagger \le \id_Y$.
  \item \label{prop:calculus-relations:counit-iso} The above inequality is an
    equality if and only if $f$ is a regular epimorphism.
  \item \label{prop:calculus-relations:unit} $\id_X \le f^\dagger \cdot f$.
  \item \label{prop:calculus-relations:unit-iso} The above inequality is an equality if and only if $f$ is a monomorphism.
  \item \label{prop:calculus-relations:map} Conversely, given any two relations
    $r: X \leftrightsquigarrow Y$ and $s: Y \leftrightsquigarrow X$ such that $r
    \cdot s \le \id_Y$ and $\id_X \le s \cdot r$, there is a $\catC$-arrow $f: X
    \to Y$ such that $r = f$ and $s = f^\dagger$.
  \end{enumerate}
\end{prop}

The above construction of the category of relations in a regular category is in
fact $2$-functorial, in the sense that certain functors and natural
transformations can be extended to categories of relations: a functor between
regular categories is said to be \emph{nearly cartesian} when it preserves near
pullbacks (the latter were defined in
\cref{prop:calculus-relations:near-pullback} above),
while a natural transformation $\alpha: \funcF \Rightarrow \funcG$ between
nearly cartesian functors is \emph{nearly cartesian} when its naturality squares
$\alpha \circ \funcF f = \funcG f \circ \alpha$ are near pullbacks.

\begin{thm}\label{thm:relational-extension}
  Let $\funcF: \catC \to \catD$ be a functor between regular categories.
  $\funcF$ has a \emph{relational extension}, i.e., there is an order-preserving
  functor $\Rel[\funcF]: \RelC \to \RelD$ such that $\Rel[\funcF] \GrphC =
  \GrphD \funcF$, if and only if $\funcF$ is nearly cartesian. In that case
  there is only one possible such $\Rel[\funcF]$, given by $\Rel[\funcF]\left(g
    \cdot f^\dagger\right) = (\funcF g) \cdot (\funcF f)^\dagger$.

  Let $\funcF, \funcG: \catC \rightrightarrows \catD$ be two such functors and
  $\alpha: \funcF \Rightarrow \funcG$ be a natural transformation between them.
  Then $\Rel[\alpha]$, given by $\Rel[\alpha] \GrphD = \GrphD \alpha$, is a
  natural transformation $\Rel[\funcF] \Rightarrow \Rel[\funcG]$, called the
  \emph{relational extension} of $\alpha$, if and only if $\alpha$ is nearly
  cartesian.
\end{thm}
Just like we omit to write $\Grph$, we will often omit to write the functor
$\Rel$, and instead write $\funcF(g \cdot f^\dagger)$ for $\Rel[\funcF](g \cdot
f^\dagger) = \funcF g \cdot (\funcF f)^\dagger$ and $\alpha$ for $\Rel(\alpha)$.

In $\Set$, relational extensions were first constructed and characterized by
Barr~\cite{barrRelationalAlgebras1970}. This result was later extended to
regular categories, resulting in \Cref{thm:relational-extension}: the case of
functors is usually attributed to Carboni, Kelly and Wood~\cite[\textsection
4.3]{carboni2categoricalApproachChange1991} although it already appears in
\cite[\textsection 2.10]{adamekAutomataAlgebrasCategories1989} in the more
general setting of relations in a category equipped with a factorization system;
the case of natural transformations is due to
Schubert~\cite[Corollary~1.5.7]{schubertLaxAlgebrasScenic2006}.

\subsection{Subcategories of Relations}\label{sec:rel-exts:transposable-subcats}

\Cref{thm:relational-extension} is particularly useful to build weak
distributive laws over monads whose Kleisli categories are subcategories of
relations -- we call these monads \emph{sub-power-object monads}:

\begin{defi}[sub-power-object monad]
  \label{def:subpowerobject-monad}
  A monad $\monadS$ on a regular category $\catC$ is a \emph{sub-power-object}
  monad when $\KlS$ is a wide\footnote{a wide subcategory is a subcategory that
    contains all objects of the bigger category} subcategory of $\RelC$ such
  that $\GrphC$ factors as $\catC \xrightarrow{\leftKlS} \KlS \hookrightarrow
  \RelC$.
\end{defi}

If $\funcS$ is a sub-power-object monad on $\catC$, an extension to $\KlS$ can
be constructed by first finding a monotone extension to $\RelC$ using
\Cref{thm:relational-extension}, and then restricting this extension to $\KlS$.
In fact, most examples of weak distributive laws in the literature are obtained
in this way:

\begin{exa}[monotone extensions]
  \label{ex:monotone-extensions}
  $\Set$ is a regular category whose regular epimorphisms are the surjections
  and such that $\RelSet = \Rel \cong \KlP$. The endofunctors and the
  multiplications of the monads $\funcP$ and $\funcD$ are nearly cartesian,
  hence $\funcP$ and $\funcD$ have monotone weak extensions to $\KlP$ (given in
  \Cref{ex:weak-extensions}): this is how the weak distributive laws
  $\distrPPtype$ and $\distrDPtype$ of \Cref{ex:weak-distr-laws} were originally
  constructed~\cite{garnerVietorisMonadWeak2019,goyCombiningProbabilisticNondeterministic2020}.

  $\KHaus$ is also a regular category: its regular epimorphisms are the
  surjective continuous functions and $\RelKHaus$ is the category of compact
  Hausdorff spaces and closed relations, i.e., closed subsets of $X \times Y$,
  while $\KlV$ is the category of compact Hausdorff spaces and continuous
  relations, i.e., closed relations $r: X \leftrightsquigarrow Y$ such that, for
  every open $u$ of $Y$ $r^{-1}[u]$ is open in $X$. The endofunctor and
  multiplication of $\funcV$ are nearly cartesian hence they have relational
  extensions, which restrict to continuous relations: this yields a monotone
  weak extension of $\funcV$ to $\KlV$, and this is how the weak distributive
  law $\distrVVtype$ of \Cref{ex:weak-distr-laws} was originally
  constructed~\cite{goyPowersetLikeMonadsWeakly2021}.
\end{exa}

However, \Cref{thm:relational-extension} can only be used to disprove the
existence of monotone weak distributive laws over a sub-power-object monad
$\funcS$ when $\KlS$ is equivalent to $\RelC$ (because otherwise there could be
weak extensions to $\KlS$ that do not extend to $\RelC$), in which case $\funcS$
is called a \emph{power-object} monad. It turns out that the existence of such a
monad holds if and only if $\catC$ is an \emph{elementary
  topos}~\cite[\textsection 1.911]{freydCategoriesAllegories1990} (this can be
taken as a definition), and thus has much more structure than that of a regular
category. For example, $\Set$ is an elementary topos, but $\KHaus$ is not:
\Cref{thm:relational-extension} is not enough to disprove the existence of
monotone weak distributive laws over $\funcV$.

In \Cref{sec:lifting-monotone-laws} we will study the existence of monotone weak
distributive laws over the weak liftings of $\funcP$ to $\JSL$ and $\Conv$,
which we will have shown in \Cref{sec:lifted-kleisli} to be sub-power-object
monads. To be able to fully characterize the existence of such monotone weak
distributive laws, we therefore need to generalize
\Cref{thm:relational-extension} to subcategories of relations. We focus on those
of the following shape.

\begin{defi}
  Let $\catGamma$ be a class of morphims of a regular category $\catC$ that is
  stable under pre-composition with isomorphisms. We define
  \begin{itemize}
  \item $\catC \cdot \catGamma^\dagger$ to be the class of $\catC$-relations
    $r: X \leftrightsquigarrow Y$ given by jointly monic spans $\langle
    r_X, r_Y \rangle$ such that $r_X$ is a $\catGamma$-arrow;
  \item $\Grph[\catGamma]: \catC \to \catC \cdot \catGamma^\dagger$ as the
    restriction of $\GrphC: \catC \to \RelC$ to $\catC \cdot \catGamma^\dagger$.
\end{itemize}
\end{defi}

\begin{lem}
  \label{lem:pullback-stable-subcategory}
  $\catC \cdot \catGamma^\dagger$ is a wide subcategory of $\RelC$ if and only
  if $\catGamma$ is a wide subcategory of $\catC$ such that
  \begin{itemize}
  \item $\catGamma$-arrows are stable under pullbacks (in $\catC$);
  \item if $f \circ e$ is a $\catGamma$-arrow and $e$ is a regular
    epimorphism (in $\catC$), $f$ is a $\catGamma$-arrow.
  \end{itemize}
  If this holds, we call $\catGamma$ a \emph{transposable} subcategory of
  $\catC$.
\end{lem}

\begin{proof}
  Suppose that $\catGamma$ is a wide subcategory of $\catC$ satisfying these
  conditions. Then $\catC \cdot \catGamma^\dagger$ contains all the identities
  of $\RelC$, because both $\catC$ and $\catGamma$ contain all the identities of
  $\catC$. Consider jointly monic spans $\langle f_1,g_1 \rangle: R_1 \subto X
  \times Y$ and $\langle f_2, g_2 \rangle: R_2 \subto Y \times Z $ such that
  $f_1$ and $f_2$ are in $\catGamma$, and let $\langle f',g' \rangle: S \subto
  R_1 \times R_2$ be the pullback of $g_1: R_1 \to Y$ and $f_2: R_2 \to X$. Then
  by the first property of $\catGamma$, $f'$ and thus $f_1 \circ f'$ are in
  $\catGamma$ ($\catGamma$ is a category hence stable under composition). Factor
  $\langle f_1 \circ f', g_2 \circ g' \rangle$ into a regular epimorphism
  followed by a jointly monic span: by the second property of $\catGamma$, the
  resulting jointly monic span is also in $\catC \cdot \catGamma^\dagger$.

  Conversely, suppose that $\catC \cdot \catGamma^\dagger$ is a wide subcategory
  of $\RelC$. $\catC \cdot \catGamma^\dagger$ contains all identities, hence so
  does $\catGamma$. Given $f: X \to Y$ and $g: Y \to Z$ in $\catGamma$,
  $f^\dagger \cdot g^\dagger$ is in $\catC \cdot \catGamma^\dagger$ (as the
  composite of two relations therein) hence $g \circ f$ is in $\catGamma$. If
  $g: Y \to Z$ is in $\catGamma$, then for any $f: X \to Z$ in $\catC$,
  $g^\dagger \cdot f$ is in $\catC \cdot \catGamma^\dagger$ (as the composite of
  two relations therein) hence the pullback of $g$ along $f$ must be in
  $\catGamma$. Finally, if $e: X \onto Y$ and $f: Y \to Z$ are such
  that $f \circ e$ is in $\catGamma$, then $e \cdot (f \circ e)^\dagger =
  f^\dagger$ is in $\catC \cdot \catGamma^\dagger$ and $f$ itself must be in
  $\catGamma$.
\end{proof}

\begin{exa}\label{ex:subpowerobject-monads}~
  Beyond $\funcP$, there are other examples of sub-power-object monads in
  $\Set$:
  \begin{itemize}
  \item For $\Surj$ the class of surjective functions, $\Set (\Surj)^\dagger$ is
    $\Kl{\funcP_*}$, the wide subcategory of $\Rel$ whose arrows are relations
    with non-empty images of elements. Because this is a category, $\Surj$ is a
    tranposable subcategory of $\Set$. In fact, in any regular category $\catC$,
    the wide subcategory whose arrows are the regular epimorphisms is
    transposable.
  \item For $\FinFib$ the class of functions with finite fibers, $\Set
    (\FinFib)^\dagger$ is $\Kl{\funcP_f}$, the wide subcategory of $\Rel$ whose
    arrows are relations with finite images of elements.
  \end{itemize}
  In \ref{sec:lifting-monotone-laws}, we will also see that being a
  sub-power-object monad with a Kleisli category of this shape is stable under
  weak lifting.
\end{exa}

\subsection{Subrelational Extensions}\label{sec:rel-exts:thms}

As a subcategory of $\RelC$, $\catC \cdot \catGamma^\dagger$ inherits its
partial ordering. Hence we now generalize \Cref{thm:relational-extension} to
characterize monotone extensions to such subcategories of relations. Those
functors and natural transformations that have monotone extensions to $\catC
\cdot \catGamma^\dagger$ are again characterized in terms of certain near
pullbacks: call \emph{near pullback along $\catGamma$} a near pullback
\[\begin{tikzcd}[sep=large]
    W & Y \\
    X & Z\arrow["{g'}", from=1-1, to=1-2]\arrow["{f'}"', from=1-1, to=2-1]\arrow["f", from=1-2, to=2-2]\arrow["g"', from=2-1, to=2-2]\end{tikzcd}\]in $\catC$ such that both $f$ and $f'$ are in $\catGamma$. Also call
\emph{$\catGamma$-sandwiched} regular epimorphisms those regular epimorphisms
$e: X \twoheadrightarrow Y$ in $\catC$ for which there are $\catGamma$-arrows
$f: X \to Z$ and $g: Y \to Z$ such that $g \circ e = f$. The functors that admit
monotone extensions to $\catC \cdot \catGamma^\dagger$ will then be exactly
those that satisfy the following three equivalent conditions.

\begin{lem}\label{lemma:nearly-cartesian}
  Let $\funcF: \catC \to \catD$ be a functor between regular categories $\catC$
  and $\catD$, and let $\catGamma$ and $\catDelta$ be transposable
  subcategories thereof. Then, the following conditions on $\funcF$ are
  equivalent:
  \begin{enumerate}
  \item $\funcF$ sends near pullbacks along $\catGamma$ to near pullbacks along
    $\catDelta$.
  \item $\funcF$ sends near pullbacks along $\catGamma$ to near pullbacks,
    and restricts to a functor $\catGamma \to \catDelta$.
  \item $\funcF$ sends pullbacks along $\catGamma$ to near pullbacks, restricts
    to a functor $\catGamma \to \catDelta$, and sends $\catGamma$-sandwiched
    regular epimorphisms to regular epimorphisms.
  \end{enumerate}
\end{lem}

\begin{proof}~
  \begin{description}
  \item[1 implies 2] If $\funcF$ sends near pullbacks along $\catGamma$
    to near pullbacks along $\catDelta$, then in particular it sends near
    pullbacks along $\catGamma$ to near pullbacks. Consider now some $f: X
    \to Y$ in $\catGamma$. The pullback
    \[\begin{tikzcd}[sep=large]
        X & Y \\
        X & Y\arrow[equals, from=1-1, to=1-2]\arrow["f"', from=1-1, to=2-1]\arrow["f", from=1-2, to=2-2]\arrow[equals, from=2-1, to=2-2]\arrow["\pullbackcorner"{anchor=center, pos=0.125}, draw=none, from=1-1,
        to=2-2]\end{tikzcd}\]is in particular a near pullback along $\catGamma$, hence its image by
    $\funcF$ is a near pullback along $\catDelta$ and $\funcF f$ is in
    $\catDelta$.
  \item[2 implies 3] Pullbacks are near pullbacks, hence, if $\funcF$ sends
    near pullbacks along $\catGamma$ to near pullbacks, then in particular
    it sends pullbacks along $\catGamma$ to near pullbacks. Consider now
    some regular epimorphisms $e: X \onto Y$ sandwiched by $f: X \to Z$ and
    $g: Y \to Z$ in $\catGamma$. Then, the diagram below on the left is a
    near pullback along $\catGamma$, hence its image, below on the right, is
    also a near pullback.
    \[\begin{tikzcd}[sep=large]
        X && \\
        & Y & Y \\
        & Z & Z\arrow["e"{description}, two heads, from=1-1, to=2-2]\arrow[curve={height=-12pt}, equals, from=1-1, to=2-3]\arrow["f"', curve={height=12pt}, from=1-1, to=3-2]\arrow[equals, from=2-2, to=2-3]\arrow["g"{description}, from=2-2, to=3-2]\arrow["g", from=2-3, to=3-3]\arrow[equals, from=3-2, to=3-3]\arrow["\pullbackcorner"{anchor=center, pos=0.125}, draw=none, from=2-2,
        to=3-3]\end{tikzcd}
      \qquad\qquad
      \begin{tikzcd}[sep=large]
        \funcF X && \\
        & \funcF Y & \funcF Y \\
        & \funcF Z & \funcF Z\arrow["\funcF e"{description}, two heads, from=1-1, to=2-2]\arrow[curve={height=-12pt}, equals, from=1-1, to=2-3]\arrow["\funcF f"', curve={height=12pt}, from=1-1, to=3-2]\arrow[equals, from=2-2, to=2-3]\arrow["\funcF g"{description}, from=2-2, to=3-2]\arrow["\funcF g", from=2-3, to=3-3]\arrow[equals, from=3-2, to=3-3]\arrow["\pullbackcorner"{anchor=center, pos=0.125}, draw=none, from=2-2,
        to=3-3]\end{tikzcd}\]Because $\funcF e$ is the unique arrow into the pullback of $\funcF g$
    and $\id_{\funcF Z}$, it is therefore a regular epimorphism.
  \item[3 implies 1] Consider a near pullback along $\catGamma$, as below: $f$
    and $f'$ are in $\catGamma$ and $e$ is a regular epimorphism.
    \[\begin{tikzcd}[sep=large]
        W && \\
        & {X \times_Z Y} & Y \\
        & X & Z\arrow["e"{description}, two heads, from=1-1, to=2-2]\arrow["{g'}", curve={height=-12pt}, from=1-1, to=2-3]\arrow["{f'}"', curve={height=12pt}, from=1-1, to=3-2]\arrow[from=2-2, to=2-3]\arrow[from=2-2, to=3-2]\arrow["f", from=2-3, to=3-3]\arrow["g"', from=3-2, to=3-3]\arrow["\pullbackcorner"{anchor=center, pos=0.125}, draw=none, from=2-2,
        to=3-3]\end{tikzcd}\]Its image by $\funcF$ is then a near pullback along $\catDelta$, because
    the image of the inner pullback is a near pullback, the image of $e$ is
    a regular epimorphism, and the images of $f$ and $f'$ are in $\catDelta$.
    \qedhere
  \end{description}
\end{proof}

When these conditions hold, we say $\funcF$ is \emph{nearly cartesian from
  $\catGamma$ to $\catDelta$} (when $\catGamma = \catC$ and $\catDelta = \catD$,
we still just say nearly cartesian). In particular, if $\funcF$ is nearly
cartesian, then it is also nearly cartesian from $\catGamma$ to $\catDelta$ if
and only if it restricts to a functor $\catGamma \to \catDelta$. Similarly, call
a natural transformation \emph{nearly cartesian from $\catGamma$} when, for
every $\catGamma$-arrow $f: X \to Y$, the commuting diagram below is a near
pullback.
\[\begin{tikzcd}[sep=large]
    {\funcF X} & {\funcG X} \\
    {\funcF Y} & {\funcG Y}\arrow["{\alpha_X}", from=1-1, to=1-2]\arrow["{\funcF f}"', from=1-1, to=2-1]\arrow["{\funcG f}", from=1-2, to=2-2]\arrow["{\alpha_Y}"', from=2-1, to=2-2]\diagramref{1-1}{2-2}{"\text{(nat.)}"}\end{tikzcd}\]

\begin{thm}\label{thm:partial-relational-extensions}
  Let $\catC$, $\catD$, $\catGamma$, $\catDelta$ and $\funcF: \catC \to \catD$
  be as in \Cref{lemma:nearly-cartesian}. $\funcF$ has a
  \emph{$(\catGamma,\catDelta)$-relational extension}, i.e., there is an
  order-preserving functor $\Rel_{\catGamma,\catDelta}(\funcF): \catC \cdot
  \catGamma^\dagger \to \catD \cdot \catDelta^\dagger$ such that the diagram
  below commutes, if and only if $\funcF$ is nearly cartesian from $\catGamma$
  to $\catDelta$.
  \[\begin{tikzcd}[column sep=huge]
      {\catC \cdot \catGamma^\dagger} & {\catD \cdot \catDelta^\dagger} \\
      \catC & \catD
      \arrow["{\Rel_{\catGamma,\catDelta}(\funcF)}", from=1-1, to=1-2]
      \arrow["{\Grph[\catGamma]}", from=2-1, to=1-1]
      \arrow["\funcF"', from=2-1, to=2-2]
      \arrow["{\Grph[\catDelta]}"', from=2-2, to=1-2]
    \end{tikzcd}\]

  Let $\alpha: \funcF \Rightarrow \funcG$ be a natural transformation between
  two functors $\catC \to \catD$ having such $(\catGamma,\catDelta)$-relational
  extensions. Then $\Rel_{\catGamma,\catDelta}(\alpha)$, given by
  $\Rel_{\catGamma, \catDelta}(\alpha) \Grph[\catDelta] = \Grph[\catDelta]
  \alpha$, is a natural transformation $\Rel_{\catGamma, \catDelta}(\funcF)
  \Rightarrow \Rel_{\catGamma, \catDelta}(\funcG)$, called the
  \emph{$(\catGamma, \catDelta)$-relational extension} of $\alpha$, if and only
  if $\alpha$ is nearly cartesian from $\catGamma$.
\end{thm}

\begin{proof}
  We start with the proof of uniqueness of $\Rel_{\catGamma,
    \catDelta}(\funcF)$, which generalizes that in~\cite[\textsection
  5.3.11]{demoorCategoriesRelationsDynamic1992} and uses
  \Cref{prop:calculus-relations:counit,prop:calculus-relations:unit,prop:calculus-relations:map}
  of \Cref{prop:calculus-relations}. Consider an arbitrary functor $\funcF':
  \catC \cdot \catGamma^\dagger \to \RelD$ such that each mapping $(\catC \cdot
  \catGamma^\dagger)(X,Y) \to \RelD(X,Y)$ is monotone. Then, for every
  $\catGamma$-arrow $f: X \to Y$,
  \[ \funcF'f \cdot \funcF' {f}^\dagger = \funcF'(f \cdot {f}^\dagger) \le
    \funcF' \id_Y = \id_{\funcF' Y} \]\[ \id_{\funcF' X} = \funcF' \id_X \le \funcF'(f^\dagger \cdot f) = \funcF'
    {f}^\dagger \cdot \funcF'f \]Hence $\funcF' {f}^\dagger = {(\funcF' f)}^\dagger$. Because it is
  functorial, $\funcF'$ is therefore uniquely determined by its value along
  $\Grph[\catGamma]$: for every $f: Z \to X$ in $\catGamma$ and $g: Z \to Y$ in
  $\catC$,
  \[ \funcF'(g \cdot {f}^\dagger) = \funcF' g \cdot {(\funcF' f)}^\dagger \]Suppose now that $\funcF': \catC \cdot \catGamma^\dagger \to \catD \cdot
  \catDelta^\dagger$ is an extension of $\funcF$ along
  $(\Grph[\catGamma],\Grph[\catDelta])$, so that $\funcF' f = \funcF f$ for
  every $\catC$-arrow $f$. Then, for every $f: Z \to X$ in $\catGamma$ and $g: Z
  \to Y$ in $\catC$,
  \[ \funcF'(g \cdot f^\dagger) = \funcF g \cdot {(\funcF
      f)}^\dagger \]It follows that $\funcF$ is nearly cartesian from $\catGamma$ to
  $\catDelta$: first,
  \begin{itemize}
  \item for every $\catGamma$-arrow $f$, ${(\funcF f)}^\dagger = \funcF'
    f^\dagger$ is in $\catD \cdot \catDelta^\dagger$, hence $\funcF f$ is in $\catDelta$ and
    $\funcF$ restricts to a functor $\catGamma \to \catDelta$.
  \end{itemize}
  Recall moreover from \Cref{prop:calculus-relations},
  \Cref{prop:calculus-relations:near-pullback}, that a commuting diagram
  \[\begin{tikzcd}[sep=large]
      W & Y \\
      X & Z\arrow["{g'}", from=1-1, to=1-2]\arrow["{f'}"', from=1-1, to=2-1]\arrow["f", from=1-2, to=2-2]\arrow["g"', from=2-1, to=2-2]\end{tikzcd}\]is a near pullback if and only if ${f}^\dagger \cdot g = g' \cdot
  {f'}^\dagger$. Hence
  \begin{itemize}
  \item if the diagram above is a near pullback along $\catGamma$, then we
    may apply $\funcF'$ to it because $f$ and $f'$ are in $\catGamma$. Thus
    \[ {(\funcF f)}^\dagger \cdot \funcF g = \funcF'({f}^\dagger
      \cdot g) = \funcF'(g' \cdot {f'}^\dagger) = \funcF g' \cdot
      {(\funcF f')}^\dagger \]and the image of the diagram by $\funcF$ is again a near pullback.
  \end{itemize}
  Conversely, suppose that $\funcF$ is nearly cartesian from $\catGamma$ to
  $\catDelta$. We now show that $\funcF': \catC \cdot \catGamma^\dagger \to \RelD$, given, for $f:
  Z \to X$ and $g: Z \to Y$ such that $\langle f,g \rangle$ is a monomorphism, by
  \[ \funcF'(g \cdot f^\dagger) \coloneq \funcF g \cdot {(\funcF f)}^\dagger \]defines a monotone extension of $\funcF: \catC \to \catD$ along
  $(\Grph[\catGamma], \Grph[\catDelta])$.
  \begin{description}
  \item[restriction to $\catD \cdot \catDelta^\dagger$] For every $f$ and $g$ as above, $\funcF
    f$ is in $\catDelta$ hence $\funcF'(g \cdot f^\dagger)$ is in
    $\catD \cdot \catDelta^\dagger$.
  \item[extension] By definition, for every $f: X \to Y$ in $\catC$,
    \[ \funcF' \Grph[\catGamma] f = \funcF'(f \cdot {\id_X}^\dagger) = \funcF f
      \cdot {\id_{\funcF X}^\dagger} = \Grph[\catDelta] \funcF f \]\item[monotonicity] If $r = \langle r_X, r_Y \rangle: R \hookrightarrow X
    \times Y$ is smaller than $r' = \langle r'_X, r'_Y \rangle: R'
    \hookrightarrow Y$ there is an $m: R \hookrightarrow R'$ making the diagram
    below on the left commute. By \Cref{prop:calculus-relations},
    \Cref{prop:calculus-relations:inequality-witness}, its image by $\funcF$,
    below on the right, witnesses that $\funcF' r \le \funcF' r'$.
    \[\begin{tikzcd}[sep=large]
        & R & \\
        X & {R'} & Y\arrow["r_X"', from=1-2, to=2-1]\arrow["m"{description}, tail, from=1-2, to=2-2]\arrow["r_Y", from=1-2, to=2-3]\arrow["r'_X", from=2-2, to=2-1]\arrow["r'_Y"', from=2-2, to=2-3]\end{tikzcd}
      \qquad
      \begin{tikzcd}[sep=large]
          & \funcF R & \\
          \funcF X & {\funcF R'} & \funcF Y\arrow["\funcF r_X"', from=1-2, to=2-1]\arrow["\funcF m"{description}, tail, from=1-2, to=2-2]\arrow["\funcF r_Y", from=1-2, to=2-3]\arrow["\funcF r'_X", from=2-2, to=2-1]\arrow["\funcF r'_Y"', from=2-2, to=2-3]\end{tikzcd}\]\item[preservation of identities] By definition, $\funcF'(\id_X \cdot
      \id_X^\dagger) = \funcF \id_X \cdot {(\funcF \id_X)}^\dagger = \id_{\funcF
        X} \cdot \id_{\funcF X}^\dagger$.
    \item[preservation of composition] Because every $\catC \cdot
      \catGamma^\dagger$-relation is of the shape $g \cdot {f}^\dagger$ for $g$
      in $\catC$ and $f$ in $\catGamma$, we only need to show that $\funcF'$
      preserves compositions of the shapes $g \cdot {f}^\dagger$ and
      ${f}^\dagger \cdot g$.
    \begin{itemize}
    \item Let $f: Z \to X$ be in $\catGamma$ and $g: Z \to Y$ be in $\catC$, and
      factor $\langle f,g \rangle$ as
      \[\begin{tikzcd}[sep=large]
          & Z & \\
          X & R & Y\arrow["f"', from=1-2, to=2-1]\arrow["e"{description}, two heads, from=1-2, to=2-2]\arrow["g", from=1-2, to=2-3]\arrow["r_X", from=2-2, to=2-1]\arrow["r_Y"', from=2-2, to=2-3]\end{tikzcd}\]Then, by definition, $\funcF'(g \cdot f^\dagger) = \funcF r_Y \cdot
      {(\funcF r_X)}^\dagger$. But $e$ is a $\catGamma$-sandwiched regular
      epimorphism ($r_X$ is in $\catGamma$ because $f$ is), hence $\funcF e$ is
      a regular epimorphism and (by \Cref{prop:calculus-relations:counit-iso} of
      \Cref{prop:calculus-relations})
      \[ \funcF'(g \cdot f^\dagger) = \funcF r_Y \cdot (\funcF r_X)^\dagger =
        \funcF r_Y \cdot \funcF e \cdot {(\funcF e)}^\dagger \cdot {(\funcF
          r_X)}^\dagger = \funcF g \cdot {(\funcF f)}^\dagger \]\item Let $f: Y \to Z$ be in $\catGamma$ and $g: X \to Y$ be in $\catC$.
      Then, the composite ${f}^\dagger \cdot g$ is the relation
      $r: X \leftrightsquigarrow Y$ given by their pullback:
      \[\begin{tikzcd}[sep=large]
          R & Y \\
          X & Z
          \arrow["r_Y", from=1-1, to=1-2]
          \arrow["r_X"', from=1-1, to=2-1]
          \arrow["f", from=1-2, to=2-2]
          \arrow["g"', from=2-1, to=2-2]
          \arrow["\pullbackcorner"{anchor=center, pos=0.125}, draw=none, from=1-1,
          to=2-2]\end{tikzcd}\]Because this pullback is along $\catGamma$, its image by $\funcF$ is a
      near pullback and hence
      \[ \funcF'(f^\dagger \cdot g) = \funcF r_Y \cdot {(\funcF r_X)}^\dagger =
        {(\funcF f)}^\dagger \cdot \funcF g \]\end{itemize}
  \end{description}
  Finally, consider some natural transformation $\alpha: \funcF \Rightarrow
  \funcG$ as in the statement of the theorem. Then
  $\Rel_{\catGamma,\catDelta}(\alpha): \Rel_{\catGamma,\catDelta}(\funcF)
  \Rightarrow \Rel_{\catGamma,\catDelta}(\funcG)$ is as a natural transformation
  if and only if, for every $\langle f,g \rangle: R \subto X \times Y$ in $\catC
  \cdot \catGamma^\dagger$, i.e., with $f$ in $\catGamma$, $\alpha_Y \cdot
  \funcF g \cdot (\funcF f)^\dagger = \funcG g \cdot (\funcG f)^\dagger \cdot
  \alpha_X$. Of course $\alpha_Y \cdot \funcF g = \funcG g \cdot \alpha_R$
  because $\alpha$ is a natural transformation $\funcF \Rightarrow \funcG$,
  hence this is true if and only if $\alpha_R \cdot (\funcF f)^\dagger = (\funcG
  f)^\dagger \cdot \alpha_X$ (consider the case $g = \id_R$), i.e., if and only
  if the naturality square $\alpha_X \circ \funcF f = \funcG f \circ \alpha_R$
  is a near pullback.
\end{proof}

Again, we simply write $\funcF$ and $\alpha$ for the unique monotone extensions
of a functor $\funcF$ or a natural transformation $\alpha$ (when they exist).

As an example, a first corollary of
\Cref{thm:partial-relational-extensions,ex:subpowerobject-monads} is the
following:
\begin{cor}
  \label{cor:monotone-weak-distr-laws-powersets}
  In $\Set$, let $\monadT$ be a monad whose endofunctor and multiplication are nearly
  cartesian. Then:
  \begin{itemize}
  \item $\monadT$ has a (necessarily unique) monotone weak distributive law over
    the monad $\funcP_*$ of non-empty subsets;
  \item $\monadT$ has a (necessarily unique) monotone weak distributive law over
    the monad $\funcP_f$ of finite subsets if and only if $\funcT$ preserves
    functions with finite pre-images of elements.
  \end{itemize}
\end{cor}

We could more generally characterize the existence of monotone weak distributive
laws over these powerset monads for any monad on $\Set$ (not just the nearly
cartesian ones). Still, \Cref{cor:monotone-weak-distr-laws-powersets} is already
enough to prove that $\funcP$, $\funcD$ and $\Ult$ have monotone weak
distributive laws over $\funcP_*$, but that $\funcP$ and $\funcD$ (and most
likely $\Ult$ too, although no proof is known to the other) do not have monotone
weak distributive laws over $\funcP_f$:
\begin{itemize}
\item the copairing $[\id_{\naturals}, \id_{\naturals}]: \naturals \sqcup
  \naturals \to \naturals$ has finite fibers (each element has two preimages)
  but $\funcP [\id_{\naturals}, \id_{\naturals}]: \funcP(\naturals \sqcup
  \naturals) \to \funcP \naturals$ does not ($\naturals \in \funcP \naturals$
  has an infinite number of preimages);
\item the unique function $2 \to 1$ has finite fibers, but its image by
  $\funcD$, the unique function $[0,1] \cong \funcD 2 \to \funcD 1 \cong 1$,
  does not.
\end{itemize}
This answers in particular a question left open in
\cite[Table~3.1]{goyCompositionalityMonadsWeak2021}, asking whether such a
monotone law $\funcPP_f \Rightarrow \funcP_f \funcP$ exists.

\section{Kleisli Categories of Weakly Lifted Monads}\label{sec:lifted-kleisli}

In \Cref{sec:lifting-monotone-laws}, we will apply
\Cref{thm:partial-relational-extensions} to characterize the existence of
monotone weak distributive laws over the weak liftings of $\funcP$ to $\JSL$ and
$\Conv$. To be able to do this, we first need to describe the Kleisli categories
of the latter as subcategories of relations: this is the objective of the
current section.

We thus start, in \Cref{sec:lifted-kleisli:}, by describing the Kleisli
categories of weak liftings in terms of the Kleisli arrows of the monad being
weakly lifted. In \Cref{sec:lifted-kleisli:relations}, we then recall that
algebras often form regular categories, and thus show that the weak lifting of a
sub-power-object monad is again a sub-power-object monad. More precisely, in
\Cref{sec:lifted-kleisli:open} we describe the resulting Kleisli categories of
relations in terms of so-called \emph{open} morphisms of algebras. Finally,
\Cref{sec:lifted-kleisli:subobject-classifier} shows that this equips categories
of algebras with additional structure resembling that of elementary topoï.

\subsection{Kleisli Categories of Weakly Lifted Monads}
\label{sec:lifted-kleisli:}

Let us forget about regular categories and relations for an instant and first
describe the Kleisli categories of a weakly lifted monad $\liftR$ in terms of
the Kleisli category of $\funcR$ itself.

\begin{prop}
  \label{prop:kleisli-category-lifted-monad}
  Let $\distrTR$ be a weak distributive law in a category $\catC$ where
  idempotents split, so that $\funcR$ has a weak lifting $\liftR$ to $\EMT$ and
  $\funcT$ a weak extension $\extT$ to $\KlR$. Then $\KlliftR$-arrows $(A,a)
  \klarrow (B,b)$ are in one-to-one correspondence with $\KlR$-arrows $f: A
  \klarrow B$ such that $f \circ \leftKlR a = \leftKlR b \circ \extT f$:
  \[\begin{tikzcd}
      {\funcT A} & {\funcT B} \\
      A & B
      \arrow["{\extT f}", kleisli, from=1-1, to=1-2]
      \arrow["a"', from=1-1, to=2-1]
      \arrow["b", from=1-2, to=2-2]
      \arrow["f"', kleisli, from=2-1, to=2-2]
      \diagramlabel{1-1}{2-2}{algebra-kleisli-morphism}
    \end{tikzcd}\]
\end{prop}

The rest of this subsection is dedicated to proving this result. To do so, we
first briefly recall, from~\cite[\textsection 3.2]{garnerVietorisMonadWeak2019},
how weak liftings can actually be constructed. In a nutshell, the idea is to try
to lift $\monadR$ to $\EMT$ as if $\distrTR$ were a distributive law, and to
notice that, because $\distrTR$ does not make \Cref{cd:unit+} commute, the
lifted functor does not send algebras to algebras: $\rightEMT \counitT \liftR
\circ \unitT \funcR$ should be an identity, but is merely an idempotent.
However, splitting this idempotent yields back algebras, and hence a monad on
$\EMT$. In \Cref{ex:weak-liftings}, this idempotent appears as the closure
operation.

\subsubsection{Weak Liftings from Semi-Algebras.}

Call a \emph{semi-monad} on a category $\catC$ the data $\semimonadT$ of an
endofunctor $\funcT: \catC \to \catC$ and a natural transformation $\multT:
\funcTT \to \funcT$ such that $\multT \circ \funcT \multT = \multT \circ \multT
\funcT$. In particular, any monad has an underlying semi-monad, weak
distributive laws are really distributive laws between a monad and a semi-monad,
and weak extensions of monads are just extensions of the underlying semi-monads.
Given a semi-monad $\funcT$, one may form the category $\SemiEMT$ of its
\emph{semi-algebras}: these are defined like the algebras of a monad, except
that the unit axiom $a \circ \unitT_A = \id_A$ is not required to hold anymore.
Similarly, a morphism of semi-$\funcT$-algebras is defined just like a morphism
of algebras is. There are functors
$\adjunction{\catC}{\SemiEMT}{\leftSemiEMT}{\rightSemiEMT}$ and natural
transformations $\unitT: \Id \Rightarrow \rightSemiEMT \leftSemiEMT$ and
$\counitT: \leftSemiEMT \rightSemiEMT \Rightarrow \Id$, constructed exactly like
those for $\EMT$ instead of $\SemiEMT$, the only difference being that they do
not form an adjunction.

If $\funcT$ is also a monad, $\EMT$ is a full subcategory of $\SemiEMT$: we
write $\SemiT$ for the inclusion $\EMT \to \SemiEMT$, and $\leftEMT$ and
$\rightEMT$ are the restrictions of $\leftSemiEMT$ and $\rightSemiEMT$ to
$\EMT$.
\[\begin{tikzcd}
    \SemiEMT & \EMT & \SemiEMT & \EMT \\
    & \catC && \catC \arrow["\SemiT"', from=1-2, to=1-1]\arrow["\rightSemiEMT"', curve={height=12pt}, from=1-3, to=2-4]\arrow["\SemiT"', from=1-4, to=1-3]\arrow["\rightEMT", from=1-4, to=2-4]\arrow["\leftSemiEMT", curve={height=-12pt}, from=2-2, to=1-1]\arrow["\leftEMT"', from=2-2, to=1-2]\end{tikzcd}\]Call \emph{semi-lifting} of an endofunctor $\funcF$ or natural
transformation $\alpha$ on $\catC$ a lifting of $\funcF$ or $\alpha$ to
$\SemiEMT$, i.e., an endofunctor $\semilift \funcF$ or natural transformation
$\semilift \alpha$ such that $\rightSemiEMT \semilift \funcF = \funcF
\rightSemiEMT$ and $\rightSemiEMT \semilift \alpha = \alpha \rightSemiEMT$.
\begin{itemize}
\item Given a functor $\funcR: \catC \to \catC$, a natural transformation
  $\distrTR$ making \Cref{cd:mult+} commute gives rise to a semi-lifting
  $\semiliftR: \SemiEMT \to \SemiEMT$ of $\funcR$: $\semiliftR$ sends
  \begin{itemize}
  \item a semi-algebra $x: \funcT X \to X$ to the composite semi-algebra
    \[ \funcTR X \xrightarrow{ \distrTRname_X } \funcRT X \xrightarrow{\funcR x}
      \funcR X \]\item and a morphism $(X,x) \to (Y,y)$ with underlying arrow $f: X
    \to Y$ to the morphism with underlying arrow $\funcR f: \funcR X
    \to \funcR Y$.
  \end{itemize}
\item If $\funcS, \funcR: \catC \to \catC$ have semi-liftings $\semiliftS,
  \semiliftR: \SemiEMT \rightrightarrows \SemiEMT$ respectively given by laws
  $\distrTS$ and $\distrTR$, a natural transformation $\alpha: \funcS
  \Rightarrow \funcR$ has a lifting $\semilift \alpha$ along $\rightSemiEMT$ if
  and only if it makes \Cref{cd:weak-lifting:cond} commute.
\end{itemize}
In particular, a semi-lifting of a monad $\monadR$ to $\SemiEMT$ can be
constructed when there is a weak distributive law
$\distrTR$~\cite[Thm.~4.1]{petrisanSemialgebrasWeakDistributive2021}.

Next, we see how semi-liftings induce weak liftings to $\EMT$ when all
idempotents split in $\catC$. Consider the composite natural transformation in
$\catC$
\[ \rightSemiEMT \xrightarrow{\unitT \rightSemiEMT} \rightSemiEMT
  \leftSemiEMT \rightSemiEMT \xrightarrow{\rightSemiEMT \counitT}
  \rightSemiEMT \]Each component of this natural transformation can be seen to actually underly
a morphism of semi-algebras: the natural transformation itself
underlies a natural transformation $\closeT: \Id_{\SemiEMT}
\Rightarrow \Id_{\SemiEMT}$.

By definition, $\closeT$ is the identity along $\funcT$-algebras. Along a
semi-algebra $(X,x)$, however, $\closeT_{(X,x)}$ can only be shown to be an
idempotent. Yet its splitting
\[ (X,x) \xrightarrow{\projT_{(X,x)}} \SplitT (X,x) \xrightarrow{\incT_{(X,x)}}
  (X,x) \]has a proper $\funcT$-algebra in the middle. In fact, this construction yields a
functor $\SplitT: \SemiEMT \to \EMT$, which is both left and right adjoint to
$\SemiT$. Indeed, $\SplitT \SemiT = \Id_{\EMT}$ (because the idempotent is the
identity on algebras), and $\projT: \Id_{\SemiEMT} \Rightarrow \SemiT \SplitT$
and $\incT: \SemiT \SplitT \Rightarrow \Id_{\SemiEMT}$ are respectively the unit
and counit of these two
adjunctions~\cite[Lemma~12]{garnerVietorisMonadWeak2019}.

If $\semiliftR: \SemiEMT \to \SemiEMT$ is a semi-lifting of a functor
$\funcR: \catC \to \catC$, then the composite
\[ \EMT \xrightarrow{\SemiT} \SemiEMT \xrightarrow{\semiliftR}
  \SemiEMT \xrightarrow{\SplitT} \EMT \]is a weak lifting $\liftR$ of $\funcR$ to $\EMT$, with
\[ \projT_{\funcS} \coloneq \rightSemiEMT \projT \semiliftR \SemiT
  \qquad \incT_{\funcS} \coloneq \rightSemiEMT \incT \semiliftR
  \SemiT \]Similarly, if $\semilift \alpha$ is the semi-lifting of a natural transformation
$\alpha$ to $\SemiEMT$, then $\lift \alpha \coloneq \SplitT \semilift \alpha
\SemiT$ is a weak lifting of $\alpha$ to $\EMT$. In particular, if
$\monadR$ has a semi-lifting $\monadsemiliftR$ to $\SemiEMT$, then it also has a
weak lifting $\monadliftR$ to $\EMT$ obtained by sandwiching
$\monadsemiliftR$ inside the adjunction $\adjunction{\SemiEMT}{\EMT}{\SemiT}{\SplitT}$~\cite[Prop.~13]{garnerVietorisMonadWeak2019}.

\subsubsection{Proof of \Cref{prop:kleisli-category-lifted-monad}.}

From the discussion above, it follows that $\KlliftR$ is precisely the full
subcategory of $\KlsemiliftR$ whose objects are $\funcT$-algebras. Hence, we
show more generally that $\KlsemiliftR$ is the category whose objects are
semi-$\funcT$-algebras and whose arrows between two semi-algebras are the
$\KlR$-arrows making \Cref{cd:algebra-kleisli-morphism} commute.

Call directly $\KlsemiliftR$ this category. We construct an adjunction
$\adjunction{\leftKlsemiliftR}{\rightKlsemiliftR}{\SemiEMT}{\KlsemiliftR}$ with
$\leftKlsemiliftR$ bijective on objects and which induces $\monadsemiliftR$,
so that $\KlsemiliftR$ must indeed be the Kleisli category of
$\semiliftR$.

\begin{description}
\item[The left adjoint] Let $\leftKlsemiliftR: \SemiEMT \to \KlsemiliftR$ be
  the functor that sends a semi-$\funcT$-algebra on itself and a morphism $f:
  (A,a) \to (B,b)$ to the $\KlR$-arrow $\leftKlR \rightEMT f: A \klarrow B$.
  Because $\extT$ is an extension of $\funcT$ ($\extT \leftKlR =
  \leftKlR \funcT$), \Cref{cd:algebra-kleisli-morphism} for this last arrow is
  the image by $\leftKlR: \catC \to \KlR$ of the diagram making $f$ a morphism
  of semi-algebras.

\item[The right adjoint] Let $\rightKlsemiliftR: \KlsemiliftR \to \SemiEMT$ be
  the functor that sends a semi-$\funcT$-algebra $(A,a)$ to the
  semi-$\funcT$-algebra $\semiliftR (A,a) = (\funcR A, \funcR a \circ
  \distrTRname_A)$, and a $\KlsemiliftR$-arrow $(A,a) \klarrow (B,b)$ with
  underlying $\KlR$-arrow $f: A \klarrow B$ to the morphism $\semiliftR (A,a)
  \to \semiliftR (B,b)$ with underlying $\catC$-arrow $\rightKlR f: \funcR A \to
  \funcR B$. This is indeed a morphism in $\SemiEMT$ because the following
  diagram commutes.
  \[\begin{tikzcd}[column sep=huge, row sep=large]
      {\funcTR A} &&& {\funcTR B} \\
      & {\funcRTR A} & {\funcRTR B} \\
      {\funcRT A} &&& {\funcRT B} \\
      {\funcR A} &&& {\funcR B}\arrow["{\funcT \rightKlR f}"{name=0}, from=1-1, to=1-4]\arrow["{\unitR_{\funcTR A}}"{description}, from=1-1, to=2-2]\arrow["{\distrTRname_A}"'{name=4}, from=1-1, to=3-1]\arrow["{\unitR_{\funcTR A}}"{description}, from=1-4, to=2-3]\arrow["{\distrTRname_B}"{name=5}, from=1-4, to=3-4]\arrow["{\funcRT \rightKlR f}"{description,name=1},
      curve={height=-18pt}, from=2-2, to=2-3]\arrow["{\rightKlR \extT \leftKlR \rightKlR f}"{description,name=2},
      curve={height=18pt}, from=2-2, to=2-3]\arrow["{\rightKlR \extT \counitR_A}"{description}, from=2-2,
      to=3-1]\arrow["{\rightKlR \extT \counitR_B}"{description}, from=2-3,
      to=3-4]\arrow["{\rightKlR \extT f}"{description, name=3}, from=3-1,
      to=3-4]\arrow["{\funcR a}"', from=3-1, to=4-1]\arrow["{\funcR b}", from=3-4, to=4-4]\arrow["{\rightKlR f}"', from=4-1, to=4-4]\diagramref{4}{2-2}{"\text{(construction)}"}\diagramref{5}{2-3}{"\text{(construction)}"}\diagramref{0}{1}{"\text{(nat.)}"}\diagramref{1}{2}{"\text{(extension)}"}\diagramref{2}{3}{"\text{(nat.)}"}\diagramref{3-1}{4-4}{"\rightEMR\ref{cd:algebra-kleisli-morphism}"}\end{tikzcd}\]

\item[The unit] For the unit, we of course take the unit of
  $\semiliftR$. This is possible because $\rightKlsemiliftR
  \leftKlsemiliftR = \semiliftR$ by construction.

\item[The counit] Let $\counitsemiliftR_{(A,a)}: (\funcR A, \funcR a
  \circ \distrTRname_A) \to (A,a)$ be the $\KlsemiliftR$-arrow with underlying
  $\KlR$-arrow $\counitR_A: \funcR A \klarrow A$.
  \Cref{cd:algebra-kleisli-morphism} is then the commuting diagram below
  (in $\KlR$).
  \[\begin{tikzcd}[column sep=huge, row sep=large]
      {\funcTR A} && {\funcTR A} & \funcT A \\
      & {\funcRTR A} \\
      {\funcRT A} &&& \funcT A \\
      {\funcR A} &&& A\arrow[""{name=0}, equals, from=1-1, to=1-3]\arrow["{\leftKlR \unitR_{\funcTR A}}"{description}, from=1-1, to=2-2]\arrow["{\leftKlR \distrTRname_A}"'{name=1}, from=1-1, to=3-1]\arrow["{\extT \counitR_A}", from=1-3, to=1-4]\arrow[""{name=2}, equals, from=1-4, to=3-4]\arrow["{\counit_{\funcTR A}}"{description}, from=2-2, to=1-3]\arrow["{\leftKlR \rightKlR \extT \counitR_{A}}"{description},
      from=2-2, to=3-1]\arrow["{\counitR_A}"{description}, from=3-1, to=3-4]\arrow["{\leftKlR \funcR a}"', from=3-1, to=4-1]\arrow["{\leftKlR a}", from=3-4, to=4-4]\arrow["{\counitR_A}"', from=4-1, to=4-4]\diagramref{0}{2-2}{"\text{(adjunction)}"}\diagramref{1}{2-2}{"\text{(construction)}"}\diagramref{2-2}{2}{"\text{(nat.)}"}\diagramref{3-1}{4-4}{"\text{(nat.)}"}\end{tikzcd}\]\end{description}

The forgetful functor $\KlsemiliftR \to \KlR$, which we write
$\rightSemiEMextT$, satisfies by construction
\[ \rightSemiEMextT \leftKlsemiliftR = \leftKlR \rightSemiEMT \qquad
  \rightSemiEMT \rightKlsemiliftR = \rightKlR \rightSemiEMextT \qquad
  \rightSemiEMextT \counitsemiliftR = \counitR \rightSemiEMextT \qquad
  \rightSemiEMT \unitsemiliftR = \unitR \rightSemiEMT \]Because $\rightSemiEMextT$ is faithful and $\counitR$ is natural, we get from
the third equation that $\counitsemiliftR$ is natural as well. Similarly, the
equations making
$\adjunction{\leftKlsemiliftR}{\rightKlsemiliftR}{\SemiEMT}{\KlsemiliftR}$ into
an adjunction are obtained from the same equations for
$\adjunction{\leftKlR}{\rightKlR}{\catC}{\KlR}$. Hence
$\adjunction{\leftKlsemiliftR}{\rightKlsemiliftR}{\SemiEMT}{\KlsemiliftR}$ is an
adjunction inducing the monad $\monadsemiliftR$ (in fact,
$(\rightSemiEMT,\rightSemiEMextT)$ is a \emph{morphism of adjunctions}).

\subsection{Weakly Lifted Sub-Power-Object Monads}
\label{sec:lifted-kleisli:relations}

For the rest of this section, we will assume that:
\begin{itemize}
\item $\catC$ is regular,
\item $\funcR$ is a sub-power-object monad on $\catC$, whose Kleisli category is
  of the shape $\catC \cdot \catGamma^\dagger$ for some transposable subcategory
  $\catGamma$,
\item the monotone weak distributive law $\distrTP$ exists, so that
  \begin{itemize}
    \item $\funcT$ and $\multT$ are nearly cartesian from $\catGamma$ to itself
    \item $\funcR$ weakly lifts to $\EMT$ (idempotents always split in a regular
      category~\cite[Lemma~5.2]{goyCompositionalityMonadsWeak2021})
  \end{itemize}
\item $\funcT$ is in fact nearly cartesian (not just along $\catGamma$), hence
  in particular preserves regular epimorphisms.
\end{itemize}
Not all of those assumptions is necessary for the intermediate results that
follow, so that in these we restate each time precisely which conditions are
needed.

It then follows from \Cref{prop:kleisli-category-lifted-monad} that $\KlliftR$
has certain $\catC$-relations for morphisms: $\KlliftR$-morphisms $(A,a)
\klarrow (B,b)$ correspond to $\catC$-relations $\psi: A \leftrightsquigarrow B$
that are in $\KlR$ and such that $\psi \cdot a = b \cdot \funcT \psi$:
\[\begin{tikzcd}
    {\funcT A} & {\funcT B} \\
    A & B
    \arrow["{\funcT \psi}", squiggly, tail reversed, from=1-1, to=1-2]
    \arrow["a"', from=1-1, to=2-1]
    \arrow["b", from=1-2, to=2-2]
    \arrow["{=}"{description, anchor=center, rotate=30}, draw=none, from=2-1, to=1-2]
    \arrow["\psi"', squiggly, tail reversed, from=2-1, to=2-2]
  \end{tikzcd}\]In this subsection we show that $\EMT$ is in fact also a regular category, and
that $\KlliftR$-arrows can also be seen as certain $\EMT$-relations.

\subsubsection{Algebras are Regular.}

It is folklore that, under mild conditions, the algebras of a monad on a regular
category form a regulary category as well. For instance on $\Set$, all finitary
monads, and even all monads if the axiom of choice is assumed to be true, have
regular categories of algebras \cite[Theorems 3.5.4 and
4.3.5(1)]{borceuxHandbookCategoricalAlgebra1994a}. The usual proof of this
however requires that regular epimorphisms be split in $\catC$, which is not
true in an abritrary regular category. Hence here we give the proof of this
result again, but using the assumption that $\funcT$ is nearly cartesian
instead.

\begin{thm}[categories of algebras are regular]
  \label{thm:algebras-regular}
  Let $\monadT$ be a monad on a regular category $\catC$ such that $\funcT$ is
  nearly cartesian. Then $\SemiEMT$ and $\EMT$ are regular, and $\rightSemiEMT$
  and $\rightEMT$ create finite limits, regular epimorphisms and near
  pullbacks\footnote{meaning a cone (resp. arrow, resp. square) is limiting
    (resp. a regular epimorphism, resp. a near pullback) if and only if its
    image by $\rightSemiEMT$ or $\rightEMT$ is.}.
\end{thm}
\begin{proof}
  We prove the result for $\EMT$ and $\rightEMT$. The proof for $\SemiEMT$ and
  $\rightSemiEMT$ is the same, with the only difference that we only need to
  show that (co)limits of semi-algebras are semi-algebras. This works because
  proving associativity for a (co)limit of a diagram $D$ only relies on
  associativity of the algebras in $D$, not on their unit axiom.

  By \cite[Exercise V.2.2]{maclaneCategoriesWorkingMathematician1978}
  $\rightEMT$ creates the limits that $\catC$ has.

  We now prove that $\EMT$ has coequalizers of kernel pairs. Consider the kernel
  pair $u, v: \alg{P}{p} \to \alg{X}{x}$ of a $\funcT$-algebra morphism $f:
  \alg{X}{x} \to \alg{Y}{y}$. Because $\rightEMT$ preserves pullbacks,
  $\rightEMT u, \rightEMT v: P \to X$ is the kernel pair of $\rightEMT f$: let
  $e: X \to Q$ be its coequalizer.

  \begin{description}
  \item[$\funcT e$ is the coequalizer of $\funcT u, \funcT v: \funcT P
    \rightrightarrows \funcT X$] Because $e \circ u = e \circ v$ is a pullback
    (see, e.g., \cite[Proposition
    2.5.8]{borceuxHandbookCategoricalAlgebra1994}), the square $\funcT e \circ
    \funcT u = \funcT e \circ \funcT v$ is a near pullback: writing $u', v': P'
    \to \funcT X$ for the kernel pair of $\funcT e$, there is a regular
    epimorphism $e': \funcT P \onto P'$ such that $u' \circ e' = \funcT u$ and
    $v' \circ e' = \funcT v$. Suppose now $g: \funcT X \to \funcT Z$ is such
    that $g \circ \funcT u = g \circ \funcT v$. Then because $e'$ is an
    epimorphism, $g \circ u' = g \circ v'$ and $g$ factors uniquely through the
    coequalizer of $(u',v')$, which is $\funcT e$ because $\funcT e$ is a
    regular epimorphism hence the coequalizer of its kernel pair.
  \item[The algebra structure on $Q$] Consider now the composite $\funcT X
    \xrightarrow{x} X \xrightarrow{e} Q$. It satisfies $e \circ x \circ \funcT u
    = e \circ u \circ p = e \circ v \circ p = e \circ x \circ \funcT v$, so it
    factors through the coequalizer $\funcT e$ of $\funcT u$ and $\funcT v$. We
    let $q: \funcT Q \to Q$ be the corresponding colimiting arrow, which
    satisfies $q \circ \funcT e = e \circ x$ by definition. Now $\alg{Q}{q}$ is
    in fact an algebra:
    \[ \begin{aligned}[c]
      q \circ \unitT_Q \circ e &= q \circ \funcT e \circ \unitT_X \\
                               &= e \circ x \circ \unitT_X \\
                               &= e \\
                               &~\\
                               &~
    \end{aligned}
    \qquad
    \begin{aligned}[c]
      q \circ \multT_Q \circ \funcTT e &= q \circ \funcT e \circ \multT_X \\
                                       &= e \circ x \circ \multT_X \\
                                       &= e \circ x \circ \funcT x \\
                                       &= q \circ \funcT e \circ \funcT x \\
                                       &= q \circ \funcT q \circ \funcTT e \\
    \end{aligned} \]
    hence $q \circ \unitT_Q = \id_Q$ and $q \circ \multT_Q = q \circ \funcT q$
    because $e$ and $\funcTT e$ are epimorphisms.
  \item[$e$ is a morphism of algebras] By construction, $q \circ \funcT e = e
    \circ x$.
  \item[$e: \alg{X}{x} \to \alg{Q}{q}$ is the coequalizer of $u,v: \alg{P}{p}
    \to \alg{X}{x}$ in $\EMT$] Consider some morphism of $\funcT$-algebras
    $g: \alg{X}{x} \to \alg{A}{a}$ such that $g \circ u = g \circ v$. Then
    $\rightEMT g \circ \rightEMT u = \rightEMT g \circ \rightEMT v$ in $\catC$,
    hence $\rightEMT g$ factors as $h \circ e$ for some $h: Q \to A$. We now
    show that $h$ is in fact a $\funcT$-algebra morphism $\alg{Q}{q} \to
    \alg{A}{a}$:
    \[
      h \circ q \circ \funcT e = h \circ e \circ x = g \circ x = a \circ \funcT
      g = a \circ \funcT h \circ \funcT e
    \]
    Because $\funcT e$ is an epimorphism, $h \circ q = a \circ \funcT h$.
  \end{description}

  By construction, $e$ is a regular epimorphism in $\EMT$ if and only if
  $\rightEMT e$ is so in $\catC$, hence $\rightEMT$ creates regular
  epimorphisms. The creation of pullbacks and regular epimorphisms implies the
  creation of near pullbacks, and that regular epimorphisms in $\EMT$ are stable
  under pullback.
\end{proof}

\begin{exa}\label{ex:relations}\hfill
  \begin{itemize}
  \item A $\JSL$-relation between two complete join-semilattices
    $(X,\vee^X)$ and $(Y,\vee^Y)$ is a "relation" $R \subseteq X \times Y$ such
    that, if $(x_i,y_i) \in R$ for every $i$ in some set $I$, then
    \[ \left({\bigvee}^X \suchthat{ x_i }{ i \in I}, {\bigvee}^Y \suchthat{ y_i }{ i
          \in I }\right) \in R \]
  \item A $\Conv$-relation between two barycentric algebras
    $X$ and $Y$ is a relation $R \subseteq X \times Y$ between their
    underlying subsets such that, if $(x_i,y_i) \in R$ for every $1 \le i \le
    n$, then, for every $p \in \funcD \{1, \ldots, n\}$, $\left( \sum_{i = 1}^n
      p(i) x_i, \sum_{i = 1}^n p(i) y_i \right) \in R$ as well.
  \item A $\Mon$-relation ($\Mon \cong \EML$) between two monoids $X$ and $Y$ is
    a relation $R \subseteq X \times Y$ between their underlying subsets such
    that $(1_X,1_Y) \in R$ and, if $(x,y) \in R$ and $(x',y') \in R$, then
    $(xx',yy') \in R$ as well.
  \item A $\CMon$-relation ($\CMon \cong \EMM$) between two commutative monoids
    is a relation between them in $\Mon$.
    In~\cite{bonchiDiagrammaticAlgebraLinear2019}, a $\CMon$-relation $R
    \subseteq \naturals^l \times \naturals^k$ which is finitely generated as an
    $\naturals$-semimodule is called an \emph{additive relation}; these
    relations are shown to give rise to a ``string-diagrammatic language for
    concurrent systems'' [ibid., Thm.~17].
  \end{itemize}
\end{exa}

\subsubsection{Weakly Lifting Relations}

By \Cref{thm:relational-extension,thm:algebras-regular}, $\rightEMT$ has a
relational extension $\Rel[\rightEMT]: \Rel[\EMT] \to \RelC$. It follows that
$\EMT$-relations also correspond to certain $\catC$-relations:

\begin{lem}
  \label{lem:relations-between-algebras}
  In the setting of \Cref{thm:algebras-regular}, a $\catC$-relation $\psi: A
  \leftrightsquigarrow B$ is the image of an $\EMT$-relation $\alg{A}{a}
  \leftrightsquigarrow \alg{B}{b}$ by the faithful functor $\Rel[\rightEMT]:
  \Rel[\EMT] \to \RelC$ if and only if $\psi \cdot a \ge b \cdot \funcT \psi$:
  \[\begin{tikzcd}
      {\funcT A} & {\funcT B} \\
      A & B
      \arrow["{\funcT \psi}", squiggly, tail reversed, from=1-1, to=1-2]
      \arrow["a"', from=1-1, to=2-1]
      \arrow["b", from=1-2, to=2-2]
      \arrow["{\ge}"{description, anchor=center, rotate=30}, draw=none, from=2-1, to=1-2]
      \arrow["\psi"', squiggly, tail reversed, from=2-1, to=2-2]
    \end{tikzcd}\]
\end{lem}
\begin{proof}
  Recall \cite[Proposition 1.8]{carboni2categoricalApproachChange1991}: consider
  $\catC$-arrows $f: C \to A$ and $g: D \to B$ and $\catC$-relations $\phi: C
  \leftrightsquigarrow D$ and $\psi: A \leftrightsquigarrow B$ respectively
  given by the jointly monic spans $\langle \phi_C, \phi_D \rangle: S
  \subto C \times D$ and $\langle \psi_A, \psi_B \rangle: R
  \subto A \times B$. Then, $g \cdot \phi \le \psi \cdot f$ if and only
  if there is a necessarily unique $t: S \to R$ such that $\psi_A \circ t = f
  \circ \phi_C$ and $\psi_B \circ t = g \circ \phi_D$:
\[\begin{tikzcd}[column sep=large]
	C & S & D && C & S & D \\
	&&& \iff \\
	A & R & B && A & R & B
	\arrow["f"', from=1-1, to=3-1]
	\arrow["{\phi_C}"', from=1-2, to=1-1]
	\arrow["{\phi_D}", from=1-2, to=1-3]
	\arrow["\ge"{description, rotate=30}, draw=none, from=1-3, to=3-1]
	\arrow["g", from=1-3, to=3-3]
	\arrow["f"', from=1-5, to=3-5]
	\arrow["{\phi_C}"', from=1-6, to=1-5]
	\arrow["{\phi_D}", from=1-6, to=1-7]
	\arrow["t"{description}, dotted, from=1-6, to=3-6]
	\arrow["g", from=1-7, to=3-7]
	\arrow["{\psi_A}", from=3-2, to=3-1]
	\arrow["{\psi_B}"', from=3-2, to=3-3]
	\arrow["{\psi_A}", from=3-6, to=3-5]
	\arrow["{\psi_B}"', from=3-6, to=3-7]
\end{tikzcd}\]

Consider now an $\EMT$-relation $\psi: \alg{A}{a} \leftrightsquigarrow
\alg{B}{b}$, given by a jointly monic span $\langle \psi_{\alg{A}{a}},
\psi_{\alg{B}{b}} \rangle: \alg{R}{r} \subto \alg{A}{a} \times \alg{B}{b}$.
Then, because $\rightEMT$ preserves monomorphisms and products (it preserves all
limits), $\left\langle \rightEMT \psi_{\alg{A}{a}}, \rightEMT \psi_{\alg{B}{b}}
\right\rangle: R \subto A \times B$ is a jointly monic span for the
$\catC$-relation $\rightEMT \psi$. Factor $\left\langle \funcT \rightEMT
  \psi_{\alg{A}{a}}, \funcT \rightEMT \psi_{\alg{B}{b}} \right\rangle: \funcT R
\subto \funcT A \times \funcT B$ as $\funcTR \onto S \subto \funcT A \times
\funcT B$, and call $e: \funcTR \onto S$ the regular epimorphism in this
factorization. Then $r$ factors through $e$ as well (in a regular category,
regular epimorphisms are strong~\cite[\textsection
1.3]{carboni2categoricalApproachChange1991}), and because $e$ is an epimorphism,
\Cref{cd:alg-relation-left,,cd:alg-relation-right} below commute.
\[\begin{tikzcd}[sep=large]
	& {\funcT R} \\
	{\funcT A} & S & {\funcT B} \\
	\\
	A & R & B
	\arrow["{\funcT \rightEMT \psi_{\alg{A}{a}}}"', from=1-2, to=2-1]
	\arrow["e"{description}, two heads, from=1-2, to=2-2]
	\arrow["{\funcT \rightEMT \psi_{\alg{B}{b}}}", from=1-2, to=2-3]
	\arrow["a"', from=2-1, to=4-1]
	\arrow[""'{name=2-left}, from=2-2, to=2-1]
	\arrow[""{name=2-right}, from=2-2, to=2-3]
	\arrow[dotted, from=2-2, to=4-2]
	\arrow["b", from=2-3, to=4-3]
	\arrow["{\rightEMT \psi_{\alg{A}{a}}}"{name=1-left}, from=4-2, to=4-1]
	\arrow["{\rightEMT \psi_{\alg{B}{b}}}"'{name=1-right}, from=4-2, to=4-3]
  \diagramlabel{2-left}{1-left}{alg-relation-left}
  \diagramlabel{2-right}{1-right}{alg-relation-right}
\end{tikzcd}\]

By \cite[Proposition 1.8]{carboni2categoricalApproachChange1991},
we get that $\rightEMT \psi \cdot a \ge b \cdot \funcT \rightEMT \psi$.

Conversely, consider a $\catC$-relation $\psi: A \leftrightsquigarrow B$ given
by a jointly monic span $\langle \psi_A, \psi_B \rangle: R \subto A
\times B$ and such that $\psi \cdot a \ge b \cdot \funcT \psi$. Factor
$\langle \funcT \psi_A, \funcT \psi_B \rangle: \funcT R \rightarrow \funcT A
\times \funcT B$ as $\funcT A \onto S \subto \funcT A
\times \funcT B$, and call $e: \funcT A \onto S$ and $\phi =
\langle \phi_{\funcT A}, \phi_{\funcT B} \rangle: S \subto \funcT A
\times \funcT B$ the regular epimorphism and monomorphism in this factorization.
We then have the inequality:
\[\begin{tikzcd}[sep=large]
	& {\funcT R} \\
	{\funcT A} & S & {\funcT B} \\
	\\
	A & R & B
	\arrow["{\funcT \psi_A}"', from=1-2, to=2-1]
	\arrow["e"{description}, two heads, from=1-2, to=2-2]
	\arrow["{\funcT \psi_B}", from=1-2, to=2-3]
	\arrow["a"', from=2-1, to=4-1]
	\arrow["\phi_{\funcT A}", from=2-2, to=2-1]
	\arrow["\phi_{\funcT B}"', from=2-2, to=2-3]
	\arrow["\ge"{description, rotate=30}, draw=none, from=2-3, to=4-1]
	\arrow["b", from=2-3, to=4-3]
	\arrow["{\psi_A}", from=4-2, to=4-1]
	\arrow["{\psi_B}"', from=4-2, to=4-3]
\end{tikzcd}\]

By \cite[Proposition 1.8]{carboni2categoricalApproachChange1991}, we get a
$\catC$-arrow $t: S \to R$ such that $\psi_A \circ t = a \circ \phi_{\funcT A}$
and $\psi_B \circ t = b \circ \phi_{\funcT B}$. Let $r = t \circ e: \funcT R \to
R$. We now show that $\alg{R}{r}$ is a $\funcT$-algebra, so that $\psi_A$ and
$\psi_B$ are actually jointly monic $\funcT$-algebra morphisms and $\psi$ is an
$\EMT$-relation:
\[
\begin{aligned}
  \psi \circ r \circ \unitT_R &= \langle a \circ \funcT \psi_A, b \circ \funcT \psi_B \rangle \circ \unitT_R \\
                              &= \langle a \circ \unitT_A \circ \psi_A, b \circ \unitT_B \circ \psi_B \rangle \\
                              &= \psi \\
                              &~\\
                              &~\\
\end{aligned}
\qquad
\begin{aligned}
  \psi \circ r \circ \multT_R &= \langle a \circ \funcT \psi_A, b \circ \funcT \psi_B \rangle \circ \multT_R \\
                              &= \langle a \circ \mult_A \circ \funcTT \psi_A, b \circ \multT_B \circ \funcTT \psi_B \rangle \\
                              &= \langle a \circ \funcT a \circ \funcTT \psi_A, b \circ \funcT b \circ \funcTT \psi_B \rangle \\
                              &= \langle a \circ \funcT \psi_A \circ \funcT r, b \circ \funcT \psi_B \circ \funcT r \rangle \\
                              &= \psi \circ r \circ \funcT r
\end{aligned}
\]
$\alg{R}{r}$ is thus indeed a $\funcT$-algebra because $\psi$ is a monomorphism.

The correspondence between $\EMT$- and $\catC$-relations we have just described
is bijective because $\rightEMT$ is faithful.
\end{proof}

Because $\funcR$ is a sub-power-object monad on $\catC$, $\liftR$ is thus a
sub-power-object monad on $\EMT$: an $\EMT$-relation $\alg{A}{a}
\leftrightsquigarrow \alg{B}{b}$, seen as a $\catC$-relation $\psi: A
\leftrightsquigarrow B$ such that $\psi \cdot a \ge b \cdot \funcT \psi$, is a
$\KlliftR$-morphism $\alg{A}{a} \klarrow \alg{B}{b}$ if and only if this
inequality is an equality.

\subsection{Open morphisms of algebras}\label{sec:lifted-kleisli:open}

We now describe in more concrete terms which $\EMT$-relations are arrows in
$\KlliftR$. \emph{Open} morphisms of $\funcT$-algebra play a central role in
this description:

\begin{defi}[open morphisms of algebra]
  \label{def:open-morphism}
  Let $\funcT$ be a monad on a regular category $\catC$. A $\funcT$-algebra
  morphism $f: (X,x) \to (Y,y)$ is called \emph{open} when the square $f \circ x
  = y \circ \funcT f$ is a near pullback:
  \[\begin{tikzcd}
      {\funcT X} & {\funcT Y} \\
      X & Y
      \arrow["\funcT f", from=1-1, to=1-2]
      \arrow["x"', from=1-1, to=2-1]
      \arrow["{=}"{description, anchor=center, rotate=-20, pos=0.45}, draw=none, from=1-1, to=2-2]
      \arrow["y", from=1-2, to=2-2]
      \arrow["f"', from=2-1, to=2-2]
    \end{tikzcd}\]

  Given a jointly monic span $\langle \psi_X, \psi_Y \rangle$ in $\EMT$, the
  corresponding relation $\psi = \psi_Y \cdot \psi_X^\dagger$ is called
  \emph{continuous} when $\psi_X$ is open.
\end{defi}

Open morphisms have been studied in \cite[Definition
3.1.1]{clementinoLaxAlgebrasSpaces2014} in the setting of \emph{monoidal
  topology}. They are called \emph{open} as they generalize open maps between
compact Hausdorff spaces, as the next example shows. In
\cite{aristoteMonotoneWeakDistributive2025} they were called ``decomposable''
instead because we focus on examples that feel more algebraic than topological.
Indeed, when $\funcT$ is a monad on $\Set$, a function $f: X \to Y$ underlying a
morphism $(X,x) \to (Y,y)$ is open when, for every $e \in X$ and $t \in \funcT
Y$ such that $f(e) = y(t)$, there is some $t' \in \funcT X$ such that $e = x(t)$
and $t = (\funcT f)(t')$. In words, this holds whenever decompositions of
elements of $Y$ with respect to $y$ can be pulled back along $f$ to
decompositions of elements of $X$. Continuous relations, on the other hand,
allow for pushing these decompositions forward along the relation.

\begin{exa}[open morphisms and continuous
  relations]\label{ex:open-morphisms-relations}\hfill
  \begin{enumerate}
  \item \label{ex:open-morphisms-relations:khaus} In $\EMUlt \cong \KHaus$, a
    continuous map is open if and only if it is open (it preserves open sets),
    and continuous relations are the continuous ones, i.e., those relations
    $\psi: X \leftrightsquigarrow Y$ such that $\psi^{-1}[u]$ is open in $X$ for
    every open subset $u$ of $Y$.

  \item \label{ex:open-morphisms-relations:jsl} In $\EMP \cong \JSL$,
    $\psi: X \leftrightsquigarrow Y$ is open if and only if for every
    family $(x_i)_{i \in I}$ of elements of $X$ and every $y \in Y$ such that
    $\left(\bigvee_{i \in I} x_i, y \right) \in \psi$, there is a family
    $(y_i)_{i \in I}$ of elements of $Y$ such that $(x_i,y_i) \in \psi$ for all
    $i \in I$ and $\bigvee_{i \in I} y_i = y$.

  \item \label{ex:open-morphisms-relations:conv} In $\EMD \cong \Conv$,
    $\psi: X \leftrightsquigarrow Y$ is open if and only if for every $x
    \in X$, every \emph{disintegration} of $x$ as a barycenter $x = \sum_{i =
      1}^n \lambda_i x_i$ and every $y \in Y$ such that $(x,y) \in \psi$, $y$
    disintegrates as a barycenter $y = \sum_{i = 1}^n \lambda_i y_i$ such that
    $(x_i,y_i) \in \psi$ for all $i \in I$.

  \item In $\EML \cong \Mon$ (the category of monoids) or $\EMM \cong \CMon$
    (the category of commutative monoids), a relation is continuous if and only
    if, for every $(x,y) \in R$, if $x = x'x''$ for some $x',x'' \in X$, then $y
    = y'y''$ for some $y', y'' \in Y$ such that $(x',y'), (x'', y'') \in R$.
    Notably, these continuous relations between monoids form a category which is
    a \emph{fully abstract model of Basic
      SCI}~\cite{mccuskerFullyAbstractRelational2002}.
  \end{enumerate}
\end{exa}

\begin{proof}[Proof of \Cref{ex:open-morphisms-relations}]\hfill
  \begin{enumerate}
  \item Open morphisms are shown to correspond to open continuous functions in
    \cite[Rem.~V.3.3.5(2)]{clementinoLaxAlgebrasSpaces2014}. The
    characterization of continuous functions follows from
    \Cref{thm:kleisli-lifted-subpowerobject} below, as $\KlV$-arrows were shown
    to be continuous relations in~\cite{goyPowersetLikeMonadsWeakly2021}.

\item A relation $\alg{A}{\vee^A} \leftrightsquigarrow \alg{B}{\vee^B}$ in
    $\JSL$ given by some $\psi: A \leftrightsquigarrow B$ in $\Set$ is
    open if and only if for every $a \in A$, $E \in \funcP A$ such that
    $a = \bigvee^A E$ and $b \in B$ such that $(a,b) \in \psi$, there is an $F
    \in \funcP B$ such that
    \[ \forall a' \in E, \exists b' \in F, (a',b') \in \psi \mathand \forall b'
      \in F, \exists a' \in E, (a',b') \in \psi \] and $\bigvee^B F = b$.

    Consider some $E = \suchthat{a_i}{i \in I}$ so that $a = \bigvee^A_{i \in I}
    a_i$ and some $b \in B$ such that $(a,b) \in \psi$. Then if $\psi$ is
    open, consider the corresponding $F \in \funcP B$ and write $F_i =
    \suchthat{b' \in F}{(a_i,b) \in \psi}$ and $b_i = \bigvee^B F_i$. By
    definition of $F$, every $F_i$ is non-empty and $\bigcup_{i \in I} F_i = F$,
    so that $\bigvee^B_{i \in I} b_i = \bigvee^B F = b$. Moreover $\psi
    \subseteq A \times B$ is a relation of join-semilattices and $(a_i,b') \in
    \psi$ for every $b' \in F_i$, hence $\left( a_i, b_i \right) = \bigvee^{A
      \times B}_{b' \in F_i} \left( a_i, b' \right) \in \psi$ as well.

    Conversely, if $\psi$ satisfies the stated property then if $\bigvee^A E =
    a$ and $(a,b) \in \psi$, $\bigvee^A_{e \in E} e = a$ so there is a family
    $(b_e)_{e \in E}$ of elements of $B$ such that $(e,b_e) \in \psi$ and
    $\bigvee^B_{e \in E} b_e = b$. Taking $F = \suchthat{b_e}{e \in E}$, we see
    that $\psi$ is open.
  \item A relation $A \leftrightsquigarrow B$ in $\Conv$ given by some $\psi: A
    \leftrightsquigarrow B$ in $\Set$ is open if and only if for every
    $a \in A$ given as $a = \sum_{i = 1}^n \lambda_i a_i$ with $\sum_{i = 1}^n
    \lambda_i = 1$, $\lambda_i \neq 0$ and $a_i \in A$ for every $1 \le i \le
    n$, and $b \in B$ such that $(a,b) \in \psi$, there are for every $1 \le i
    \le n$ some $m_i \ge 1$, $(\mu_{i,j})_{1 \le j \le m_i}$ such that $\sum_{j
      = 1}^{m_i} \mu_{ij} = \lambda_i$ and $(b_{ij})_{1 \le j \le m_i}$ such
    that $(a_i,b_{ij}) \in \psi$ for all $1 \le j \le m_i$ and $\sum_{i = 1}^n
    \sum_{j = 1}^{m_i} \mu_{ij} b_{ij} = b$.

    Suppose this holds. For every $1 \le i \le n$, let $b_i =
    \frac{1}{\lambda_i} \sum_{j = 1}^{m_i} \mu_{ij} b_{ij}$. Then $\sum_{i =
      1}^n \lambda_i b_i = \sum_{i = 1}^n \sum_{j = 1}^{m_i} \mu_{ij} b_{ij} =
    b$, and for every $1 \le i \le n$, $(a_i, b_i) = \frac{1}{\lambda_i} \sum_{i
      = j}^{m_i} \mu_{ij} (a_i,b_{ij}) \in \psi$ because $\psi \subseteq A
    \times B$ is a relation of barycentric algebras.

    Conversely, consider some $a = \sum_{i = 1}^n \lambda_i a_i$ and $b$ such
    that $(a,b) \in \psi$ as above. If there is always some $(b_i)_{1 \le i \le
      n}$ such that $\sum_{i = 1}^n \lambda_i b_i = b$ and $(a_i,b_i) \in \psi$,
    then taking $m_i = 1$, $\mu_{i1} = \lambda_i$ and $b_{i1} = b_i$ for every
    $1 \le i \le n$ shows that $\psi$ is open. \qedhere
  \item As for the case of $\KHaus$, it has already been shown
    in~\cite{mccuskerFullyAbstractRelational2002} that the relations described
    above are the Kleisli arrows of the lifting of $\funcP$ to $\EML$. They are
    thus exactly the continuous relations by
    \Cref{thm:kleisli-lifted-subpowerobject}.
  \end{enumerate}
\end{proof}

A systematic way to construct open morphisms is the following.

\begin{lem}
  \label{lem:free-morphisms-are-open}
  In the setting of \Cref{def:open-morphism}, suppose also that $\rightEMT$
  creates near pullbacks. Then $\multT$ is nearly cartesian if and only if, for
  every $f: X \to Y$ in $\catC$, the free $\funcT$-algebra morphism $\leftEMT f:
  \alg{\funcT X}{\multT_X} \to \alg{\funcT Y}{\multT_Y}$ is open.
\end{lem}
\begin{proof}
  By definition.
\end{proof}

We are now able to state the main result of this section. We first state it in
full generality (\Cref{thm:kleisli-lifted-subpowerobject}), but in practice we
will be especially concerned with the case $\funcR = \funcP$ on $\catC = \Set$
(\Cref{cor:kleisli-lifted-powerset}).

\begin{thm}
  \label{thm:kleisli-lifted-subpowerobject}
  Let $\monadT$ be a monad on a regular category $C$. When the endofunctor
  $\funcT$ is nearly cartesian, an $\EMT$-relation $\psi: \alg{A}{a}
  \leftrightsquigarrow \alg{B}{b}$ is open if and only if $\rightEMT
  \psi \cdot a = b \cdot \funcT \rightEMT \psi$.

  If $\funcR$ is a sub-power-object monad on $\catC$ and $\multT$ extends to
  $\KlR$ (for instance if $\multT$ is also nearly cartesian), then the
  corresponding weakly lifted monad $\liftR$ on $\EMT$ is a sub-power-object
  monad and its Kleisli category $\KlliftR$ has for arrows $\alg{A}{a} \klarrow
  \alg{B}{b}$ the continuous relations $\psi: \alg{A}{a} \leftrightsquigarrow
  \alg{B}{b}$ in $\EMT$ such that $\rightEMT \psi$ is in $\KlR$.
\end{thm}
\begin{proof}
  Consider some $\EMT$-relation $\psi: \alg{A}{a} \leftrightsquigarrow
  \alg{B}{b}$. By \Cref{lem:relations-between-algebras}, it corresponds to the
  $\catC$-relation $\rightEMT \psi: A \leftrightsquigarrow B$, which
  satisfies $\rightEMT \psi \cdot a \ge b \cdot \funcT \rightEMT
  \psi$. By \Cref{prop:kleisli-category-lifted-monad}, $\psi$ is in $\KlliftR$
  if and only if we have moreover that $\rightEMT \psi$ is in $\KlR$ and
  $\rightEMT \psi \cdot a = b \cdot \funcT \rightEMT \psi$. We now
  show that this last condition is equivalent to $\psi$ being continuous.

  If $\psi$ is continuous, it is given by a jointly monic span $\langle
  \psi_{\alg{A}{a}}, \psi_{\alg{B}{b}} \rangle: \alg{R}{r} \subto \alg{A}{a}
  \times \alg{B}{b}$ such that $\psi_{\alg{A}{a}}$ is an open morphism.
  Writing $\psi_A = \rightEMT \psi_{\alg{A}{a}}$ and $\psi_B = \rightEMT
  \psi_{\alg{B}{b}}$, this means that $\psi_A \circ r = a \circ \funcT \psi_A$
  is a near pullback in $\catC$, or equivalently that $\psi_A^\dagger \cdot a =
  r \cdot (\funcT \psi_A)^\dagger$. But because $\psi_{\alg{B}{b}}$ is a
  morphism of $\funcT$-algebras, we also have that $\psi_B \cdot r = b \cdot
  \funcT \psi_B$, and so
  \begin{align*}
    \rightEMT \psi \cdot a &= \psi_B \cdot \psi_A^\dagger \cdot a \\
                                 &= \psi_B \cdot r \cdot (\funcT \psi_A)^\dagger \\
                                 &= b \cdot \funcT \psi_B \cdot (\funcT \psi_A)^\dagger \\
                                 &= b \cdot \funcT \rightEMT \psi
  \end{align*}

  Conversely, suppose that $\psi: \alg{A}{a} \leftrightsquigarrow \alg{B}{b}$
  given by $\langle \psi_{\alg{A}{a}}, \psi_{\alg{B}{b}} \rangle: \alg{R}{r}
  \hookrightarrow \alg{A}{a} \times \alg{B}{b}$ is such that $\rightEMT \psi
  \cdot a = b \cdot \funcT \rightEMT \psi$. We now show that $\psi_A = \rightEMT
  \psi_{\alg{A}{a}}$ is open: consider the pullback
 \[\begin{tikzcd}
      && {\funcT R} \\
      {\funcT A} & P \\
      A & R
      \arrow["{\funcT \psi_A}"', curve={height=6pt}, from=1-3, to=2-1]
      \arrow["e"{description}, dashed, from=1-3, to=2-2]
      \arrow["r", curve={height=-6pt}, from=1-3, to=3-2]
      \arrow["a"', from=2-1, to=3-1]
      \arrow["p_{\funcT A}"{description}, from=2-2, to=2-1]
      \arrow["\pullbackcorner"{anchor=center, pos=0.125, rotate=-90}, draw=none, from=2-2, to=3-1]
      \arrow["p_R"{description}, from=2-2, to=3-2]
      \arrow["\psi_A", from=3-2, to=3-1]
    \end{tikzcd}\] To prove that $\psi_A$ is open, we need to prove that
  $e: \funcT R \to P$ in the diagram above is a regular epimorphism. By
  definition of the composition of relations, $\psi_A^\dagger \cdot a = p_R
  \cdot p_{\funcT A}^\dagger$, so that $b \cdot \funcT \rightEMT \psi =
  \rightEMT \psi \cdot a = \psi_B \cdot p_R \cdot p_{\funcT A}^\dagger$. Factor
  $\langle p_{\funcT A}, \psi_B \circ p_R \rangle: P \to \funcT A \times B$ into
  the regular epimorphism $e': P \onto P'$ and the jointly monic span $\langle
  p_{\funcT A}', p_B' \rangle: P' \subto \funcT A \times B$. Then $p_B' \cdot
  (p_{\funcT A}')^\dagger = \psi_B \cdot p_R \cdot p_{\funcT A}^\dagger = b
  \cdot \funcT \rightEMT \psi = \psi_B \cdot r \cdot (\funcT \psi_A)^\dagger$,
  therefore $e' \circ e: \funcT R \to P'$, as in the diagram below, must be a
  regular epimorphism.
  \[\begin{tikzcd}
      & {\funcT R} \\
      & P \\
      {\funcT A} & {P'} & R & B
      \arrow["e", from=1-2, to=2-2]
      \arrow["{\funcT \psi_A}"', curve={height=12pt}, from=1-2, to=3-1]
      \arrow["r", curve={height=-12pt}, from=1-2, to=3-3]
      \arrow["{p_{\funcT A}}"{description}, curve={height=6pt}, from=2-2, to=3-1]
      \arrow["{e'}", two heads, from=2-2, to=3-2]
      \arrow["{p_R}"{description}, curve={height=-6pt}, from=2-2, to=3-3]
      \arrow["{p_{\funcT A}'}", from=3-2, to=3-1]
      \arrow["{p_B'}"', curve={height=12pt}, from=3-2, to=3-4]
      \arrow["{\psi_B}", from=3-3, to=3-4]
    \end{tikzcd}\]

  We will now show that $e'$ is a monomorphism and thus an isomorphism, and it
  will follow that $e$ must be a regular epimorphism as well. We have the
  commuting diagram
  \[\begin{tikzcd}[column sep=large]
      P & {P'} & {\funcT A \times B} \\
      {\funcT A \times R} && {\funcT A \times A \times B}
      \arrow["e'", two heads, from=1-1, to=1-2]
      \arrow["{\langle p_{\funcT A}, p_R \rangle}"', hook, from=1-1, to=2-1]
      \arrow["{\langle p_{\funcT A}', p_B' \rangle}", hook, from=1-2, to=1-3]
      \arrow["{\langle \id_{\funcT A}, a \rangle \times \id_B}", hook, from=1-3, to=2-3]
      \arrow["{\id_{\funcT A} \times \langle \psi_A, \psi_B \rangle}"', hook, from=2-1, to=2-3]
    \end{tikzcd}\]where all the arrows except $e'$ are monomorphisms: $e'$ is also a
  monomorphism.
\end{proof}

\begin{cor}
  \label{cor:kleisli-lifted-powerset}
  If $\funcT$ has a monotone weak distributive law over $\funcP$ in $\Set$, the
  Kleisli category of the lifted powerset monad $\liftP$ on $\EMT$ is the
  category of $\funcT$-algebras and continuous relations between them.
\end{cor}

\begin{exa}
  The Kleisli arrows of $\liftP$ in $\KHaus$, $\JSL$, $\Conv$, $\Mon$ and
  $\CMon$ are the continuous relations therein, as described in
  \Cref{ex:open-morphisms-relations}. As explained in the proof of the latter,
  this was already known in the cases of $\KHaus$ and $\Mon$; for these the new
  result is that these arrows may be described as spans of morphisms whose left
  legs are open.

  In the case of $\CMon$, we note that another characterization of continuous
  additive relations (recall \Cref{ex:relations}) is given in
  \cite[Prop.~82]{piedeleuPicturingResourcesConcurrency}.

  The non-empty version of these five lifted powerset monads, obtained via the
  weak distributive laws over $\funcP_*$ in $\Set$ (nearly cartesian functors
  in $\Set$ automatically preserve surjections) similarly have continuous
  relations with non-empty image as Kleisli arrows; alternatively, these
  are relations whose left branches are open surjective morphisms.
\end{exa}

\subsection{Open Subobject Classifiers}
\label{sec:lifted-kleisli:subobject-classifier}

\Cref{cor:kleisli-lifted-powerset} above holds more generally in any
\emph{elementary topos}. Recall that an elementary topos is a regular category
$\catE$ such that $\GrphE: \catE \to \RelE$ has a right
adjoint~\cite[\textsection 1.911]{freydCategoriesAllegories1990}. The monad
induced by this adjunction, also written $\funcP$, is called the
\emph{power-object} monad, and is trivially a sub-power-object monad since
$\GrphE$ factors as $\catE \xrightarrow{\leftKlP} \KlP \cong \RelE$. The fact
that $\KlP \cong \Rel$ in $\Set$ makes the latter an elementary topos.

In \Cref{thm:kleisli-lifted-subpowerobject}, we may thus take $\catC$ to be an
elementary topos and $\funcR$ to be its power-object monad, and, just as in
\Cref{cor:kleisli-lifted-powerset}, get:
\begin{cor}
  \label{cor:kleisli-lifted-powerobject}
  If $\funcT$ has a monotone weak distributive law over $\funcP$ in an
  elementary topos, the Kleisli category of the lifted power-object monad
  $\liftP$ on $\EMT$ is the category of $\funcT$-algebras and open
  relations between them.
\end{cor}

Elementary topoï turn out to have a lot of structure: an elementary topos has
finite limits, is cartesian-closed \cite[1.92]{freydCategoriesAllegories1990}
and has a \emph{subobject classifier}
\cite[1.912]{freydCategoriesAllegories1990} -- an object classifying all
monomorphisms:

\begin{defi}[{$\catM$-subobject classfier~\cite[Definition~14.1]{wylerLectureNotesTopoi1991}}]
  Let $\catC$ be a regular category and $\catM \subseteq \Mono$ be a class of
  monomorphisms in $\catC$. An \emph{$\catM$-subobject classifier} is an object
  $\Omega$ that \emph{classifies} $\catM$-subobjects: for every object $A$,
  $\catM$-monomorphisms $m: A' \hookrightarrow A$ are in one-to-one
  correspondence with morphisms $\chi: A \to \Omega$, the former being obtained
  by pulling back $\top: 1 \hookrightarrow \Omega$ along the latter.
  \[\begin{tikzcd}
      {A'} & 1 \\
      A & \Omega
      \arrow[from=1-1, to=1-2]
      \arrow["m"', hook, from=1-1, to=2-1]
      \arrow["\pullbackcorner"{anchor=center, pos=0.125}, draw=none, from=1-1, to=2-2]
      \arrow["\top", hook, from=1-2, to=2-2]
      \arrow["\chi"', from=2-1, to=2-2]
    \end{tikzcd}\]
\end{defi}

Intuitively, $\Omega$ is an object of truth values and an $M$-subobject $A'
\hookrightarrow A$ is the pre-image of $1 \hookrightarrow \Omega$ by the
corresponding morphism $A \to \Omega$. The rich structure of elementary topoï
allows for interpreting higher-order logic within them
\cite[B.315]{freydCategoriesAllegories1990}.

Let $\funcT$ be a monad on $\Set$. Its category of algebras $\EMT$ is often
regular (as discussed above~\Cref{thm:algebras-regular}), but it is seldom an
elementary topos because $\Grph[\EMT]$ need not have a right adjoint, and in
particular it seldom has a subobject classifier. But if $\funcT$ and $\multT$
are nearly cartesian, $\EMT$ is not devoid of additional structure:
\Cref{cor:kleisli-lifted-powerobject} tells us that $\Grph[\EMT]$ has a right
adjoint when co-restricted to the category of continuous relations. In
particular, this entails the existence of a \emph{open subobject
  classifier} in $\EMT$:

\begin{prop}[open subobject classifiers]
  \label{prop:open-subobject-classifiers}
  In an elementary topos $\catE$, let $\funcT$ be a monad that has a
  monotone weak distributive law over $\funcP$. Then $\liftP 1$ classifies
  open subobjects.
\end{prop}

\begin{proof}
  Let $\liftP$ be the weak lifting of $\funcP$ to $\EMT$. The Kleisli adjunction
  for $\liftP$ yields a bijection between $\funcT$-algebra morphisms $\chi: A
  \to \liftP 1$ and $\KlliftP$-arrows $m: A \klarrow 1$, by means of $\chi
  \mapsto \counitliftP_1 \circ \leftKlliftP \chi$.

  Recall from \Cref{cor:kleisli-lifted-powerobject} that $\KlliftP$ is the
  category of continuous relations. In particular, the bijection above says
  that $\EMT$-morphisms $\chi: A \to \liftP 1$ are in one-to-one correspondence
  with continuous relations $A \leftrightsquigarrow 1$. The latter are given
  by jointly monic spans $\langle m, !_{A'}\rangle: A' \hookrightarrow A \times 1$
  such that $m$ is open, or equivalently by open monomorphisms
  $m: A' \hookrightarrow A$.

  $\langle m, !_{A'} \rangle$ is obtained from $\chi$ by writing $\counitliftP_1$
  as a jointly monic span $\langle \top, !_1 \rangle: R \hookrightarrow \liftP 1
  \times 1$ and pulling $\top: R \hookrightarrow \liftP 1$ back along $\chi: A
  \to \Omega$:
  \[\begin{tikzcd}
      {A'} & R & 1 \\
      A & \liftP 1
      \arrow[from=1-1, to=1-2]
      \arrow["{!_{A'}}", curve={height=-12pt}, from=1-1, to=1-3]
      \arrow["m"', hook, from=1-1, to=2-1]
      \arrow["\pullbackcorner"{anchor=center, pos=0.125}, draw=none, from=1-1, to=2-2]
      \arrow["{!_1}"', from=1-2, to=1-3]
      \arrow["\top", hook, from=1-2, to=2-2]
      \arrow["\chi"', from=2-1, to=2-2]
    \end{tikzcd}\] ($m$ must then already be a monomorphism because $\top$ is
  one and pullbacks preserve monomorphisms, hence $\langle m, !_{A'} \rangle$ is
  indeed a jointly monic span). It follows from \cite[Proposition
  14.2]{wylerLectureNotesTopoi1991} that $R \cong 1$.
\end{proof}

We end this subsection with some examples on $\Set$ that follow from
\Cref{cor:kleisli-lifted-powerobject}. A first consequence of this corollary is
that if all $\funcT$-algebra morphisms are open, then $\Grph[\EMT]: \EMT
\to \Rel[\EMT]$ has a right adjoint and $\EMT$ is therefore an elementary topos.

Note that all $\funcT$-algebra morphisms being open is not a necessary
condition for $\EMT$ to be an elementary topos, as the next example shows. We
still retrieve a classic example of elementary topoï, namely the categories of
group actions.

\begin{exa}
  Let $(M, {\cdot}, 1)$ be a monoid. $M \times -$ is a monad with unit $\eta_X:
  x \mapsto (1_M, x)$ and multiplication $\mu_X: (m, (n, x)) \mapsto (m \cdot n,
  x)$. The endofunctor, multiplication and unit of this monad are cartesian,
  hence nearly cartesian. Its algebras are \emph{$M$-sets}, i.e., sets equipped
  with a left-action of $M$.

  By definition, a morphism $f: X \to Y$ of $M$-sets is open if and only if for
  every $x \in X$, $y \in Y$ and $m \in M$ such that $f(x) = m \cdot y$, there
  is an $x' \in X$ such that $f(x') = y$ and $m \cdot x' = x$. In particular,
  every morphism of $M$-set is open if and only if every element of $m$ is
  invertible (take $x' = m^{-1} \cdot x$ in one direction, and consider the
  morphism of $M$-sets $m \cdot -: M \to M$ for some non-invertible $M$ in the
  other direction).

  By \Cref{cor:kleisli-lifted-powerobject}, $\EM{M \times -}$ is therefore an
  elementary topos as soon as $M$ is a group. It is in fact well known that
  $\EM{M \times -}$ is always an elementary topos, no matter whether $M$ is a
  group or not \cite[\textsection 27.3]{wylerLectureNotesTopoi1991}.
\end{exa}

For our running examples $\KHaus$, $\JSL$ and $\Conv$,
\Cref{prop:open-subobject-classifiers} and the characterization of
open morphisms in \Cref{ex:open-morphisms-relations} yield:

\begin{exa} \hfill
  \begin{enumerate}
  \item In $\KHaus$, clopen subspaces are classified by $\funcV 1$, i.e., the set
    $2$ equipped with the discrete topology (for which all subsets are open).

  \item In $\JSL$, $\liftP 1 = (2, \max)$ classifies downwards-closed sets.

  \item In $\Conv$, $\liftP 1$ has carrier set $2 = \{\bot, \top\}$ and
    barycenters $\lambda \cdot \bot + (1 - \lambda) \cdot \top = \bot$ for
    $\lambda \in [0,1)$. It classifies \emph{walls}, i.e., subsets $E$ of convex
    sets $C$ such that if $e \in E$ and $e = \lambda \cdot c + (1 - \lambda)
    \cdot c'$ for $\lambda \in (0,1)$ and $c, c' \in C$, then $c, c' \in E$ as
    well. Walls appear for instance in the structure theorem for convex
    algebras, which state that every convex algebra is a subalgebra of the
    \emph{P\l{}onka sum} of its walls~\cite[Theorem
    4.5]{romanowskaStructureBarycentricAlgebras1990}.
  \end{enumerate}
\end{exa}

\section{Monotone Weak Distributive Laws in Categories of
  Algebras}\label{sec:lifting-monotone-laws}

In \Cref{sec:lifted-kleisli} we have seen that if there is a monotone weak
distributive law $\distrTPtype$ in $\Set$, than the Kleisli category of $\liftP$
to $\EMT$ is of the shape $\EMT \cdot (\Open_\funcT)^\dagger$, where
$\Open_\funcT$ denotes the subcategory of open $\funcT$-algebra morphisms. From
\Cref{thm:partial-relational-extensions} it follows that monotone weak
distributive laws over $\liftP$ are fully characterized: for a monad $\funcS$ on
$\EMT$, there is at most one monotone weak distributive law $\distrSliftPtype$,
and it exists if and only if $\funcS$ and $\multS$ are nearly cartesian from
$\Open_\funcT$ to itself. We now apply these results to build monotone weak
distributive laws in categories of algebras, or conversely show that monotone
weak distributive laws between certain monads cannot exist.

In \Cref{sec:lifting-monotone-laws:lifting-near-cartesian}, we start with a
generic argument that will make studying concrete examples easier. In
\Cref{sec:lifting-monotone-laws:khaus}, we show how the results from
\Cref{sec:lifted-kleisli} may be used to exhibit monotone weak distributive laws
in $\KHaus$. Finally, in \Cref{sec:lifting-monotone-laws:nogo-theorems}, we show
conversely that in other categories of algebras, e.g. $\JSL$ and $\Conv$, some
monotone weak distributive laws involving the weakly lifted powerset monad
cannot exist.

\subsection{Weakly Lifting Preserves Near Cartesianness}
\label{sec:lifting-monotone-laws:lifting-near-cartesian}

In this subsection, we put ourselves again (as in \Cref{sec:lifted-kleisli}) in
the general setting where $\catC$ is a regular category, $\funcR$ a
sub-power-object monad on $\catC$, and $\funcT$ a monad on $\catC$ with nearly
cartesian endofunctor and multiplication, whose relational extension $\extT$
restricts to $\KlR$.

For a monad $\funcS'$ on $\EMT$ to have a monotone weak extension to $\KlliftR$
(equivalently a monotone weak distributive law over $\liftR$),
\Cref{thm:kleisli-lifted-subpowerobject,thm:partial-relational-extensions} tell
us in particular that the endofunctor $\funcS'$ should preserve a certain
subcategory $\catGamma$ of morphisms, as well as preserve near pullbacks
involving $\catGamma$-morphisms, and the multiplication $\mult[\funcS']$ should
have near pullbacks for its naturality squares along morphisms in $\catGamma$.

Recall now that our motivating examples were possible monotone weak distributive
laws $\distrliftPliftPtype$ in $\JSL$ or $\Conv$: in these cases, $\funcS' =
\liftP$ is itself the weak lifting of a monad whose endofunctor and
multiplication are nearly cartesian. It turns out this is enough for $\funcS'$
itself to have nearly cartesian endofunctor and multiplication:

\begin{lem}
  \label{lem:weak-lifting-nearly-cartesian}
  Let $\catC$ be a regular category, and suppose $\funcT$ is nearly cartesian.
  If $\funcF: \catC \to \catC$ is nearly cartesian and weakly lifts to $\EMT$,
  then its weak lifting is nearly cartesian. If $\alpha: \funcF \Rightarrow
  \funcG$ between two such functors is nearly cartesian and weakly lifts to
  $\EMT$, then its weak lifting is nearly cartesian as well.
\end{lem}
\begin{proof}
  Suppose $\funcF$ is nearly cartesian, i.e., preserves near pullbacks, and that
  it has a weak lifting $\lift{\funcF}$ to $\EMT$. Then it also has a
  semi-lifting $\semilift{\funcF}$ to $\SemiEMT$, and $\funcF \rightSemiEMT =
  \rightSemiEMT \semilift{\funcF}$ is nearly cartesian as well. Since
  $\rightSemiEMT$ creates near pullbacks by \Cref{thm:algebras-regular},
  $\semilift{\funcF}$ is nearly cartesian. Now $\lift{\funcF} \cong \SplitT
  \semilift{\funcF} \SemiT$, but $\SplitT$ and $\SemiT$ are both left and right
  adjoints, hence they preserve both limits and colimits, and thus near
  pullbacks (regular epimorphisms are preserved as coequalizers of their kernel
  pairs). It follows that $\lift{\funcF}$ preserves near pullbacks.

  Suppose $\alpha: \funcF \Rightarrow \funcG$ is a nearly cartesian natural
  transformation between two such functors that has a weak lifting
  $\lift{\alpha}: \lift{\funcF} \Rightarrow \lift{\funcG}$ to $\EMT$. Then it
  also has a semi-lifting $\semilift{\alpha}$ to $\SemiEMT$, and $\alpha
  \rightSemiEMT = \rightSemiEMT \semilift{\alpha}$ is nearly cartesian. Since
  $\rightSemiEMT$ creates near pullbacks, $\semilift{\alpha}$ is nearly
  cartesian. Now $\lift{\alpha} \cong \SplitT \semilift{\alpha} \SemiT$, hence
  since $\SplitT$ and $\SemiT$ preserve near pullbacks $\lift{\alpha}$ is also
  nearly cartesian.
\end{proof}

In other words, if $\funcS$ has nearly cartesian endofunctor and multiplication,
then this is also the case of $\funcS' = \liftS$: $\liftS$ and $\multliftS$ have
relational extensions, and $\liftS$ admits a monotone weak distributive law over
$\liftR$ if and only if $\Rel[\liftS]$ restricts to $\KlliftR$. If $\catC =
\Set$, this means for instance:

\begin{cor}
  \label{cor:monotone-weak-distr-laws-lifted-powerset}
  Let $\funcT$ be a monad on $\Set$ equipped with a monotone weak distributive
  law $\distrTPtype$, and let $\funcS'$ be a monad on $\EMT$. If there is a
  monotone weak distributive law $\funcS' \liftP \Rightarrow \liftP \funcS'$
  (resp. $\funcS' \lift{\funcP_*} \Rightarrow \lift{\funcP_*} \funcS'$), $\funcS'$
  preserves open (resp. open surjective) $\funcT$-algebra
  morphisms. Moreover if $\funcS'$ is itself the weak lifting of a monad on
  $\Set$ that has a monotone weak distributive law over $\funcP$, then the
  previous condition is not only necessary, but also sufficient.
\end{cor}
\begin{proof}
  By \Cref{cor:kleisli-lifted-powerset}, open $\EMT$-relations form a
  wide subcategory of all $\EMT$-relations, hence the conditions in
  \Cref{lem:pullback-stable-subcategory} hold for the class $\catGamma$ of
  open morphisms and we may thus apply
  \Cref{thm:partial-relational-extensions} to characterize monotone extensions
  to $\KlliftP$. In particular if $\funcS$ is the weak lifting of a monad on
  $\Set$ whose underlying endofunctor and multiplication are nearly cartesian,
  then this is also true of $\funcS$ itself by
  \Cref{lem:weak-lifting-nearly-cartesian}: the monotone weak distributive law
  $\liftS \liftP \Rightarrow \liftP \liftS$ exists if and only if $\liftS$
  preserves open morphisms.

  By \Cref{thm:kleisli-lifted-subpowerobject}, the Kleisli category of
  $\lift{\funcP_*}$ has for arrows $X \klarrow Y$ the $\funcT$-relations given
  as jointly monic spans $\langle \psi_X, \psi_Y \rangle: R \subto X \times Y$
  such that $\psi_X$ is open and $\rightEMT \psi_X$ is surjective.
  $\Kl{\lift{\funcP_*}}$ is a category, so these relations form a wide
  subcategory of all $\EMT$-relations, hence the conditions of
  \Cref{lem:pullback-stable-subcategory} hold and we may thus apply
  \Cref{thm:partial-relational-extensions} to characterize monotone extensions
  to $\Kl{\lift{\funcP_*}}$.
\end{proof}

\Cref{cor:monotone-weak-distr-laws-lifted-powerset} makes it much simpler to
check whether $\funcS'$ has a monotone weak distributive law over $\liftP$ or
$\lift{\funcP_*}$: there is no need to check that some near pullback squares are
preserved anymore.

\subsection{Monotone Weak Distributive Laws in \texorpdfstring{$\KHaus$}{KHaus}}
\label{sec:lifting-monotone-laws:khaus}

A first use of \Cref{cor:monotone-weak-distr-laws-lifted-powerset} lies in
retrieving the monotone weak distributive law $\distrVVtype$: the latter is
exhibited in \cite{goyPowersetLikeMonadsWeakly2021} by showing that $\funcV$ and
$\multV$ are nearly cartesian, and that $\Rel[\funcV]$ preserves continuous
relations. Thanks to \Cref{cor:monotone-weak-distr-laws-lifted-powerset}, it is
now only necessary to show that $\funcV$ preserves open morphisms.

\begin{thm}[{Vietoris weakly distributes over itself~\cite[Theorem 24]{goyPowersetLikeMonadsWeakly2021}}]
  \label{thm:vietoris-monotone-weak-distr-law}
  There are monotone weak distributive laws $\distrVVtype$ and
  $\distrVVstartype$.
\end{thm}

We are also able to answer quite easily \cite[Conjecture
7.31]{goyCompositionalityMonadsWeak2021} on the weak distributivity of the Radon
monad -- the monad of Radon probability measures on a compact Hausdorff space --
over the Vietoris monad:

\begin{thm}
  \label{thm:radon-monotone-weak-distr-law}
  The Radon monad $\Rad$ does not have a monotone weak distributive law over the
  Vietoris monad $\funcV$, but it has (a unique) one over the non-empty Vietoris
  monad $\funcV_*$.
\end{thm}

Before giving the proof, let us first recall how the Radon monad is defined.
Given a compact Hausdorff space $X$, let $C(X)$ denote the Banach space of
continuous functions $X \to \complices$ equipped with the uniform norm
($\norm{f} = \sup_{x \in X} \abs{f(x)}$). A continuous function $C(X) \to
\complices$ is called a \emph{functional}, and it is called \emph{positive} when
it sends functions with positive real domains to positive real numbers. The
space of functionals $C(X) \to \complices$ is equipped with the \emph{operator
  norm}, defined by $\norm{\varphi} = \sup_{\norm{f} = 1} \abs{\varphi(f)}$.

$\Rad$ sends a compact Haudorff space $X$ on the space $\Rad X$ of \emph{Radon}
probability measures on $X$, or equivalently by the Riesz-Markov-Kakutani
theorem \cite[Theorem 2.14]{rudinRealComplexAnalysis1987} to the space of
positive linear functionals $C(X) \to \complices$ with operator norm $1$: for
every such functional $\varphi$ there is thus a Radon measure $m$ on $X$ such
that $\varphi(f) = \int_X u\,\mathrm{d}m$ for any $f \in C(X)$, and conversely
integrating along a measure defines an operator with the aforementioned
properties. $\Rad X$ (seen as the space of functionals) is equipped with the
vague topology, i.e., the weakest topology that makes each $ev_f: \Rad X \to
\complices$ continuous, where $f$ ranges in $C(X)$ and $ev_f$ is the function
that evaluates an operator at $f$.

Given $f: X \to Y$ in $\KHaus$, $\varphi \in \Rad X$ and $g \in C(Y)$, $(\Rad
f)(\varphi)(g) = \varphi(g \circ f)$. $\unitRad(x) \in \Rad X$ is the operator
that evaluates a continuous function at $x \in X$, and for $\Phi \in \RadRad X$
and $f \in C(X)$, $\multRad(\Phi)(f) = \Phi(ev_f)$.

\begin{proof} \hfill
  \begin{description}
  \item[Radon over Vietoris] Consider the inclusion $f: 1 \to 2$. $1$ and $2$
    are sets equipped with the discrete topology, so that $f$ is an open
    continuous function. But $\Rad 1 = 1$ and $\Rad 2$ is the unit interval: a
    point is never open in the unit interval, hence $\Rad f$ cannot be open.
    $\Rad$ does not preserve open maps, hence by
    \Cref{cor:monotone-weak-distr-laws-lifted-powerset} it does not have a
    monotone weak distributive law over $\funcV$.
  \item[Radon over non-empty Vietoris] We use again that $\Kl{\funcV_*}$ is the
    category of relations, which, when seen as jointly monic spans $\langle
    \psi_X, \psi_Y \rangle$, are such that $\psi_X$ is both open and surjective.
    Because $\Kl{\funcV_*}$ is a category, these relations form a wide
    subcategory of the relations in $\KHaus$, hence the conditions in
    \Cref{lem:pullback-stable-subcategory} hold and
    \Cref{thm:partial-relational-extensions} applies: $\Rad$ is a nearly
    cartesian endofunctor -- it preserves regular epimorphisms
    \cite[Propositions 7.26]{goyCompositionalityMonadsWeak2021} and sends
    pullbacks on near pullbacks \cite[Proposition
    7.28]{goyCompositionalityMonadsWeak2021} -- and it preserves open
    surjections \cite[Theorem 4.4]{ditorOpenMappingTheorems1972}, therefore it
    has a monotone extension to $\Kl{\funcV_*}$.

    We now show that the naturality squares $\multRad_Y \circ \RadRad f = \Rad f
    \circ \multRad_X$ of $\multRad$ along surjections $f: X \onto Y$ are near
    pullbacks, so that $\multRad$ also extends to $\Kl{\funcV_*}$ and we get a
    weak distributive law $\distrRadVstartype$.

    Consider thus a continuous surjection $f: X \to Y$ in $\KHaus$, and recall
    that $\Rad f$ is also surjective \cite[Proposition
    7.26]{goyCompositionalityMonadsWeak2021} (in fact we will simply adapt the
    proof of preservation of surjectivity to show our result). Consider some
    $\Phi \in \RadRad Y$ and $\psi \in \Rad X$ such that $\multRad_Y(\Phi) =
    (\Rad f)(\psi)$, i.e., $\Phi(ev_g) = \psi(g \circ f)$ for every $g \in C(Y)$.

    Consider a first subspace $V_1$ of $C(\Rad X)$ consisting of continuous
    functions of the shape $g \circ \Rad f$ for $g \in C(\Rad Y)$: $g \circ \Rad
    f \mapsto \Phi(g)$ defines a linear functional $\Psi_1$ of norm $1$ on $V_1$
    (the definition does not depend on the choice of $g$ because there is only
    one such choice as $\Rad f$ is surjective).

    Similarly, consider a second subspace $V_2$ of $C(\Rad X)$ consisting of
    continuous functions of the shape $ev_g$ for some $g \in C(X)$: then $ev_g
    \mapsto \psi(g)$ defines a linear functional $\Psi_2$ of norm $1$ on $V_2$.

    If a continuous function in $C(\Rad X)$ is in $V_1 \cap V_2$, it is of the
    shape $ev_{g \circ f} = ev_g \circ \Rad f$ for some $g \in C(Y)$, and
    $\Phi(ev_g) = \psi(g \circ f)$ by assumption: $\Psi_1$ and $\Psi_2$ agree on
    $V_1 \cap V_2$. This defines a linear functional of operator norm $1$ on
    $V_1 + V_2$, which extends to a linear functional $\Psi$ of operator norm
    $1$ on $C(\Rad X)$ by the Hahn-Banach theorem~\cite[Theorem
    5.16]{rudinRealComplexAnalysis1987}. A linear functional of operator norm
    $1$ is positive if and only if it sends the constant function $1$ to $1 \in
    \complices$ (as argued for instance in~\cite[\textsection
    5.22]{rudinRealComplexAnalysis1987}), hence $ev_1 = 1$ on $\Rad X$ and thus
    $\Psi(1) = \Psi(ev_1) = \psi(1) = 1$: this implies in turn that $\Psi$ is
    positive.

    We thus found some $\Psi \in \RadRad X$ such that $(\RadRad f)(\Psi) = \Phi$
    and $\multRad_X(\Psi) = \psi$. Because $\rightEMUlt$ creates near pullbacks,
    this proves that the commuting square $\Rad f \circ \multRad_X = \multRad_Y
    \circ \RadRad f$ is a near pullback. \qedhere
  \end{description}
\end{proof}

Let us stress the importance of this new weak distributive law: the question of
how to combine probability and non-determinism has been the topic of numerous
works (again, see the introduction
of~\cite{keimelMixedPowerdomainsProbability2017}), and this law provides an
answer in $\KHaus$ that is derived from a generic construction and thus comes
with generic tools, e.g. generalized determinization and up-to techniques
\cite{goyCombiningProbabilisticNondeterministic2020,goyCompositionalityMonadsWeak2021}.
In the pre-print~\cite{goubault-larrecqWeakDistributiveLaws2024} (which appeared
online two weeks before \cite{aristoteMonotoneWeakDistributive2025} was
submitted for review), Goubault-Larrecq also constructs this law
$\distrRadVstartype$ as an instance of weak distributive laws between monads of
continuous valuations and non-deterministic choice in more general categories of
topological spaces: our result is more restricted, but we derive the law from
generic categorical principles instead of building it by hand, exhibit its
canonicity (it comes from a relational extension), and show why the non-empty
version of the Vietoris monad is needed.

\subsection{No-go Theorems for Monotone Weak Distributive Laws in Categories of
  Algebras}\label{sec:lifting-monotone-laws:nogo-theorems}

A second use of \Cref{cor:monotone-weak-distr-laws-lifted-powerset} is in
proving the absence of monotone weak distributive laws. In
\Cref{sec:lifting-laws} we were able to prove that the law $\distrPPtype$ does
not weakly lift to $\EMP$ nor $\EMD$, but we were not able to say anything about
the existence of other monotone weak distributive laws $\distrliftPliftPtype$.
Now, thanks to the framework developed above, we are able to prove that such
laws cannot exist.

\subsubsection{Complete join-semilattices.} We start with $\EMP$, where
\begin{itemize}
\item $\liftP$ is the monad of subsets closed under non-empty joins: the join of
  a family $(E_i)_{i \in I}$ of non-empty-joins-closed subsets of $X$ is the
  non-empty-joins-closed subset \[ \suchthat{\bigvee_{i \in I} x_i }{ \forall i
      \in I, x_i \in E_i } \]
\item $f: (X, \vee^X) \to (Y, \vee^Y)$ is open if and only if, for every $x \in
  X$ and $(y_i)_{i \in I}$ in $Y$ such that $f(x) = \bigvee^Y \suchthat{ y_i }{
    i \in I }$, there is some $(x_i)_{i \in I}$ in $X$ such that $f(x_i) = y_i$
  and $x = \bigvee^X \suchthat{ x_i }{ i \in I }$.
\end{itemize}

\begin{exa}
  \label{ex:no-go-monotone-weak-law:jsl}
  In $\AlgP \cong \JSL$, let $f: 4 \to 2$ (where $2 = \{0,1\}$ and $4 =
  \{0,1,2,3\}$) be the parity function given by $f(0) = f(2) = 0$ and $f(1) =
  f(3) = 1$. Then $\funcP f$ is an open morphism $(\funcP 4, \cup) \to (\funcP
  2, \cup)$ (by \Cref{lem:free-morphisms-are-open}), but $\liftP \funcP f$ is
  not.
\end{exa}
\begin{proof}
  Let
  \[ x \coloneq \{ \{0,1\}, \{0,1,2,3\} \} \subseteq \funcP 4 \qquad y \coloneq
    \{ \{ 0, 1 \} \} \subseteq \funcP 2 \]The two subsets $x$ and $y$ are both closed under non-empty unions, hence are
  respectively elements of $\liftP \funcP 4$ and $\liftP \funcP 2$, and
  satisfy that $(\liftP \funcP f)(x) = y$.

  Let moreover
  \[ y_0 \coloneq \{\{ 0 \}\} \subseteq \funcP 2 \qquad y_1 \coloneq \{\{ 1 \}\}
    \subseteq \funcP 2 \]
  These are again elements of $\liftP \funcP 2$, and they satisfy that
  \[ y = y_0 \vee y_1 \]

  If $\liftP \funcP f$ is open, then there are some $x_0, x_1 \in \liftP \funcP
  X$ such that $x = x_0 \vee x_1$ as well as both
  $(\liftP \funcP f)(x_0) = y_0$ and $(\liftP \funcP f)(x_1) = y_1$. We
  prove by contradiction that there are no such $x_0$ and $x_1$.

  Suppose indeed $x_0$ and $x_1$ exist. Then $(\liftP \funcP f)(x_0) = y_0$
  implies that $x_0 \subseteq \{ \{0\}, \{2\},\allowbreak \{0,2\} \}$, and
  $(\liftP \funcP f)(x_1) = y_1$ implies that $x_1 \subseteq \{ \{1\}, \{3\},
  \{1,3\} \}$. Because $\{0,1\} \in x = x_0 \vee x_1$, it follows that $\{0\}
  \in x_0$ and $\{1\} \in x_1$. Because $\{0,1,2,3\} \in x = x_0 \vee x_1$, it
  follows that $\{0,2\} \in x_0$ and $\{1,3\} \in x_1$. Hence $\{0,1,3\} = \{0\}
  \cup \{1,3\} \in x_0 \vee x_1 = x$, which is a contradiction.
\end{proof}

$\liftP$ does not preserve open $\funcP$-algebra morphisms, and thus there is no
monotone weak distributive law $\distrliftPliftPtype$. In fact because the
counter-example open morphism is surjective and has finite pre-images, this also
proves that there are no monotone weak distributive laws $\liftP \lift{\funcP_*}
\Rightarrow \lift{\funcP_*} \liftP$ or $\liftP \lift{\funcP_f} \Rightarrow
\lift{\funcP_f} \liftP$ in $\EMP$.

\subsubsection{Convex spaces.} Next, we consider the case of $\EMD$, where
\begin{itemize}
\item $\liftP$ is the monad of subsets closed under convex combinations: in a
  barycentric algebra $X$, for every $p \in [0,1]$ and $E, E' \in \liftP X$, $p
  E + (1-p) E' = \suchthat{ p x + (1-p) y }{ x \in E, y \in E' }$.
\item $f: X \to Y$ is open if and only if, for every $x \in X$ and $p \in \funcD
  \{1, \ldots, n\}$ and $y_1, \ldots, y_n \in Y$ such that $f(x) = \sum_{i =
    1}^n p(i) y_i$, there are some $x_1, \ldots, x_n \in X$ such that $f(x_i) =
  y_i$ and $\sum_{i = 1}^n x_i = x$.
\end{itemize}

Because there is a morphism of monads $\funcD \Rightarrow \funcP$ (that sends a
probability distribution to its support), there is a functor $\EMP \to \EMD$
which allows us to transfer \Cref{ex:no-go-monotone-weak-law:jsl} to $\EMD$:
there are no monotone weak distributive laws $\distrPPtype$ or $\liftP
\lift{\funcP_*} \Rightarrow \lift{\funcP_*} \liftP$ in $\EMD$. But this argument
is not entirely satisfying, as the resulting example is that of a morphism of
convex algebras with very unnatural structures, namely that of complete
join-semilattices: one could imagine restricting to a full subcategory of $\EMD$
that does not contain these semilattices, and perhaps $\liftP$ would preserve
open morphisms there. As shown by the following example, due to Harald Woracek
and Ana Sokolova (private communication), this cannot be the case as soon as
free convex algebras come in the picture.

\begin{figure}[h]
  \centering
  \caption{A counterexample to preservation of openness in $\Conv$. }
\label{fig:counterexamples:convex-algebras}
    \includegraphics{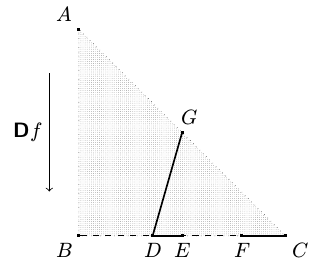}
\end{figure}

\begin{exa}[{\cite{woracekReDecomposingConvex2024}}]
  \label{ex:no-go-monotone-weak-law:conv}
  In $\EMD \cong \Conv$, let $\liftP$ be the monad of convex subsets: a convex
  combination of some convex subsets of $X$ is the convex set of the
  corresponding convex combinations of their points (in $X$). Let $f: \{A,B,C\}
  \to \{B,C\}$ be the function given by $f(A) = f(B) = B$ and $f(C) = C$. Then
  $\leftEMD f$ is open (by \Cref{lem:free-morphisms-are-open})
  but $\liftP \leftEMD f$ is not.
\end{exa}
\begin{proof}
  Indeed, depicting $\leftEMD \{A,B,C\}$ as the triangle depicted in
  \Cref{fig:counterexamples:convex-algebras}
  (page~\pageref{fig:counterexamples:convex-algebras}), $\leftEMD \{B,C\}$ is
  the line segment $[BC]$ and $\leftEMD f$ is the vertical projection. Now
  $\frac{1}{2} \{B\} + \frac{1}{2} [FC] = [DE] = \left(\liftP \leftEMD
    f\right)([GD])$, but $[GD]$ itself cannot be disintegrated as the mean of
  two convex subsets of $ABC$, one above $B$ and the other above $[FC]$: if such
  a disintegration existed, then the convex subset above $B$ would contain both
  $B$ (because $D \in [DG]$) and $A$ (because $G \in [DG]$), hence would be
  $[AB]$. The subset above $[FC]$ would contain at least one point, and hence
  the mean of these two subsets would have to contain a non-trivial vertical
  line segment, which is not the case of $[DG]$.
\end{proof}

\subsubsection{Et c\ae{}tera.}

Using similar arguments, we are able to prove the existence or absence of
monotone weak distributive laws over lifted powerset monads in several
categories of algebras: these results are gathered in
\Cref{tab:monotone-weak-distr-laws-weakly-lifted-powerset-algebras}. All the
negative results in this table come from the non-preservation of open morphisms,
and thus the absence of monotone extensions of the endofunctors themselves. The
topmost row indicates in which category we work. A monad in the second topmost
row has a monotone weak distributive law over a monad in the left column if the
corresponding cell is filled with \cmark{} --- otherwise it is filled with
\xmark{}. In $\Set$, $\liftP$ and $\lift{\funcP_*}$ are the usual powerset
monads $\funcP$ and $\funcP_*$. $\funcL$ is the monad of lists, $\funcM$ that of
multisets and $\Ms$ that of modules for a semiring $S$ satisfying the conditions
of \cite[Theorem 3.1]{bonchiConvexityWeakDistributive2022}. $\Mon \cong \EML$ is
the category of monoids and $\CMon \cong \EMM$ that of commutative monoids.
Linear theories distribute over commutative monads
\cite{manesMonadCompositionsGeneral2007}, hence $\funcL$ and $\funcM$ distribute
over $\funcM$, $\funcP$, $\funcD$ and $\Ms$ when $S$ is commutative, and so
these four monads have liftings (in particular weak liftings) to $\Mon$ and
$\CMon$.

\begin{table}[h]
  \centering
  \begin{tabular}{| c | c c c c c c | c c | c | c | c c c c | c c c | }
    \hline
    & \multicolumn{6}{c|}{$\Set$} & \multicolumn{2}{c|}{$\KHaus$} & $\JSL$ & $\Conv$ & \multicolumn{4}{c|}{$\Mon$} & \multicolumn{3}{c|}{$\CMon$} \\
    \cline{2-18}
    & $\funcL$ & $\funcM$ & $\funcD$ & $\funcP$ & $\Ult$ & $\Ms$ & $\funcV$ & $\Rad$ & $\liftP$ & $\liftP$ & $\liftM$ & $\liftD$ & $\liftP$ & $\liftMs$ & $\liftM$ & $\liftD$ & $\liftP$ \\
    \hline
    $\liftP$ & \cmark & \cmark & \cmark & \cmark & \cmark & \cmark & \cmark & \xmark & \xmark & \xmark & \xmark & \xmark & \xmark & \xmark & \xmark & \xmark & \xmark \\
    $\lift{\funcP_*}$ & \cmark & \cmark & \cmark & \cmark & \cmark & \cmark & \cmark & \cmark & \xmark & \xmark & \xmark & \xmark & \xmark & \xmark & \xmark & \xmark & \xmark \\
    \hline
  \end{tabular}
  \vspace{10pt}
  \caption{Existence or absence of monotone weak distributed laws over weakly
    lifted powerset monads in categories of algebras.}
  \label{tab:monotone-weak-distr-laws-weakly-lifted-powerset-algebras}
\vspace{-10pt}
\end{table}

\begin{proof}[Proof of \Cref{tab:monotone-weak-distr-laws-weakly-lifted-powerset-algebras}]
  The cases of $\KHaus$, $\JSL$ and $\Conv$ have already been treated in
  \Cref{thm:radon-monotone-weak-distr-law,ex:no-go-monotone-weak-law:jsl,ex:no-go-monotone-weak-law:conv}.
\begin{description}
\item[In $\Mon$] By definition, a morphism of monoids $f: A \to B$ is open if
  and only if for every $f(a) = b_1 \cdots b_n$ with $a \in A$ and $b_i \in B$,
  there are some $a_1, \ldots, a_n \in A$ such that $f(a_i) = b_i$ and $a = a_1
  \cdots a_n$.
  \begin{description}
  \item[$\distrPPtype$] Here $\funcP = \liftP$ is the monad of subsets of
    monoids: if $A$ is a monoid with unit $e$, $\funcP A$ is a monoid with unit
    $\{ e \}$ and multiplication $\alpha_1 \alpha_2 = \{ a_1 a_2 \mid a_i \in
    \alpha_i \}$.

    Consider the constant function $f : \{a,b\} \rightarrow \{a\}$. It extends
    to a monoid morphism $f^* : \{a,b\}^* \rightarrow \{a\}^*$. Let $\beta = \{
    aa \}$, $\beta_1 = \beta_2 = \{ a \}$ and $\alpha = \{ ab, ba \}$: $\beta =
    \beta_1 \beta_2$ and $\beta = f^*[\alpha]$. Suppose $\alpha_1, \alpha_2 \in
    \liftP(\{a,b\}^*)$ are such that $f^*[\alpha_i] = \beta_i$. Then all the
    words in $\alpha_i$ must have length $1$. Hence if $\alpha_1 \alpha_2
    \supseteq \alpha$, $\alpha_1$ and $\alpha_2$ must both contain $a$ and $b$,
    which implies that $\{ aa, bb \} \subset \alpha_1 \alpha_2 \neq \alpha$.

  \item[$\distrMsPtype$] Let $S$ be a commutative (unital) semiring. Here $\Ms$
    is the monad of $S$-linear combinations: if $A$ is a monoid with unit $e$,
    $\Ms A$ is a monoid with unit $1 \cdot e$ and multiplication $\left(\sum_i
      \alpha_i \cdot a_i\right)\left(\sum_j \beta_j \cdot b_j\right) =
    \sum_{i,j} \alpha_i \beta_j \cdot a_i b_j$.

    The counter-example for $\funcP$ can be adapted: $(1 \cdot a)((1 + 1) \cdot
    a) = (1 + 1) \cdot aa = f(1 \cdot ab + 1 \cdot ba)$. If there are $\alpha_1,
    \alpha_2 \in \Ms\left(\{a,b\}^*\right)$ such that $(\Ms f^*)(\alpha_1) = 1
    \cdot a$, $(\Ms f^*)(\alpha_2) = (1+1) \cdot a$ and $\alpha_1 \alpha_2 = 1
    \cdot ab + 1 \cdot ba$, then $\alpha_1 = w \cdot a + x \cdot b$ and
    $\alpha_2 = y \cdot a + z \cdot b$ where $w,x,y,z \in S$ are such that $w +
    x = 1$, $y + z = 1 + 1$, $wy = 0$, $wz = 1$, $xy = 1$ and $xz = 0$. This
    implies that $z = (w+x)z = wz + xz = 1$ and similarly $y = (w+x)y = wy + xy
    = 1$, hence that $w = 0$ and $x = 0$ and thus $1 = w + x = 0$: $S$ is the
    trivial semiring and $\Ms$ is the constant monad $\{ 0 \}$.

  \item[$\distrMPtype$] This is the instance $S = \naturals$ of the case
    $\distrMsPtype$.

  \item[$\distrDPtype$] Here $\funcD = \liftD$ is the monad of finitely
    supported probability distributions on a monoid: if $A$ is a monoid with
    unit $e$, $\funcD A$ is a monoid with unit $\delta_e$ and multiplication
    $(\alpha_1 \alpha_2)(a) = \sum_{a_1 a_2 = a} \alpha(a_1) \beta(a_2)$.

    The counter-example for $\funcP$ can be adapted: \[ \delta_{aa} = \delta_a
    \delta_a = (\funcD f)(\delta_{ab}/2 + \delta_{ba}/2) \]
  \end{description}

\item[In $\CMon$]
  By definition, a morphism of commutative
  monoids $f: A \to B$ is open if and only if for every $f(a) = m_1 b_1
  + \cdots + m_n b_n$ with $a \in A$, $b_i \in B$ and $m_i \in \naturals$, there
  are some $a_1, \ldots, a_n \in A$ such that $f(a_i) = b_i$ and $a = m_1 a_1 +
  \cdots + m_n a_n$.

  \begin{description}
  \item[$\distrPPtype$] Here $\funcP = \liftP$ is the monad of subsets of
    monoids: if $A$ is a commutative monoid with unit $0$, $\funcP A$ is a
    monoid with unit $\{ e \}$ and addition $\alpha_1 + \alpha_2 = \{ a_1 + a_2
    \mid a_i \in \alpha_i \}$.

    Consider the constant function $f : \{a,b\} \rightarrow \{a\}$. It extends
    to a commutative monoid morphism $f^{\circledast} : \{a,b\}^{\circledast}
    \rightarrow \mathbb{N}$ (where $\Sigma^\circledast$ is the free commutative
    monoid on $\Sigma$). Let $\beta = \{ 2a \}$, $\beta_1 = \beta_2 = \{ a \}$
    and $\alpha = \{ 2a, 2b \}$: $\beta = \beta_1 + \beta_2$ and $\beta =
    f^{\circledast}[\alpha]$. Suppose $\alpha_1, \alpha_2$ are such that
    $f^{\circledast}[\alpha_i] = \beta_i$. Then all the words in $\alpha_i$ must
    have length $1$. Hence if $\alpha_1 + \alpha_2 = \alpha$, both of them must
    contain $a$ and $b$: then $a + b \in \alpha_1 + \alpha_2$ and $\alpha_1 +
    \alpha_2 \neq \alpha$.

  \item[$\distrMPtype$] Here $\funcM = \liftM$ is the monad of (finite)
    multisets, also written as finite $\naturals$-linear combinations: if $A$ is
    a commutative monoid with unit $0$ and addition $\oplus$, $\funcM A$ is a
    commutative monoid with unit $1 \cdot 0$ and addition $\left(\sum_i \alpha_i
      \cdot a_i\right) \oplus \left(\sum_j \beta_j \cdot b_j\right) = \sum_{i,j}
    \alpha_i \beta_j \cdot (a_i \oplus b_j)$.

    The counter-example for $\funcP$ can be adapted: $(1 \cdot a) \oplus (2
    \cdot a) = 2 \cdot (a \oplus a) = (\funcM f^{\circledast})(1 \cdot (a \oplus
    a) + 1 \cdot (b \oplus b))$. If there are $\alpha_1, \alpha_2 \in \funcM
    \{a,b\}^{\circledast}$ such that $(\funcM f^{\circledast})(\alpha_1) = 1
    \cdot a$, $(\funcM f^{\circledast})(\alpha_2) = 2 \cdot a$ and $\alpha_1
    \oplus \alpha_2 = 1 \cdot (a \oplus a) + 1 \cdot (b \oplus b)$, then
    $\alpha_1 = x_a \cdot a + x_b \cdot b$ and $\alpha_2 = y_a \cdot a + y_b
    \cdot b$ where $x_a,x_b,y_a,y_b \in S$ are such that $x_a + x_b = 1$, $y_a +
    y_b = 1 + 1$, $x_a y_a = x_b y_b = 1$ and $x_a y_b + x_b y_a = 0$. These
    equations do not have common solutions in $\naturals$, and so such
    $\alpha_1$ and $\alpha_2$ cannot exist.

  \item[$\distrDPtype$] Here $\funcD = \liftD$ is the monad of
    finitely-supported probability distributions on a monoid: if $A$ is a
    commutative monoid with unit $0$, $\funcD A$ is a commutative monoid with
    unit $\delta_0$ and addition $(\alpha_1 + \alpha_2)(a) = \sum_{a_1 + a_2 =
      a} \alpha(a_1) \beta(a_2)$.

    The counter-example for $\funcP$ can be adapted: \[ \delta_{2a} = \delta_a +
    \delta_a = (\funcD f)(\delta_{2a}/2 + \delta_{2b}/2) \qedhere \]
  \end{description}
\end{description}
\end{proof}

\section{Conclusion}
\label{sec:conclusion}

Noticing the similarity between the laws $\distrVVtype$ and $\distrPPtype$, we
developed the theory for weakly lifting weak distributive laws and showed that
it only applied partially in this case. We then focused on the monotonicity of
the laws, and gave full characterizations for the existence of monotone weak
distributive laws over weakly lifted powerset monads in categories of algebras,
by characterizing the Kleisli categories of the latters -- a key notion
appearing then being that of open morphisms. We finally applied this
result to exhibit a new law $\distrRadVstartype$ for combining probability and
non-determinism in $\KHaus$, but also to show that in general these monotone
weak distributive laws seem to be quite rare.

We leave for further work the development of a full 2-categorical theory for
iterating weak distributive laws (in the vein of~
\cite{chengIteratedDistributiveLaws2011,bohmIterationWeakWreath2012}), which
would complete \Cref{sec:lifting-laws} but would likely be scarce in new
examples. With monotone laws over powerset-like monads fully characterized,
another natural question is now whether this can be done in other settings, e.g.
in $\Pos$-regular categories~\cite{kurzQuasivarietiesVarietiesOrdered2017} or
over other monads: the multiset monad for instance is a good candidate as its
Kleisli category can also be described through spans. Finally, the author
believes it would be interesting to tranpose the results of
\Cref{sec:lifting-monotone-laws} in the setting of monoidal
topology~\cite{hofmannMonoidalTopologyCategorical2014}, where categories of
algebras for nearly cartesian monads generalize the category of compact
Hausdorff spaces in a formal sense: perhaps for instance the weakly lifted
powerset monads we study enjoy some properties that are true of the Vietoris
monads of topological spaces.

\bibliographystyle{alphaurl}
\bibliography{bibtex}

\newpage

\end{document}